\documentclass[12pt,a4paper]{article}

\usepackage{geometry}
\usepackage{amsmath}
\usepackage{amsfonts}
\usepackage{amssymb}
\usepackage{xcolor}
\usepackage{graphicx}
\DeclareGraphicsExtensions{.pdf, .jpg, .jpeg, .png}
\graphicspath{{.}}
\usepackage{hyperref}
\hypersetup{
	colorlinks=true,
	linkcolor=blue,
	citecolor=blue,
	urlcolor=red
                   }
\usepackage{subfigure}
\usepackage{topcapt}
\newcounter{foo}
\date{}

\begin{document}

\title{A physics-driven study of dominance space in soccer}
\author{Costas J. Efthimiou$^1$\footnote{costas@physics.ucf.edu, costasatucf@gmail.com},
Gregory DeCamillis$^1$\footnote{gregdecamillis@knights.ucf.edu}, Indranil Ghosh$^2$\footnote{i.ghosh@massey.ac.nz}  \\
\small $^1$Department of Physics, University of Central Florida,\\
\small                 Orlando, FL 32816, United States \\
\small $^2$School of Mathematical and Computational Sciences,\\
\small                   Massey University, Palmerston North, New Zealand}
\maketitle

\begin{abstract}
In \cite{efthimiou}, the concept of the  Voronoi diagram was investigated closely
from a theoretical point of view. Then, a physics-driven kinematical method was introduced to produce an improved model for dominance space in soccer. Contrary to other similar attempts, the model maintains the deterministic nature of the Voronoi diagram by considering the mechanism behind the concept, thus forgoing any probabilistic notions. The author coined the term \textit{soccerdynamics} for the interdisciplinary area (the overlap of physics and sports science) which builds models for soccer performance using physical laws.

With few exceptions, probabilistic models make arbitrary assumptions in order to fit with previously studied distributions,  bypassing poorly understood mechanisms in the sport. Consequently, even the best fitted data have limited applicability in the soccer environment.
Hence, there is no known performance index which can predict the outcome of a game, or even provide a reliable high probability for the outcome.
Remaining faithful to the deterministic approach, we extend  the work of \cite{efthimiou}  by the introduction of (a) an asymmetric influence of the players in their surrounding area, (b)  the frictional forces to the players' motion, and (c) the simultaneous combination of both effects.
The asymmetric influence is fairly intuitive;  players have more control in the direction they are running than any other direction. The sharper the turn they must make to reach a point on the pitch, the weaker their control of that point will be. From simple kinematical laws, this effect can be quantified explicitly.
For the frictional force, a portion comes from air resistance, and so will be proportional to $v^2$, where $v$ is the velocity of the player, as is well known from fluid dynamics. There are no other external frictional forces, but, at the suggestion of
biokinematics,  there is an internal frictional force, relating to the consumption of energy by the muscles, which is proportional to $v$.

Although these additions are intuitively understood, mathematically they introduce many analytical complexities. We establish exact analytical solutions of the dominance areas of the pitch by introducing a few reasonable simplifying assumptions. Given these solutions the new Voronoi diagrams are drawn for the publicly available data  by Metrica Sports.
In general, it is not necessary anymore for the dominance  regions to be convex, they might contain holes, and may be disconnected. The fastest player may dominate points far away from the rest of the players.

\end{abstract}

\section{Introduction}

In the article \cite{efthimiou},  a theoretical study to construct a reasonable model for allocating dominance space to each player during a soccer match
was initiated. There was one fundamental requirement in the quest for the model: it had to be driven by the formulas which are relevant to the underlying dynamics of the activity. This is a common practice in natural phenomena but it is often conveniently ignored by other practitioners --- mathematicians and computer scientists --- who often build models by fitting arbitrary curves to data. In order to derive a meaningful curve fitting, the dynamical principles which generate the observed behavior must necessarily be used. Hence, in \cite{efthimiou}, starting with the kinematical equations which describe the motion of the players, we were able to conclude that the standard Voronoi diagram is not the accurate concept for computing dominance space. Instead, it must be replaced by a compoundly weighted Voronoi diagram which takes into account the maximal speeds that can be reached by the players, as well as their time delays in response to an event.

Despite the progress in the original article, there are many additional details which must be understood. In particular, two assumptions stand out as the most important to be removed: for each player, the model assumed an isotropic coverage of space and constant acceleration to a maximum speed. However, it is intuitively obvious that players already in motion find it easier to challenge the ball on points belonging in the `forward cone'  of their motion and find it more difficult to challenge the ball at points which lie inside the `backward cone' of  their motion. Also, we know that players do not maintain a constant acceleration while sprinting. A drag force from the air will resist players' acceleration proportionally to the square of their speed \cite{Hill}. Also, there will be some internal friction limiting acceleration within the player, as even in the absence of air drag the players will reach a maximum speed. At the suggestion of biomechanics \cite{Furu}, we take this to be proportional to speed. Those two effects --- directional bias and friction --- are added to the model in this paper.

With the improved model, we are able to create diagrams which more accurately depict the dominance regions of players. These diagrams often carry conceptually interesting aspects which are not present in the standard Voronoi model. These include, but are not limited to, disconnected and non-convex regions for individual players.

The new additions to the model further demonstrate an appeal of approaching soccer analysis with a theoretical mind set: The assumptions of the model are physically founded.
They are neither hypothetical probabilistic guesses, nor designed to produce a certain result. Instead, they are designed to accurately depict the basic actions of soccer. Further research into the abilities of individual players or the interactions between players can be naturally included into the model, and so the model can match the continual growth in our understanding of soccer.
This further refinement of the assumptions will lead to increasingly complex and enlightening models of the sport.

It is important to stress that our results are novel and quite distinct from other similar attempts.
Taki and Hasegawa had been thinking about the player dominance region in soccer as far back as 1996. In 2000, they proposed \cite{TH} what is probably the first attempt to quantify the dominance area with the player's motion taken into account. Their proposal has the same motivation as our original article \cite{efthimiou} but, unfortunately, besides stating that the dominance region should be based on a shortest time principle and proposing two possibilities for the time-dependence of the player's acceleration, the authors do not work out an explicit analytical model. In particular, their proposals for the acceleration are as follows: (a) It has the same value in all directions and (b) the acceleration is isotropic at low speeds but as the speed of the player increases, the acceleration becomes anisotropic.
Taki and Hasegawa applied their ideas to scenes of games that had been recorded by conventional cameras. Hence, they were lacking precise tracking data and thus accurate Voronoi diagrams. However, for the data they had,  they have done a decent job and motivated many other researchers (including us) to seriously think about the problem.\footnote{For a summary of what the Voronoi diagrams are and what they are not, as well as a review of related efforts, please see \cite{efthimiou}.}
 One important deficiency in Taki and Hasegawa's idea is that allows the players to reach unreasonable high speeds if we allow time to run long enough. In the models of \cite{efthimiou}, this difficulty was bypassed by arguing that the players reach a target point with their highest speed (characteristic speed).

Our models include several parameters. Two parameters already present in \cite{efthimiou}, are the maximal speeds
(characteristic speeds) and the reaction time delays of the players. Unfortunately, we have not found any source which records them for each player. EA Sports' FIFA game
rates the players based on these (and other data) but it provides no exact values. Due to this drawback, in this paper we set all time delays to zero. However, we do not do the same for the characteristic speeds. Although the same difficulty exists for them, there is extensive literature discussing maximal speeds of soccer players without identifying the players explicitly. In particular, the paper \cite{Djaoui} reports maximal speeds per position. Hence, in our paper, we adopt that wing defenders have a characteristic speed of $8.62 \text{m/s}$, central defenders have a characteristic speed of $8.50 \text{m/s}$ , wing midfielders have a characteristic speed of $8.76 \text{m/s}$, central midfielders have a characteristic speed of $8.58 \text{m/s}$, and forwards have a characteristic speed of $8.96
\text{m/s}$. We were unable to find any studies of the maximal speeds of goalkeepers, so we took their characteristic speed to be $8.25 \text{m/s}$. Although this is not optimal, it provides a good approximation and a step forward.

To produce pitch diagrams based on our models,
we have used the open data set (both tracking and event data) provided by Metrica Sports \cite{Metrica}. The data consists of three games in the standard CSV format set by Metrica. There are no references to the players' names, teams or competitions. In each axis the positioning data goes from 0 to 1 with the top left corner of the pitch having coordinates $(0, 0)$, the bottom right corner having coordinates $(1, 1)$ and the kick off point having coordinates $(0.5, 0.5)$. For our presentation, the pitch dimensions have been set to $105\,\text{m} \times 68\,\text{m}$. Finally, the tracking and the event data are synchronized. We have used Python 3+ for our analysis.

\section{Construction of the Dominance Regions}

Consider two players P$_1$ and P$_2$ on an infinite plane and let $r_1, r_2$ be their respective  distances from any point P on the plane.
Also, let $\vec v_1, \vec v_2$ be the initial velocities and $\vec V_1, \vec V_2$ be the final velocities when they reach the point P.
As in \cite{efthimiou}, we will assume that the players, when they compete to reach the point P, place their maximum effort to reach
it with their characteristic speeds. Finally, we assume that each player reacts at different time to make a run towards the point P. Player one reacts at time $t_1$ and player two at time $t_2$ after a `signal' was given. Depending on the specific details for each player's motion, the points that each can reach will be given by the functions $r_1=r_1(t,t_1)$ and $r_2=r_2(t,t_2)$.

Our goal is to determine the locus of all points which the two players can reach simultaneously,
$$
      r_1(t,t_1) = r_2(t,t_2),
$$
for any later time $t$ after the `signal'. Knowing all such loci for the 22 players on the pitch allows us to construct the dominance areas (Voronoi regions) of the players.
%

{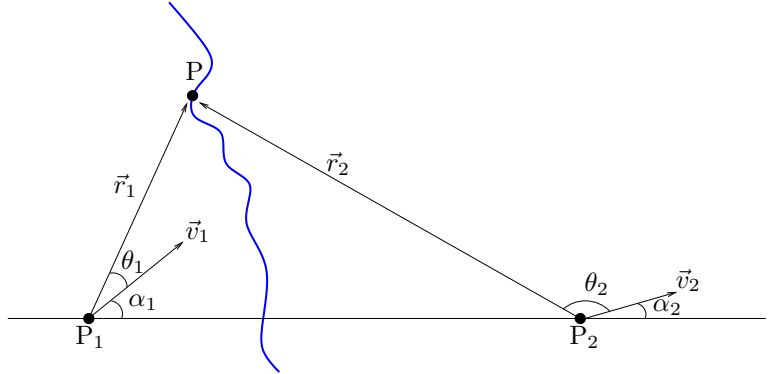
\begin{figure}[h!]
\centering
\setlength{\unitlength}{1mm}
\begin{picture}(100,45)
\put(0,-3){\includegraphics[width=10cm]{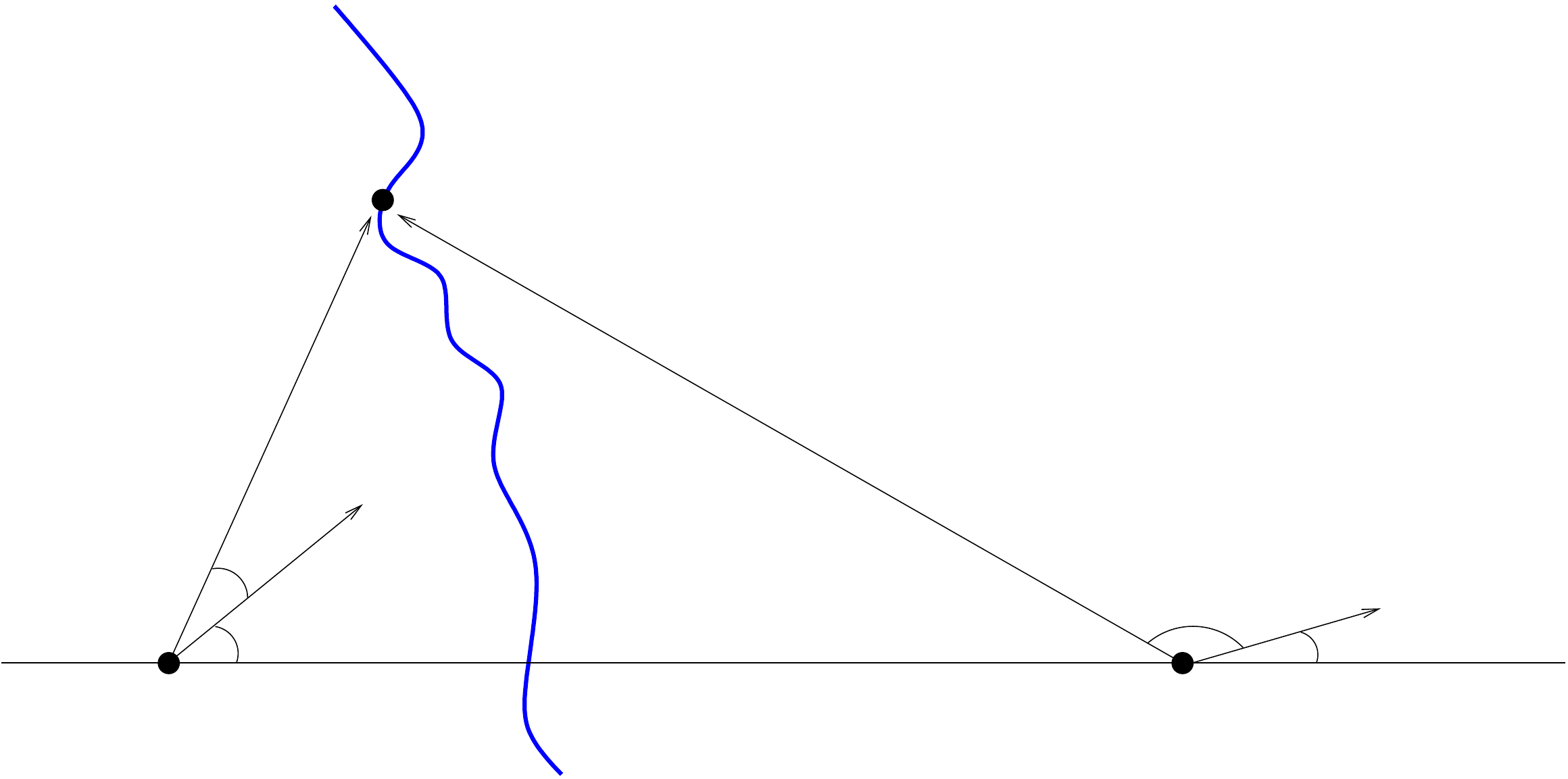}}
\put(42,24){\footnotesize $\vec r_2$}
\put(14,21){\footnotesize $\vec r_1$}
\put(88,8.5){\footnotesize $\vec v_2$}
\put(23.5,15){\footnotesize $\vec v_1$}

\put(23.5,36){\footnotesize P}
\put(9,1){\footnotesize P$_1$}
\put(74,1){\footnotesize P$_2$}

\put(85,5.4){\footnotesize $\alpha_2$}
\put(16,6){\footnotesize $\alpha_1$}

\put(76,8){\footnotesize $\theta_2$}
\put(15,11){\footnotesize $\theta_1$}
\end{picture}
\caption{\footnotesize Bipolar coordinates for the calculation of the borderline of the two Voronoi regions. Player 1 uses the polar coordinates $(r_1,\theta_1)$ while player 2 uses the polar coordinates $(r_2,\theta_2)$. The polar angles $\theta_1, \theta_2$ are computed from the corresponding directions of motion $\vec v_1$ and $\vec v_2$. These directions do not necessarily coincide with the line P$_1$P$_2$ joining the two players.}
\label{fig:BipolarCoordinates}
\end{figure}}

\subsection*{\normalsize Past Results}

Although the  standard Voronoi diagram was not constructed with motion in mind, its adoption is equivalent to the assumption that all players
can react without any delay and run with uniform speed V along any direction. Hence $r(t)=V\, t$ for all players. Consequently, the boundary
 between any two players is given by the equation
$$
     {r_1(t)\over r_2(t)} = 1
$$
and it is therefore the perpendicular bisector of the segment that joins the two players. Examples of  the standard Voronoi diagrams
are shown in Figure \ref{fig:standardVoronoi}.
%
{\begin{figure}[h!]
    \centering
    \subfigure[Frame: 98202 ]{\includegraphics[width=12cm]{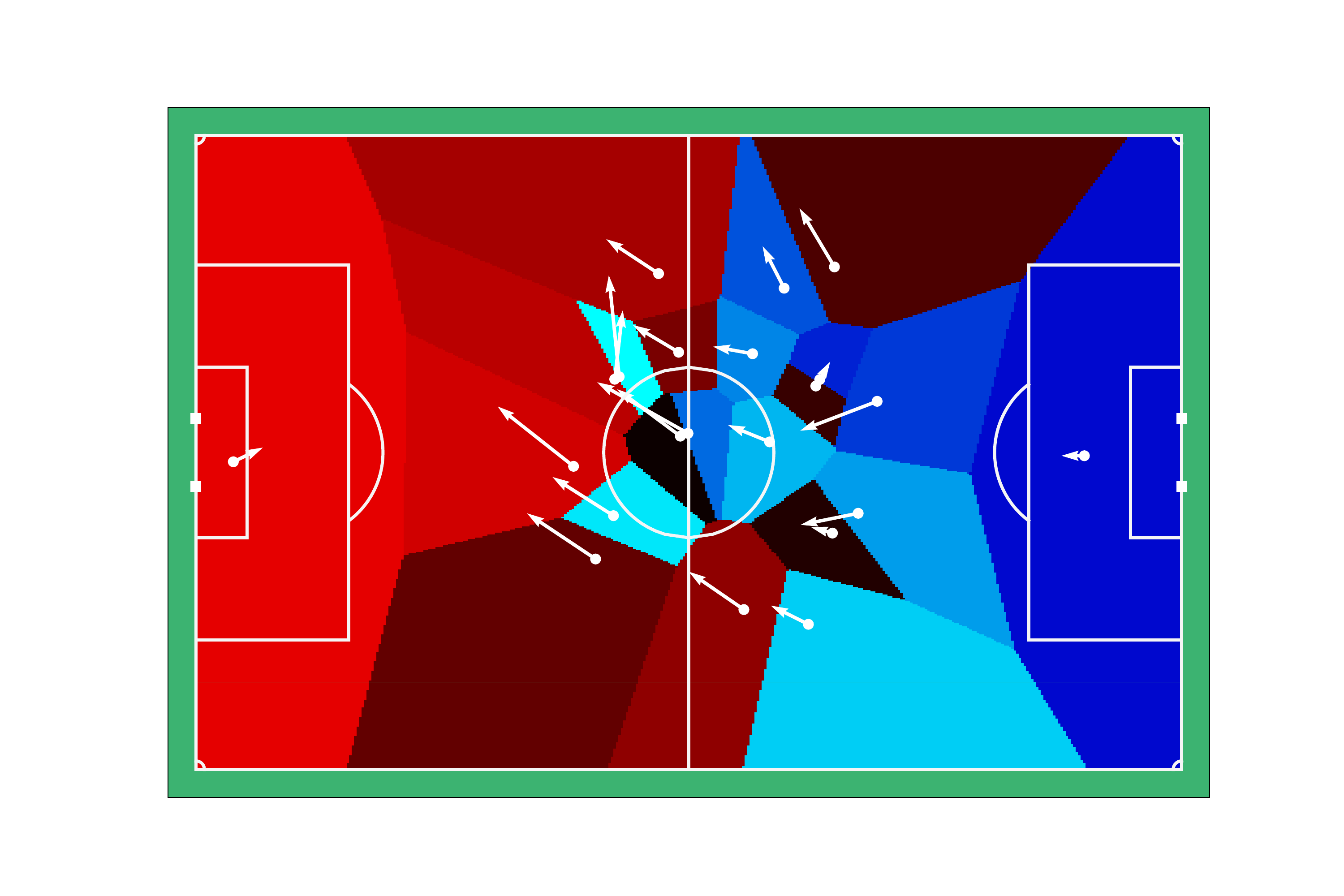}}\\
    \subfigure[Frame: 123000]{\includegraphics[width=12cm]{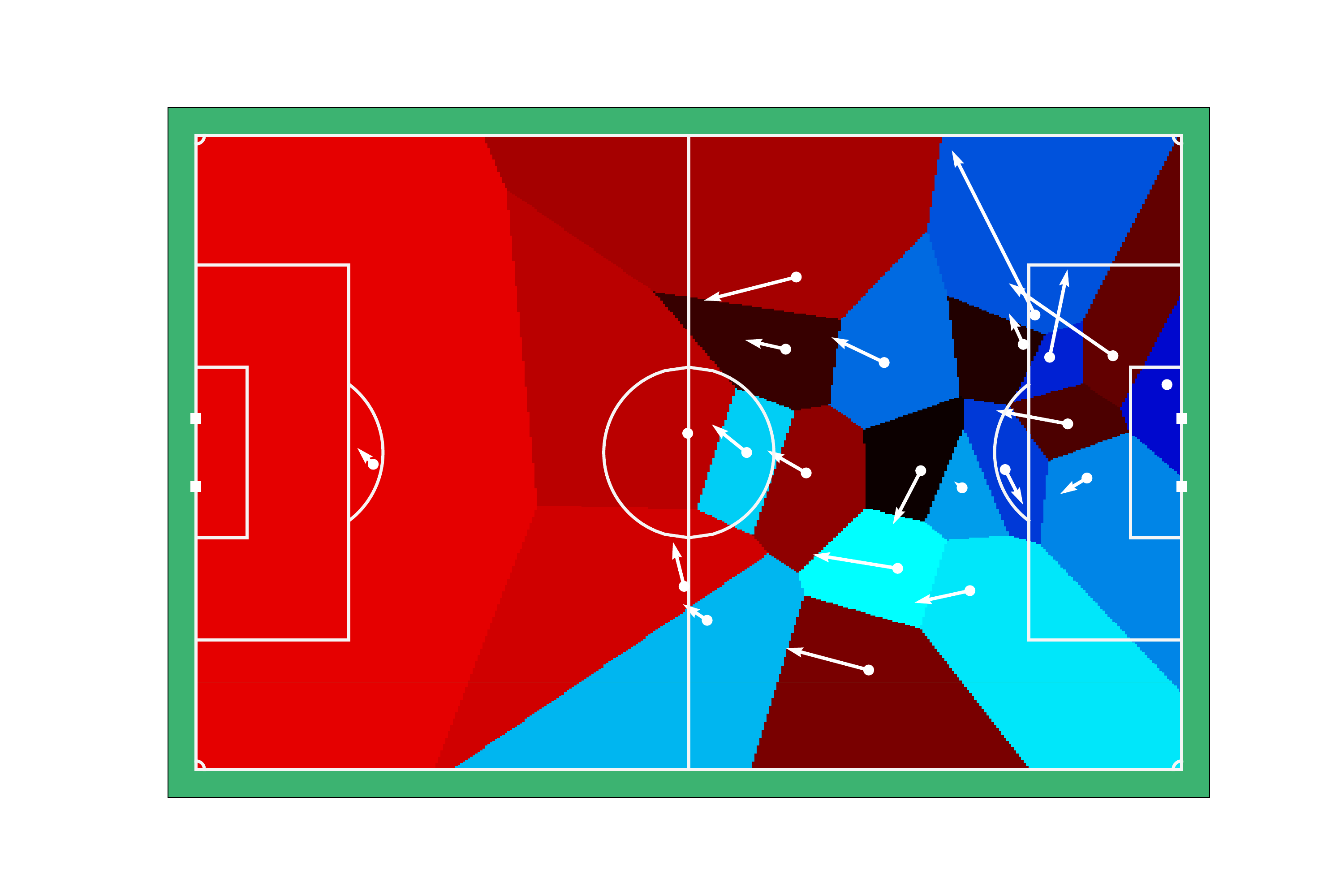}}
    \caption{\footnotesize Sketches of standard Voronoi diagrams representing dominance areas for players adopted from Metrica Sports open
                  tracking data \cite{Metrica}.   It is customary that, for the standard Voronoi diagrams, the velocities not to be shown since they are
                   not used for any calculation. However, the above diagrams do show the instantaneous velocities of the players for the moment
                   the diagram was created. }
    \label{fig:standardVoronoi}
\end{figure}}

When the initial speeds and the characteristic speeds of the players are taken into account, it was shown in \cite{efthimiou} that
\begin{equation}
     {r_1(t)\over r_2(t)} = {A_1\over A_2} = \text{const.}
\label{eq:O}
\end{equation}
where $A_i=(v_i+V_i)/2$, $i=1,2$.
This equation identifies the boundary between any two players as an Apollonius circle\footnote{The Apollonius circle reduces to the perpendicular
bisector when $A_1=A_2$.} surrounding the slower player. A curious
and demanding reader should refer to the original article to fully understand the underlying details of the derivation. For the current work those details are
not necessary.
Figure \ref{fig:BifocalOriginal} shows examples of boundaries (and hence their dominance regions) between two players placed in a pitch 212 meters by 136 meters\footnote{The size of this pitch is much larger than a standard soccer pitch. However, this choice is useful so the reader better evaluates the way boundaries curve. With not much effort, they can mentally truncate the diagrams down to the real size.} for eight different scenarios. As expected, the boundaries are either straight lines or circular arcs. The size of the circle is determined by the ratio $A_1/A_2$.
%
{\begin{figure}[h!]
    \centering
    \subfigure[\parbox{5cm}{$v_1 = 8\, \text{m/s}, v_2 = 6\, \text{m/s},\\  \alpha_1 = -135^{\circ}, \alpha_2 = 45^{\circ}$.}]%
                    {\includegraphics[width=6cm]{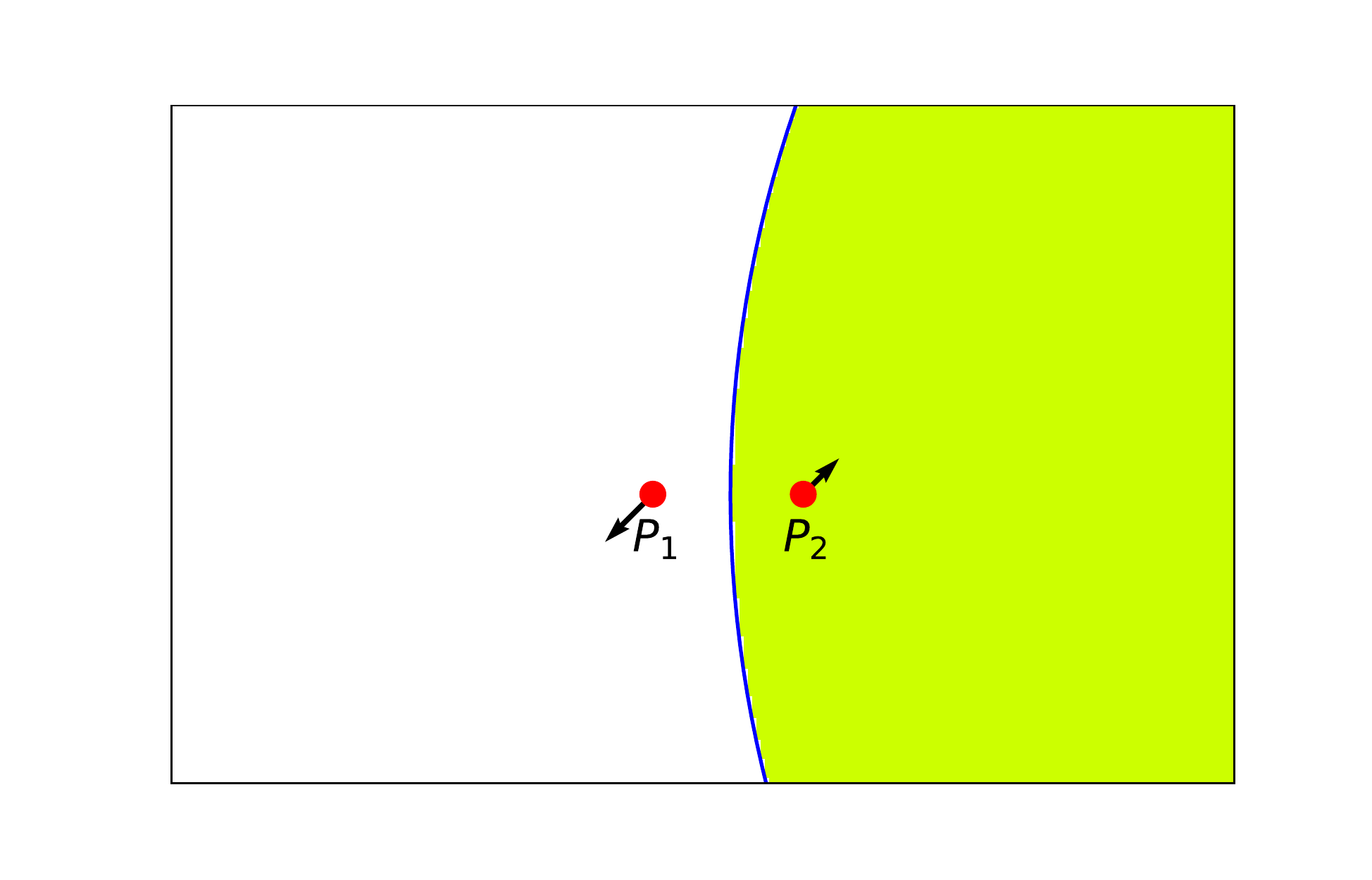}}
    \subfigure[\parbox{5cm}{$v_1 = 8\, \text{m/s}, v_2 = 3\, \text{m/s},\\ \alpha_1 = 0^{\circ}, \alpha_2 = 180^{\circ}$.}]%
                   {\includegraphics[width=6cm]{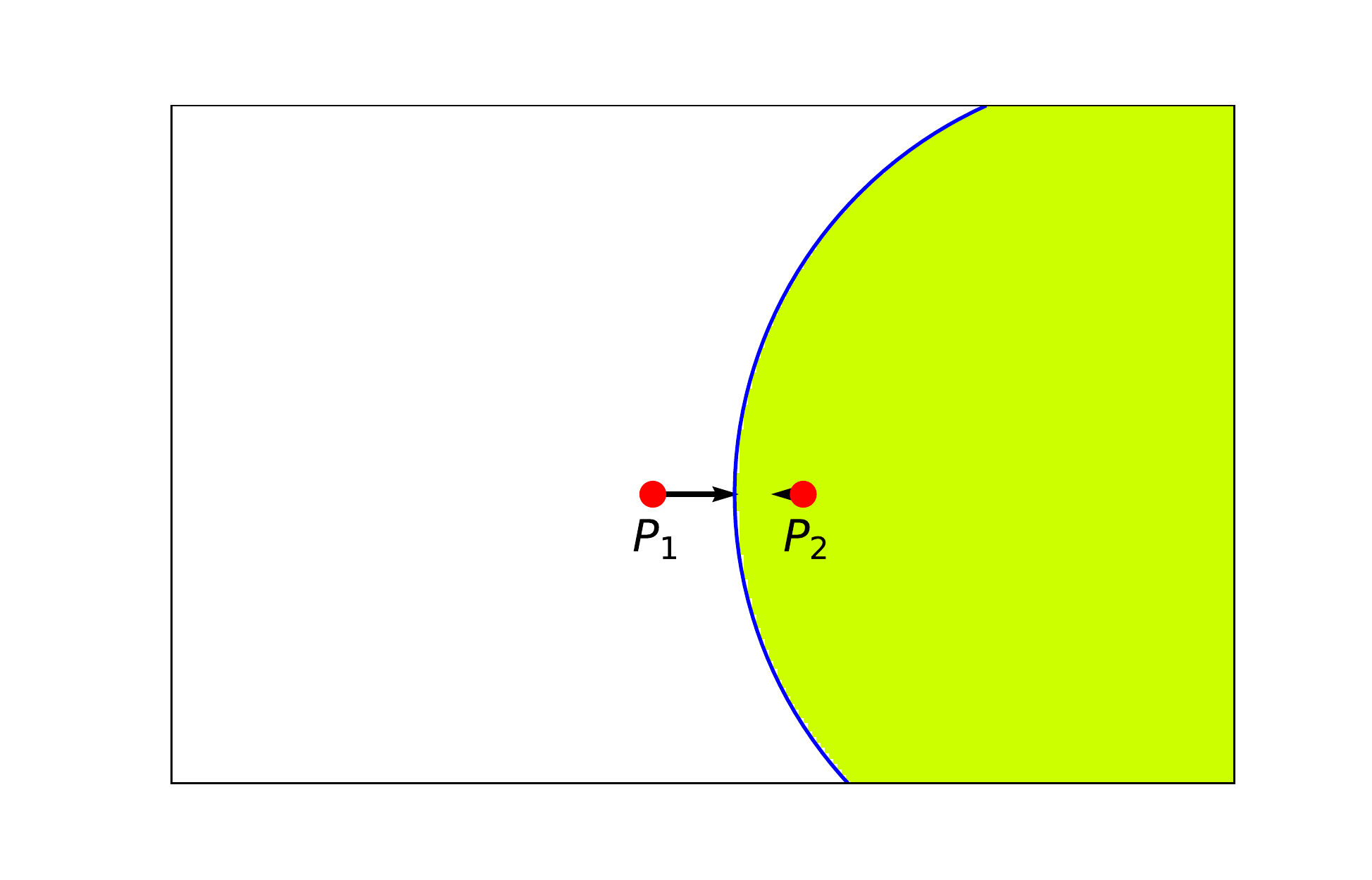}}\\
    \subfigure[\parbox{5cm}{$v_1 \sim 9\, \text{m/s}, v_2 = 0\, \text{m/s},\\ \alpha_1 = 45^{\circ}, \alpha_2 = -50^{\circ}$.}]%
                   {\includegraphics[width=6cm]{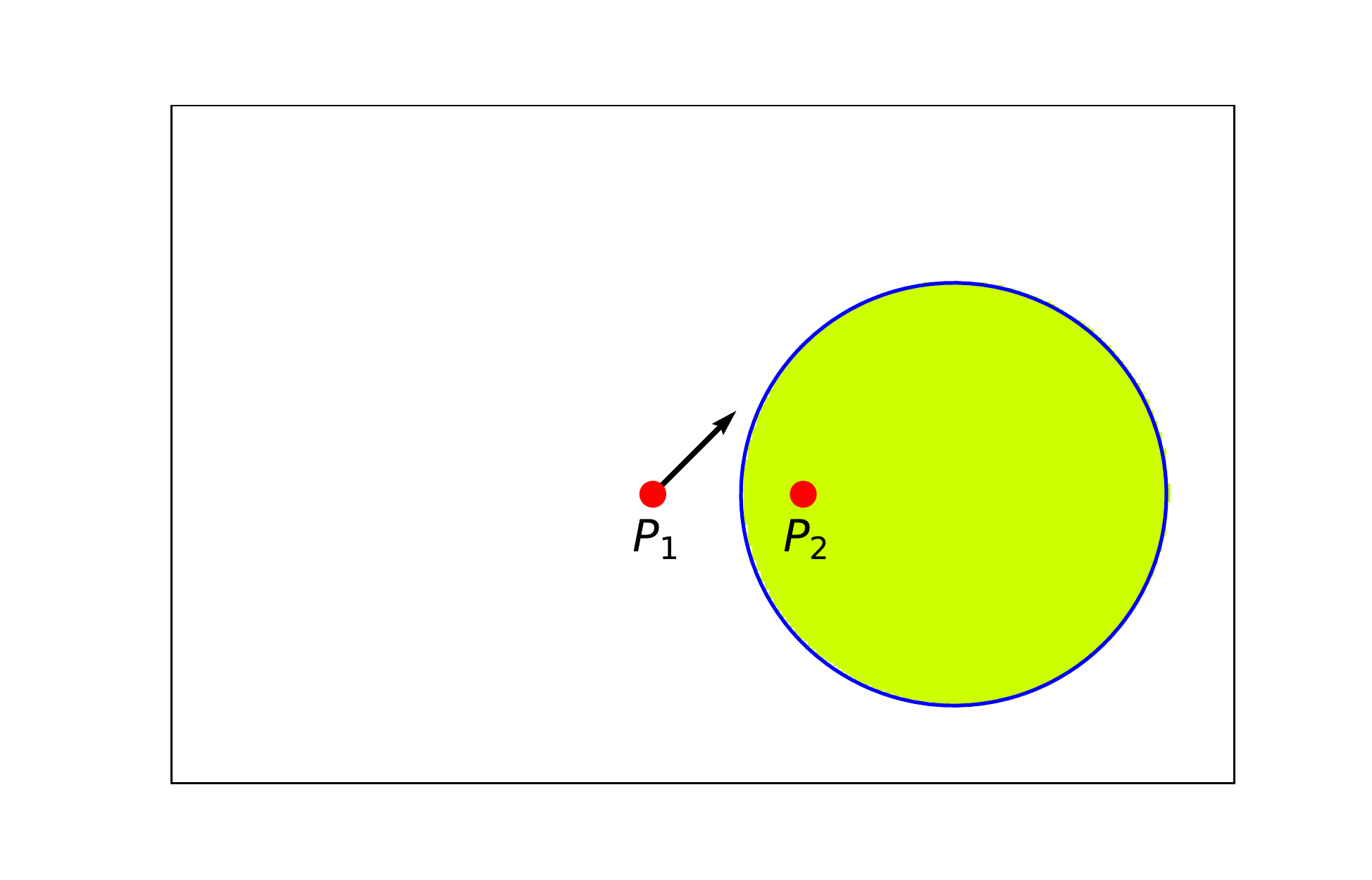}}
    \subfigure[\parbox{5cm}{$v_1 \sim 9\,\text{m/s}, v_2 \sim 9\,\text{m/s},\\ \alpha_1 = 90^{\circ}, \alpha_2 = 90^{\circ}$.}]%
                   {\includegraphics[width=6cm]{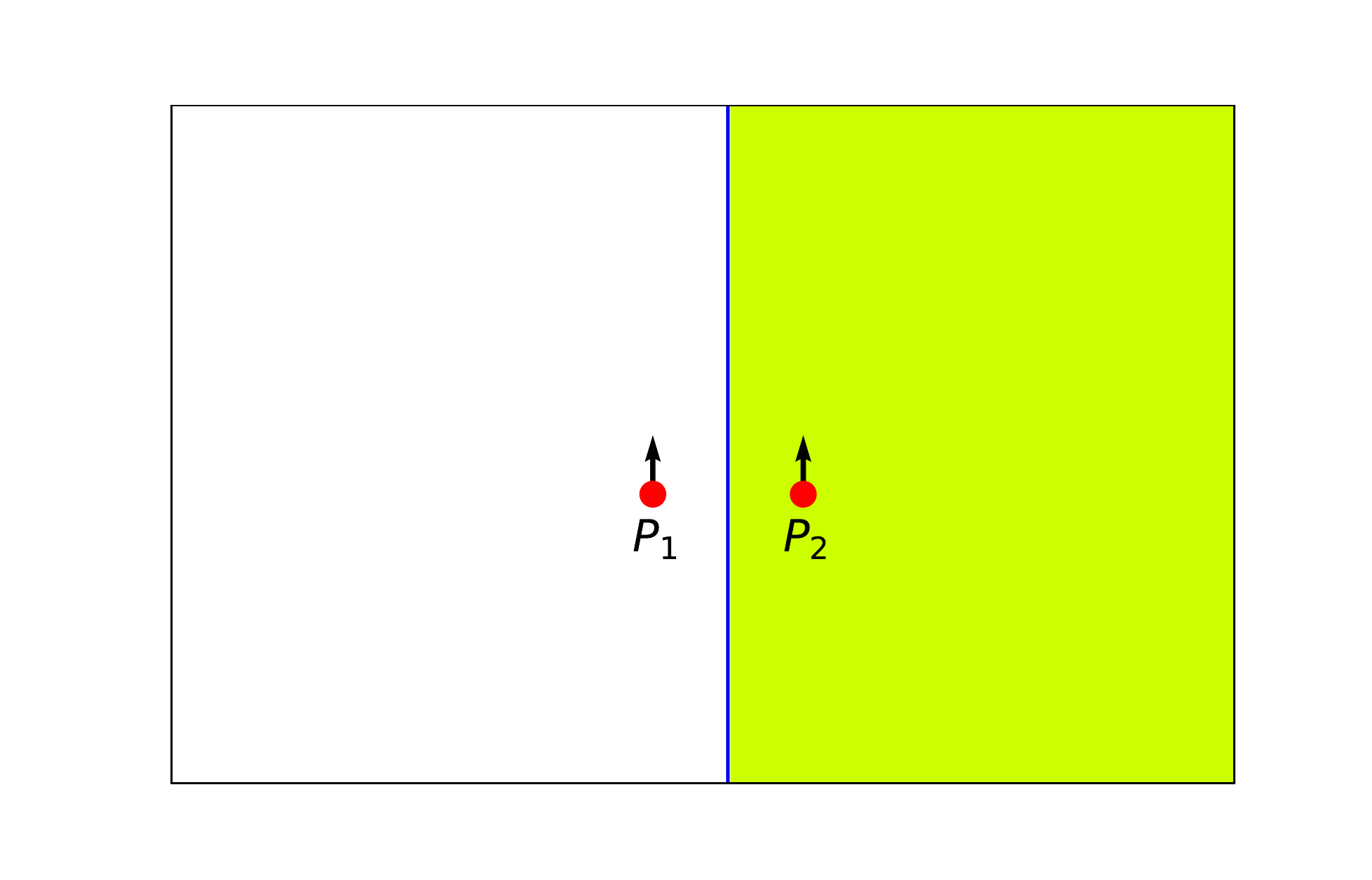}}\\
    \subfigure[\parbox{5cm}{$v_1 \sim 9\, \text{m/s}, v_2 \sim 9\, \text{m/s},\\  \alpha_1 = 45^{\circ}, \alpha_2 = 135^{\circ}$.}]%
                    {\includegraphics[width=6cm]{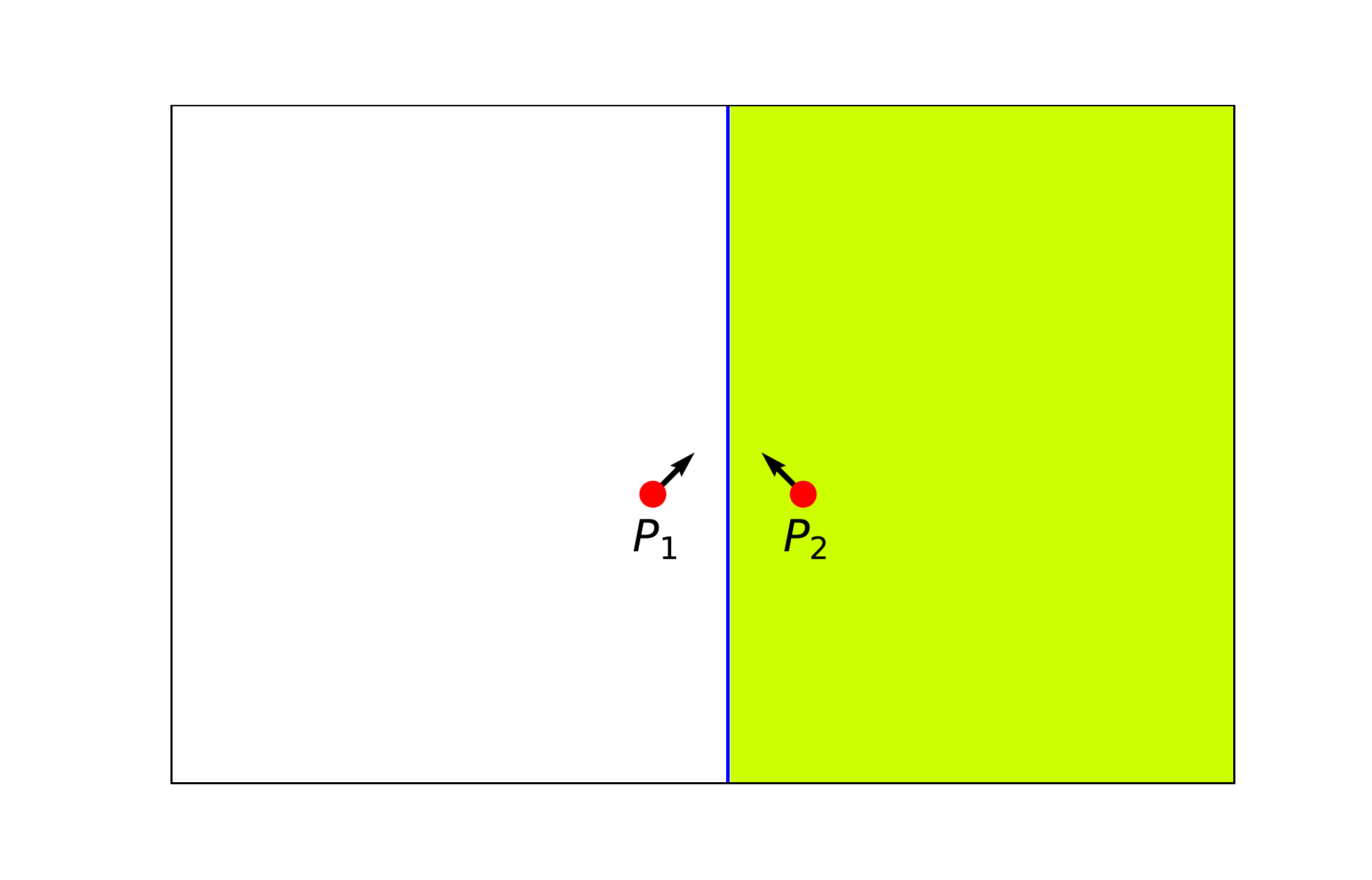}}
    \subfigure[\parbox{5cm}{$v_1 \sim 9\, \text{m/s}, v_2 \sim 9\, \text{m/s},\\ \alpha_1 = 90^{\circ}, \alpha_2 = 180^{\circ}$.}]%
                   {\includegraphics[width=6cm]{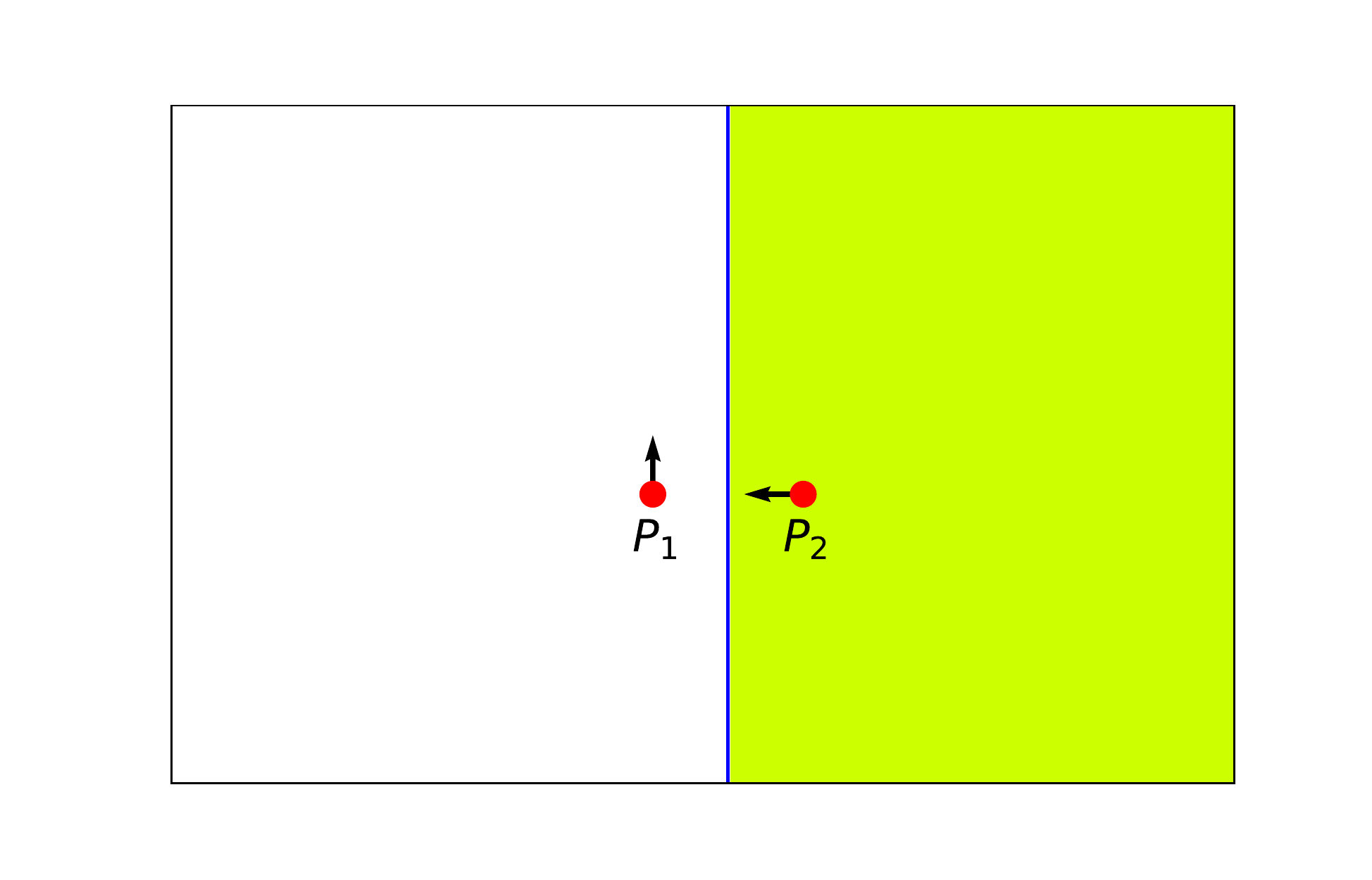}}\\
    \subfigure[\parbox{5cm}{$v_1 = 6\, \text{m/s}, v_2 = 6\, \text{m/s},\\ \alpha_1 = -45^{\circ}, \alpha_2 = 135^{\circ}$.}]%
                   {\includegraphics[width=6cm]{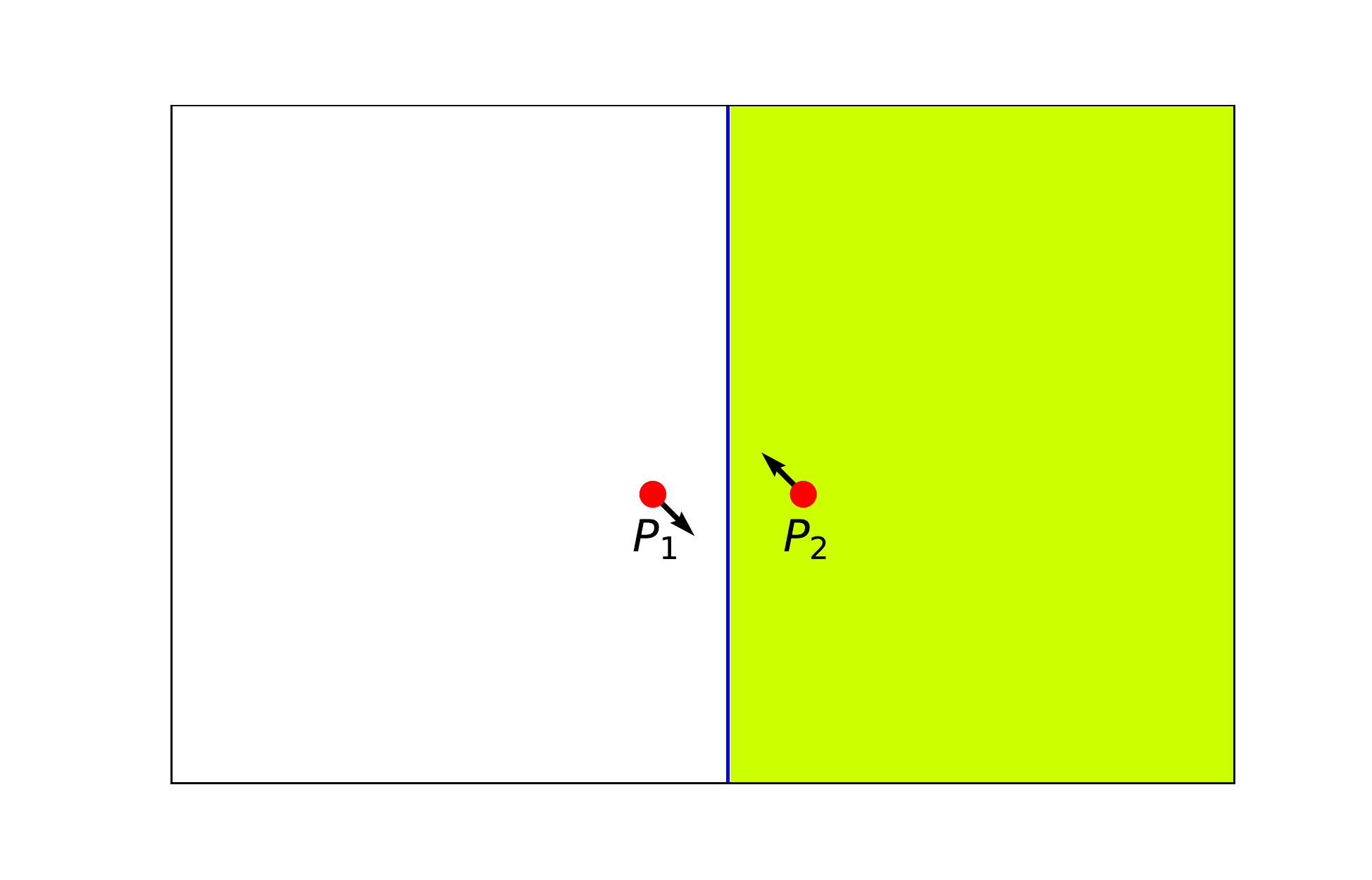}}
    \subfigure[\parbox{5cm}{$v_1=3\,\text{m/s}, v_2=8\,\text{m/s},\\ \alpha_1 = -45^{\circ}, \alpha_2 = -45^{\circ}$.}]%
                   {\includegraphics[width=6cm]{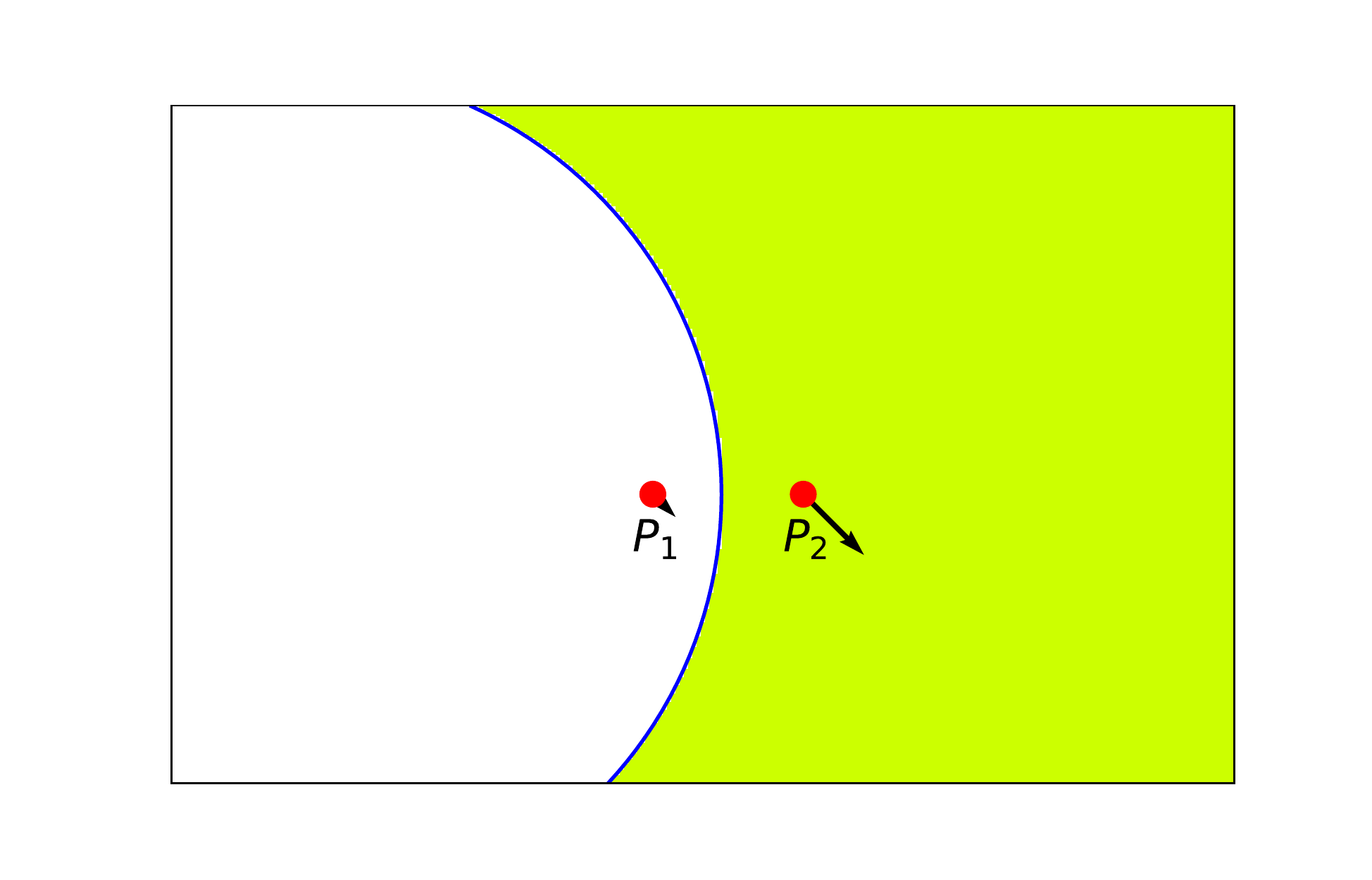}}
 \caption{\footnotesize Sketch of the boundary between two players on a $212$ meters by $136$ meters pitch according to \cite{efthimiou}.
               The bottom left corner and the upper right corner have coordinates $(-106,-68)$ and $(106,68)$ respectively.
               The two players P$_1$ and P$_2$ are located at the points $(-10, -10)$ and $(20, -10)$. We assume the players have a negligible
               difference in reaction time. In this model, only the magnitudes of the velocities
               are assumed to have an effect; their directions do not.  Finally, as there is a division by $v-9$ in our code, whenever the speeds are $v=\text{9 m/s}$ we use $v=\text{8.9999 m/s}$ to avoid a \texttt{ZeroDivisionError}. Therefore, we use $v \sim \text{9 m/s}$ rather than $v=\text{9 m/s}$ in our figure captions.}
\label{fig:BifocalOriginal}
\end{figure}}

When equation \eqref{eq:O} is used to determine the dominance regions for the 22 players of a game,
the resulting diagram for the dominance regions was called Apollonius diagram in \cite{efthimiou}. It turns out that this diagram is a variation of the standard Voronoi diagram which mathematicians refer to  as the multiplicatively weighted Voronoi diagram.
 The dominance areas might not always be convex, might contain holes, and might be disconnected. Examples of Apollonius diagrams are shown in
 Figure \ref{fig:apollonius}.
%
{\begin{figure}[h!]
\centering
    \subfigure[Frame: 98202]{\includegraphics[width=12cm]{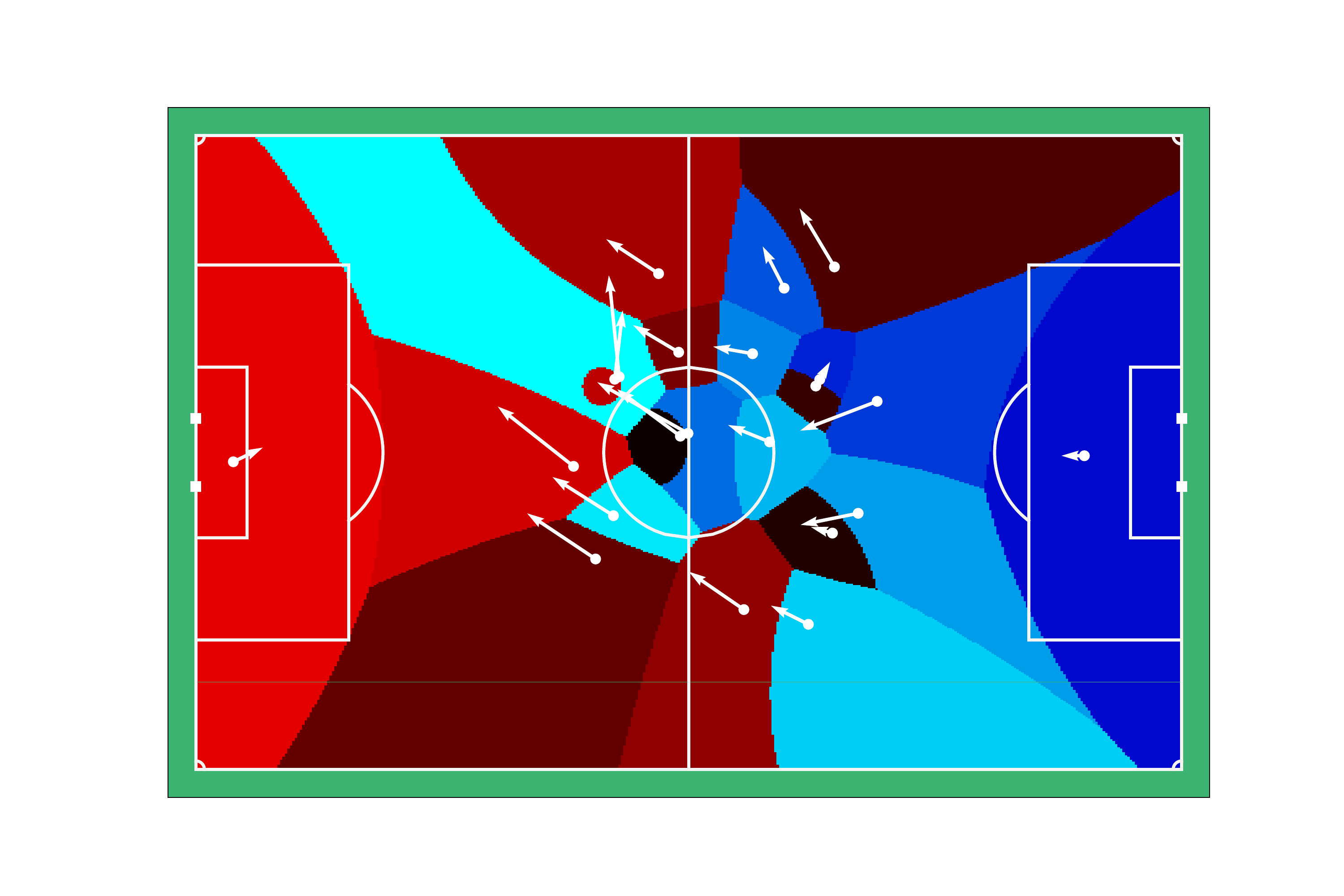}}\\
    \subfigure[Frame: 123000]{\includegraphics[width=12cm]{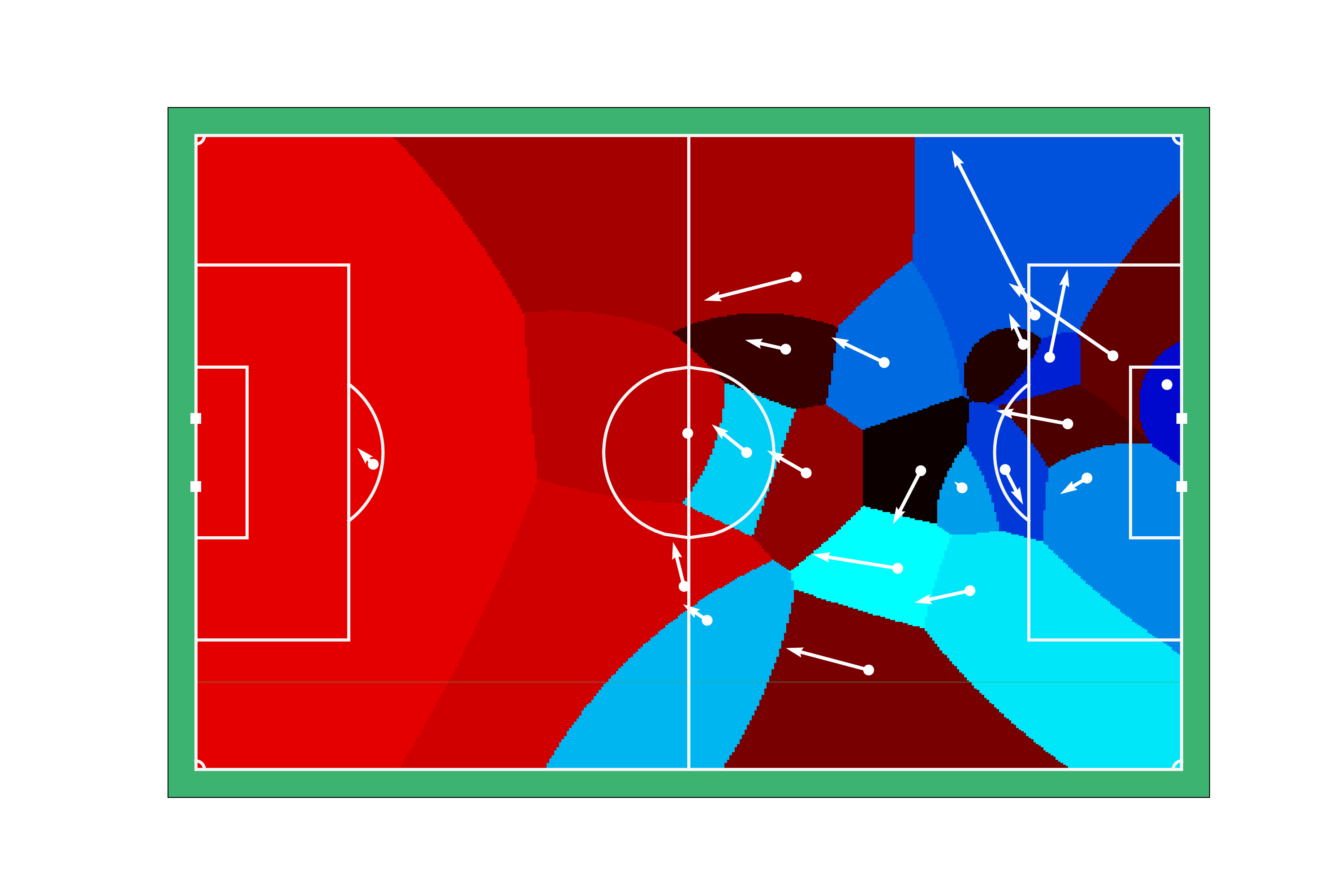}}
\caption{\footnotesize Sketches of Apollonius diagrams representing dominance areas for players adopted from Metrica Sports open data.}
\label{fig:apollonius}
\end{figure}}

When the delayed reactions and the speeds are taken into account, it was shown \cite{efthimiou} that the boundary between any two players is
a Cartesian oval. Although our theoretical models can accommodate such a parameter easily, Metrica Sports does not provide any reaction time data for the players in their open data \cite{Metrica}.  Hence, when we draw diagrams in this article, we assume the difference betwwen reaction times of the players is negligible.

\subsection*{\normalsize Inclusion of Air Drag}

In this section, we present the first extension of the model presented in \cite{efthimiou}.
When there is no air drag, the player is propelled by the frictional force $a$ acting on their feet from the  ground:\footnote{Please note that we use Newton's law of motion in the form $F'=a$ instead of $F=ma$. In other words,
our force $F'$ is really force per unit mass $F/m$.}
$$
    {dv\over dt}  = a .
$$
Here $a$  is the value of the acceleration that we used in \cite{efthimiou}. However, when an object is moving
inside a fluid, there is a frictional force --- drag force --- from the fluid which is proportional to the square of the velocity also acting
on the object:
\begin{align}
   {dv\over dt} =   a- k\, v^2 , \quad a,k>0  .
\label{eq:9}
\end{align}
In general, the coefficient $k$ does not have to be constant but we will assume that this is the case.
Notice that, as the speed increases, the air drag increases too. The velocity can increase only up to the point that the air drag  balances the propelling force from the ground. At that point, $dv/dt=0$ and hence
$$
     v_\text{limit}^2 = {a\over k}.
$$

Equation \eqref{eq:9} can be integrated easily by separation of variables:
\begin{align}
   \int_{v_0}^v  {dv\over k\left( v_\text{limit}^2- v^2 \right)} = \int_{t_0}^t dt .
\label{eq:10}
\end{align}
It is known that, if $v^2<b^2$, then
$$
  \int {du\over b^2-v^2} = {1\over 2b} \, \ln{b+v\over b-v} = {1\over b} \, \tanh^{-1}{v\over b}.
$$
Since $v \le v_\text{limit}$, equation \eqref{eq:10} gives
\begin{align*}
   \tanh^{-1}{v\over  v_\text{limit}}  - \tanh^{-1} {  v_0 \over v_\text{limit} } = k v_\text{limit}\,(t-t_0) ,
\end{align*}
or
\begin{align}
        \tanh^{-1}{v\over  v_\text{limit}} = kv_\text{limit}\,(t-t_0) + \delta ,
\label{eq:11}
\end{align}
where we have set
$$
    \delta =   \tanh^{-1} {v_0\over v_\text{limit} } .
$$
 Equation \eqref{eq:11} can easily be solved for $v$.
\begin{equation}
    v = v_\text{limit} \, \tanh[kv_\text{limit}  (t-t_0)+\delta] .
\label{eq:12}
\end{equation}

The previous equation can be integrated further, also by separation of variables, to find the distance travelled by the player:
\begin{align*}
   \int_{r_0}^ rdr=&  {1\over k} \,  \int_{t_0}^t {d\cosh[kv_\text{limit} (t-t_0)+\delta] \over \cosh[kv_\text{limit} (t-t_0)+\delta]} ,
\end{align*}
or
\begin{align*}
    r-r_0=& {1\over k} \,   \,  \ln {\cosh[kv_\text{limit}(t-t_0)+\delta] \over \cosh\delta} .
\end{align*}
In bipolar coordinates, the constant $r_0$ vanishes.
And since
\begin{align*}
   \cosh u = {1\over\sqrt{1-\tanh^2u}} ,
\end{align*}
we have
\begin{equation}
     r = -{1\over 2k}\, \ln\left[ \cosh^2\delta \, \Big(1-{v^2\over v_\text{limit}^2}\Big)\right] .
\label{eq:14}
\end{equation}

The last equation expresses the distance run by a player assuming that their speed has the value $v$.
Our assumption is that, when a player competes to reach a point, they do so by placing maximum effort and reaching it with their characteristic speed $V$.
Hence at point P:
\begin{align}
    V &=  v_\text{limit} \, \tanh[kv_\text{limit}  (t_\text{P}-t_0)+\delta] ,                                               \label{eq:17a} \\
    r_\text{P}  &=   -{1\over 2k}\, \ln\left[ \cosh^2\delta \, \Big(1- {V^2\over v_\text{limit}^2}\Big)\right] .            \label{eq:17b}
\end{align}
Here $t_\text{P}$ is the time the player needs to reach the point P starting from their position with time delay $t_0$.
We can now use equation \eqref{eq:17a} to eliminate  $k$ from equation \eqref{eq:17b}.
In particular:
$$
             k = {1\over  v_\text{limit} (t_\text{P}-t_0)}   \,   (\Delta - \delta),
$$
where we have set
$$
         \Delta =  \tanh^{-1} {V\over v_\text{limit}} .
$$
So,
\begin{equation}
     r_\text{P} = {v_\text{limit}  \,
                       \ln\left[ \cosh^2\delta \, \Big(1-{V^2\over v_\text{limit}^2}\Big) \right]  \over 2(\delta - \Delta) }\,  (t_\text{P}-t_0) .
\label{eq:19}
\end{equation}
This result is striking: It has the form
$$
     r_\text{P} = A\, (t_\text{P}-t_0)
$$
with the factor $A$ dependent on the characteristic speed  of the player.
Hence, when air drag is included but the remaining premises remain the same, one could rush to state that all the results of \cite{efthimiou} are transferable in this case too.  However, this is not as automatic as it appears to be!
In addition to $V$, the factor $A$ depends on $v_\text{limit}$ which, in turn, depends on $a$ and $k$. Since the speed $V$ has a fixed value at any point P on the boundary of the dominance regions, the acceleration $a$ has different values along different rays (although on each ray it is  constant).
This implies that $v_\text{limit}$ has a different value along each ray and complicates our analysis. Although, numerically, this issue is not a big obstacle as a computer program can be set to compute  $v_\text{limit}$ ray by ray, for an analytical analysis it creates complications.  To remove this difficulty,
we will assume that $k$ has also a different value along each ray such that $v_\text{limit}$ maintains a constant value for each player.
This constant value may be assumed to be different from player to player or might be universal. Having a universal value is more appealing --- an ultimate speed that no runner can surpass. From \cite{Ryu}, we see that the maximal sprinting speeds of elite sprinters approach 12 $\text{m/s}$. We assume that these atheletes have not reached the theoretical maximum yet, and so take $v_\text{limit}$ to be 13 $\text{m/s}$.
Hence,  we will assume that the coefficient $k$ is proportional to $a$ with the proportionality constant  being  the fixed number $1/v^2_\text{limit}$. With  this assumption,
 equation \eqref{eq:9} has the form:
$$
   {dv\over dt} =   a \left( 1 -  {v^2\over v^2_\text{limit}} \right)  .
$$
Given the previous discussion, when air drag is included then indeed all results of \cite{efthimiou} (subjected to all underlying assumptions for the original model) are transferable to this case too.
This is clear in Figure \ref{fig:BifocalAirDrag}, which demonstrates the effects of a drag force by plotting the boundaries between two players for the
same scenarios as Figure \ref{fig:BifocalOriginal}.  By comparing the two figures, we notice immediately that, when the boundary is straight in the original model,  it remains straight when air drag is added if the players have the same initial speeds and same characteristic speeds.\footnote{Since the functional forms of $A=A(v,V)$ in the original model and the model with air drag are different, the ratio $A_1/A_2$ can be different in the two models
if this condition  is not true even if $v_1+V_1=v_2+V_2$. In particular, when the condition is not true, a straight boundary in the original model becomes an Apollonius arc in the new model.}
The times required for the players to reach each point do change, but since the factors $A$ remain equal, there is no observable change in the dominance regions. When the conditions require an Apollonius circle in the original case, in general this will be the case in the new model but the radius of the circle changes due to the air drag.
%
{\begin{figure}[h!]
    \centering
    \subfigure[\parbox{5cm}{$v_1 = 8\, \text{m/s}, v_2 = 6\, \text{m/s},\\  \alpha_1 = -135^{\circ}, \alpha_2 = 45^{\circ}$.}]%
                    {\includegraphics[width=6cm]{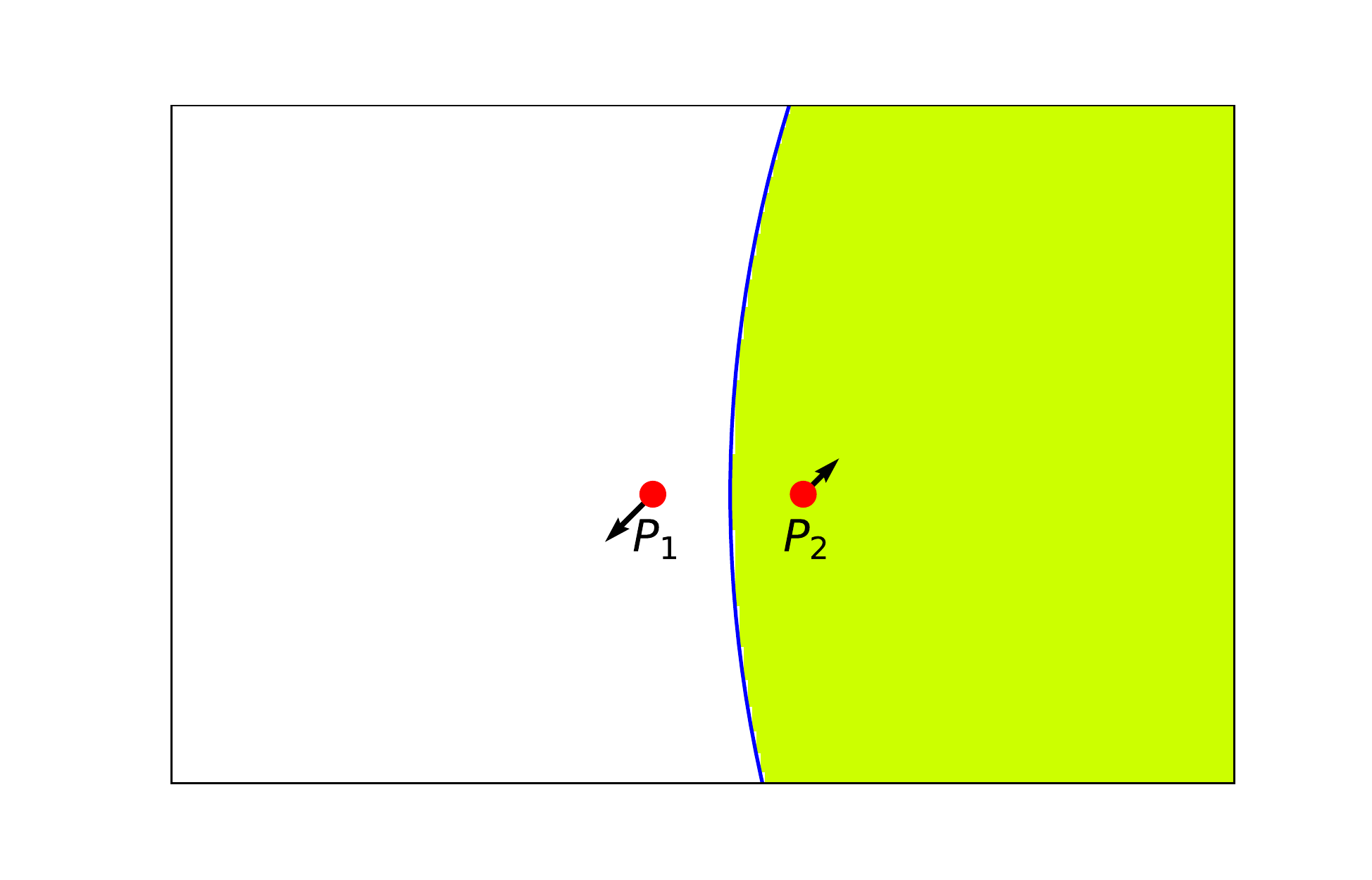}}
    \subfigure[\parbox{5cm}{$v_1 = 8\, \text{m/s}, v_2 = 3\, \text{m/s},\\ \alpha_1 = 0^{\circ}, \alpha_2 = 180^{\circ}$.}]%
                   {\includegraphics[width=6cm]{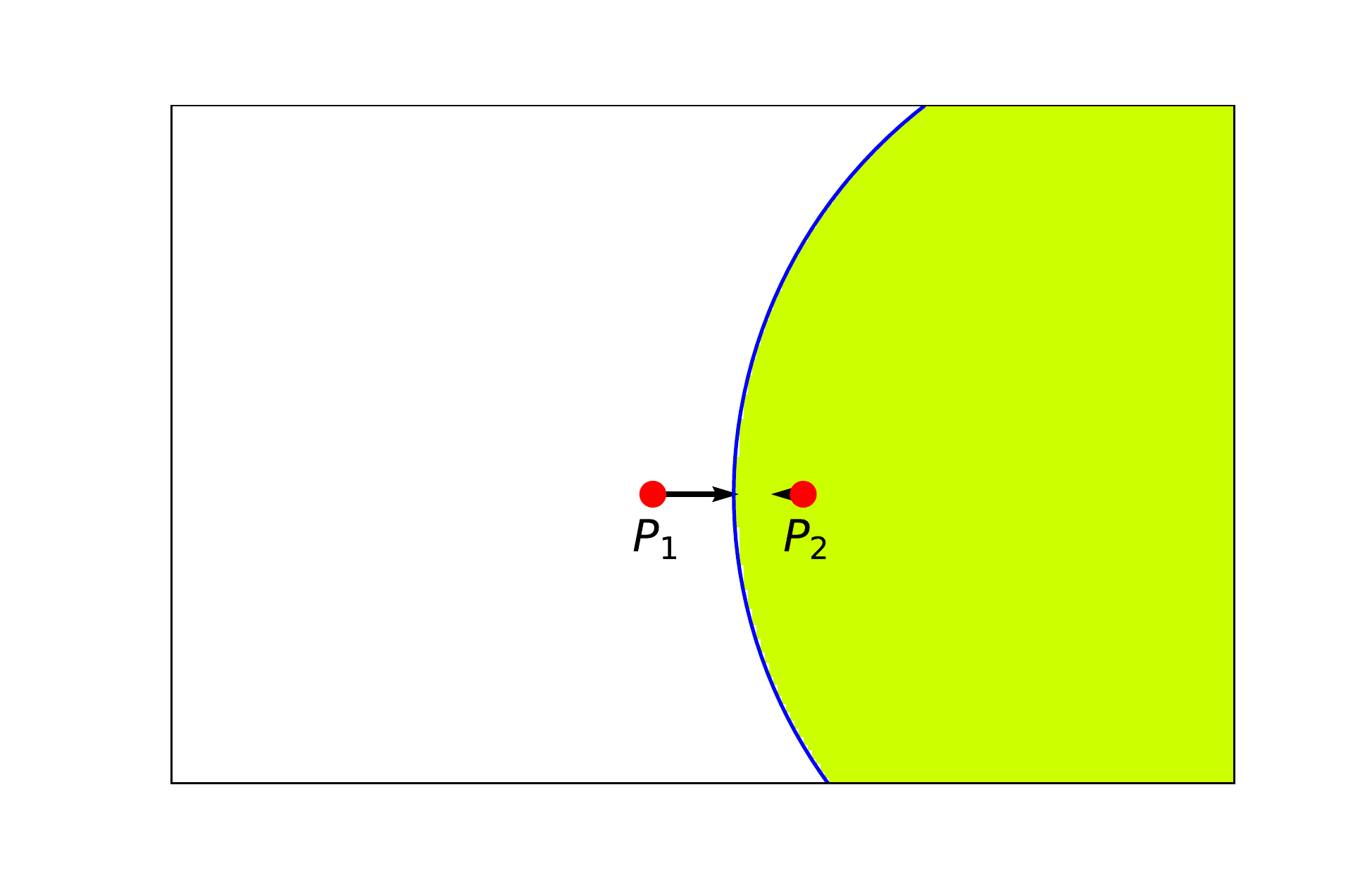}}\\
    \subfigure[\parbox{5cm}{$v_1 \sim 9\, \text{m/s}, v_2 = 0\, \text{m/s},\\ \alpha_1 = 45^{\circ}, \alpha_2 = -50^{\circ}$.}]%
                   {\includegraphics[width=6cm]{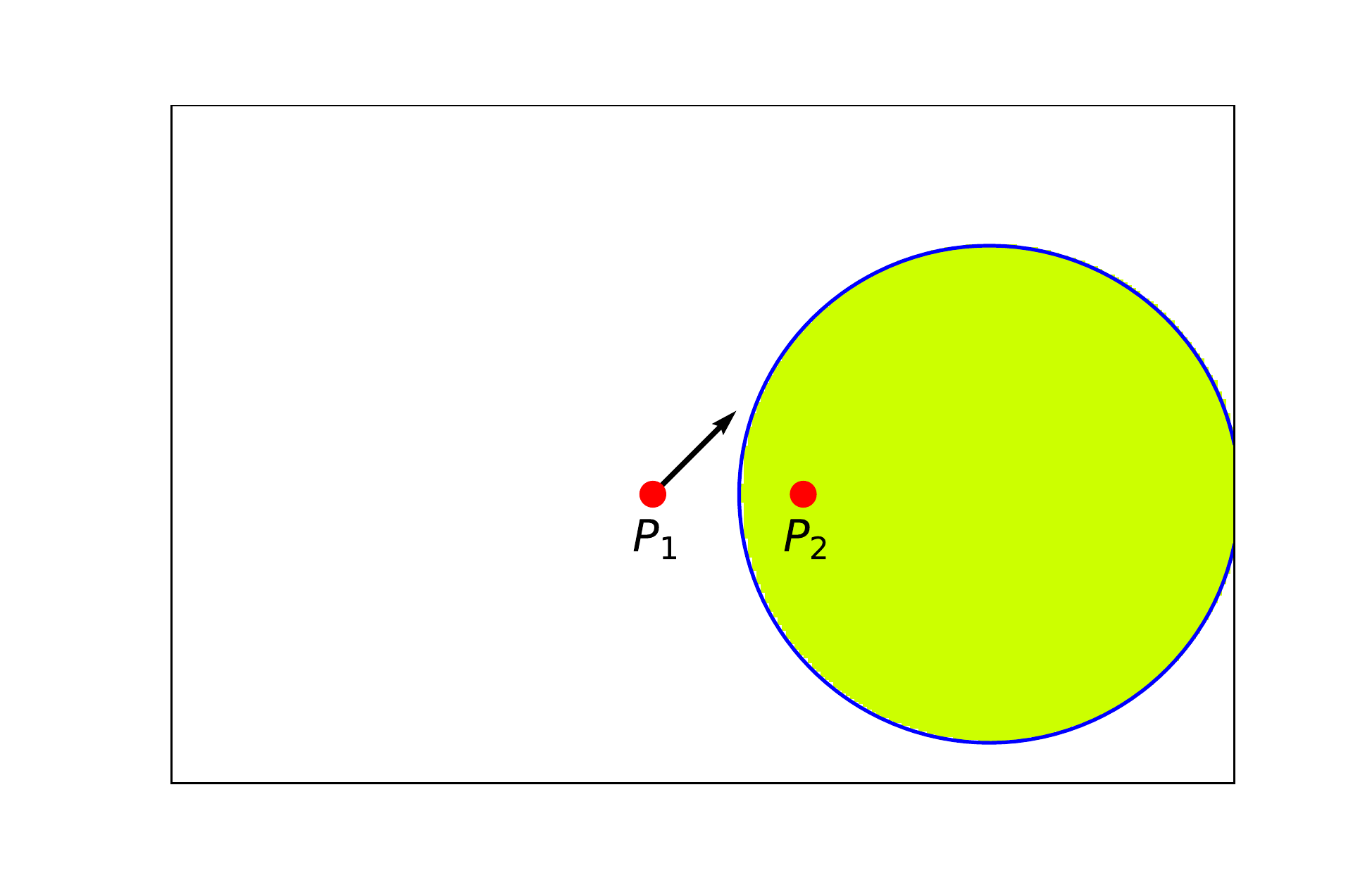}}
    \subfigure[\parbox{5cm}{$v_1 \sim 9\,\text{m/s}, v_2 \sim 9\,\text{m/s},\\ \alpha_1 = 90^{\circ}, \alpha_2 = 90^{\circ}$.}]%
                   {\includegraphics[width=6cm]{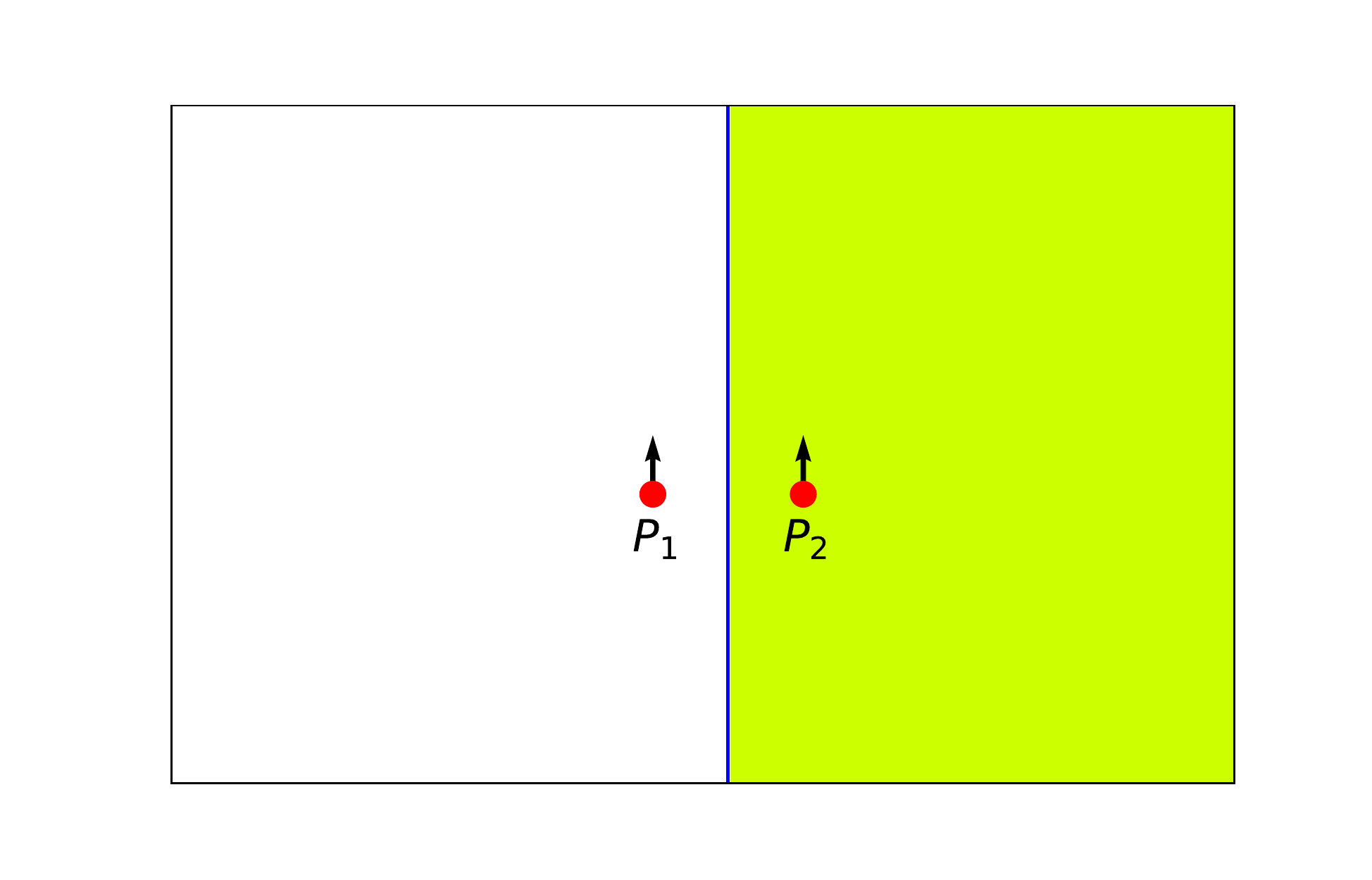}}\\
    \subfigure[\parbox{5cm}{$v_1 \sim 9\, \text{m/s}, v_2 \sim 9\, \text{m/s},\\  \alpha_1 = 45^{\circ}, \alpha_2 = 135^{\circ}$.}]%
                    {\includegraphics[width=6cm]{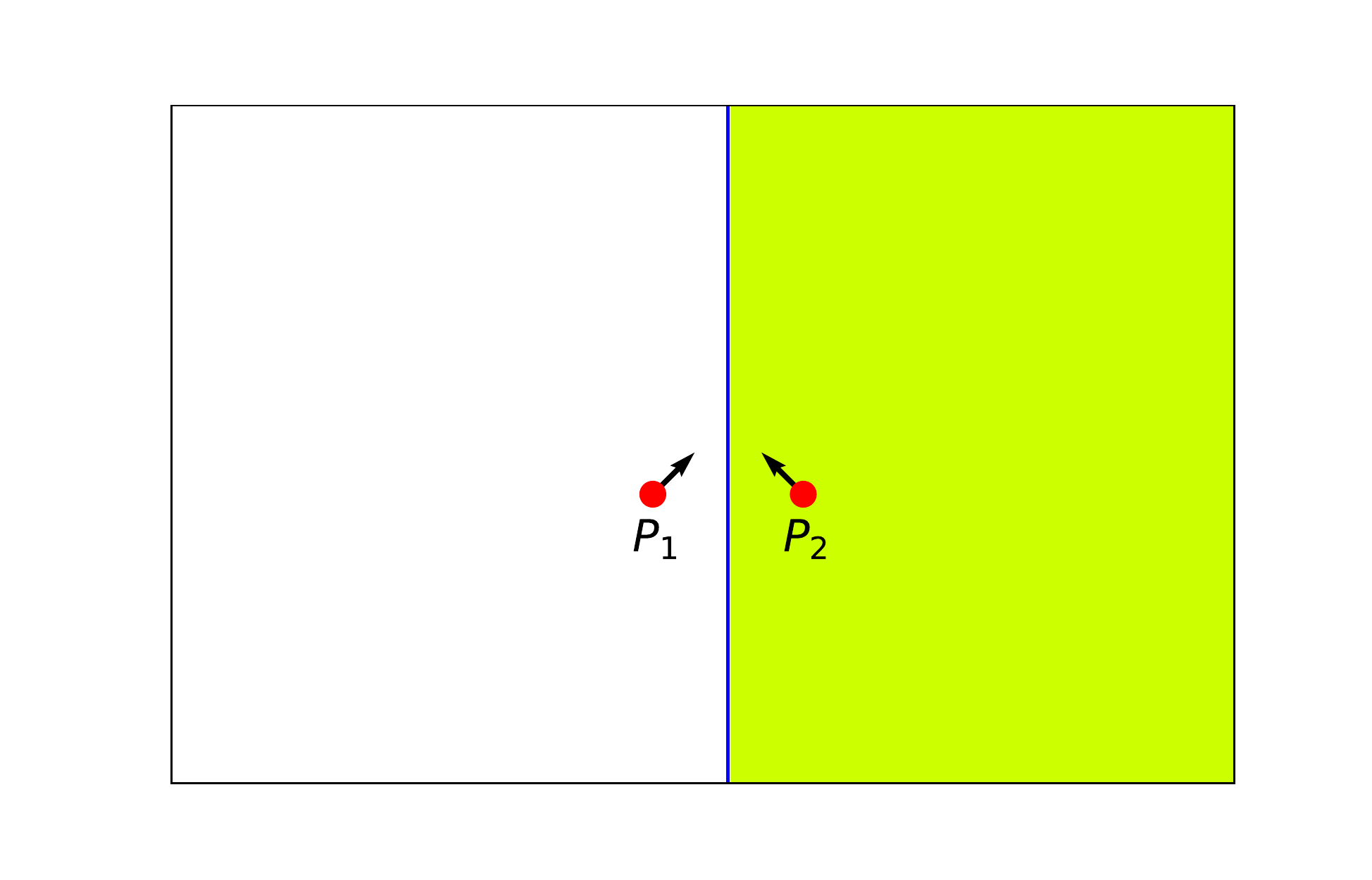}}
    \subfigure[\parbox{5cm}{$v_1 \sim 9\, \text{m/s}, v_2 \sim 9\, \text{m/s},\\ \alpha_1 = 90^{\circ}, \alpha_2 = 180^{\circ}$.}]%
                   {\includegraphics[width=6cm]{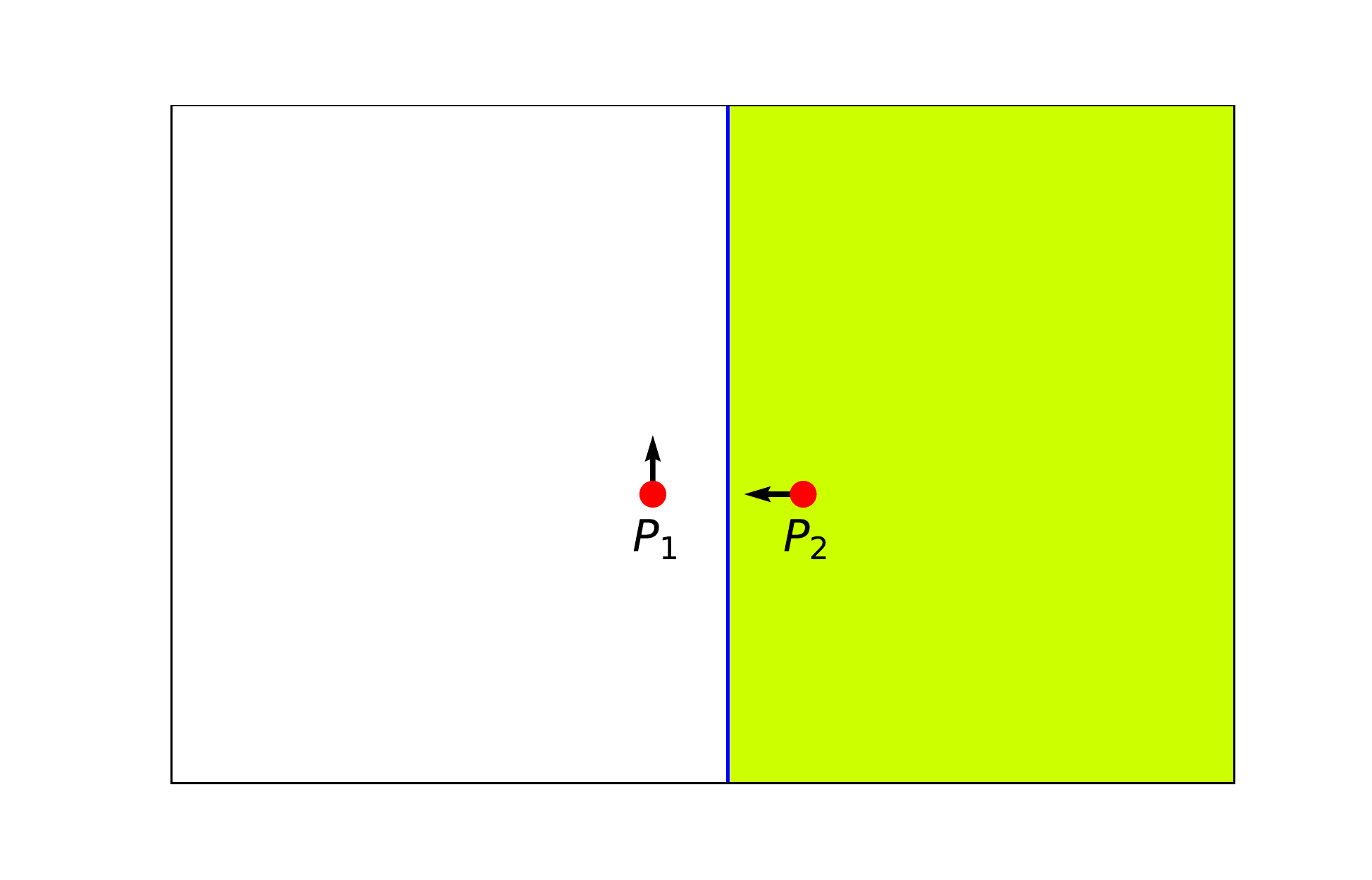}}\\
    \subfigure[\parbox{5cm}{$v_1 = 6\, \text{m/s}, v_2 = 6\, \text{m/s},\\ \alpha_1 = -45^{\circ}, \alpha_2 = 135^{\circ}$.}]%
                   {\includegraphics[width=6cm]{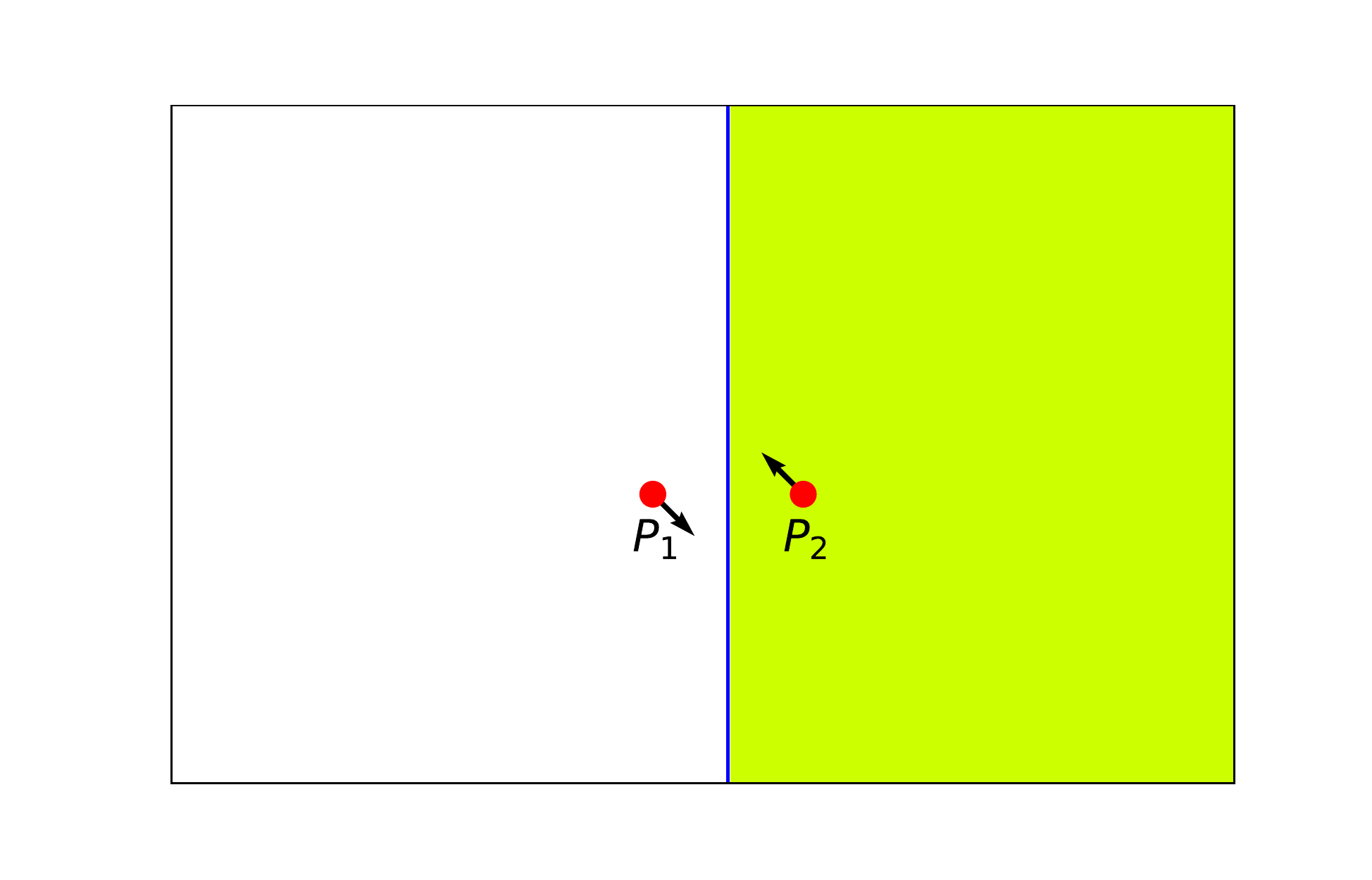}}
    \subfigure[\parbox{5cm}{$v_1=3\,\text{m/s}, v_2=8\,\text{m/s},\\ \alpha_1 = -45^{\circ}, \alpha_2 = -45^{\circ}$.}]%
                   {\includegraphics[width=6cm]{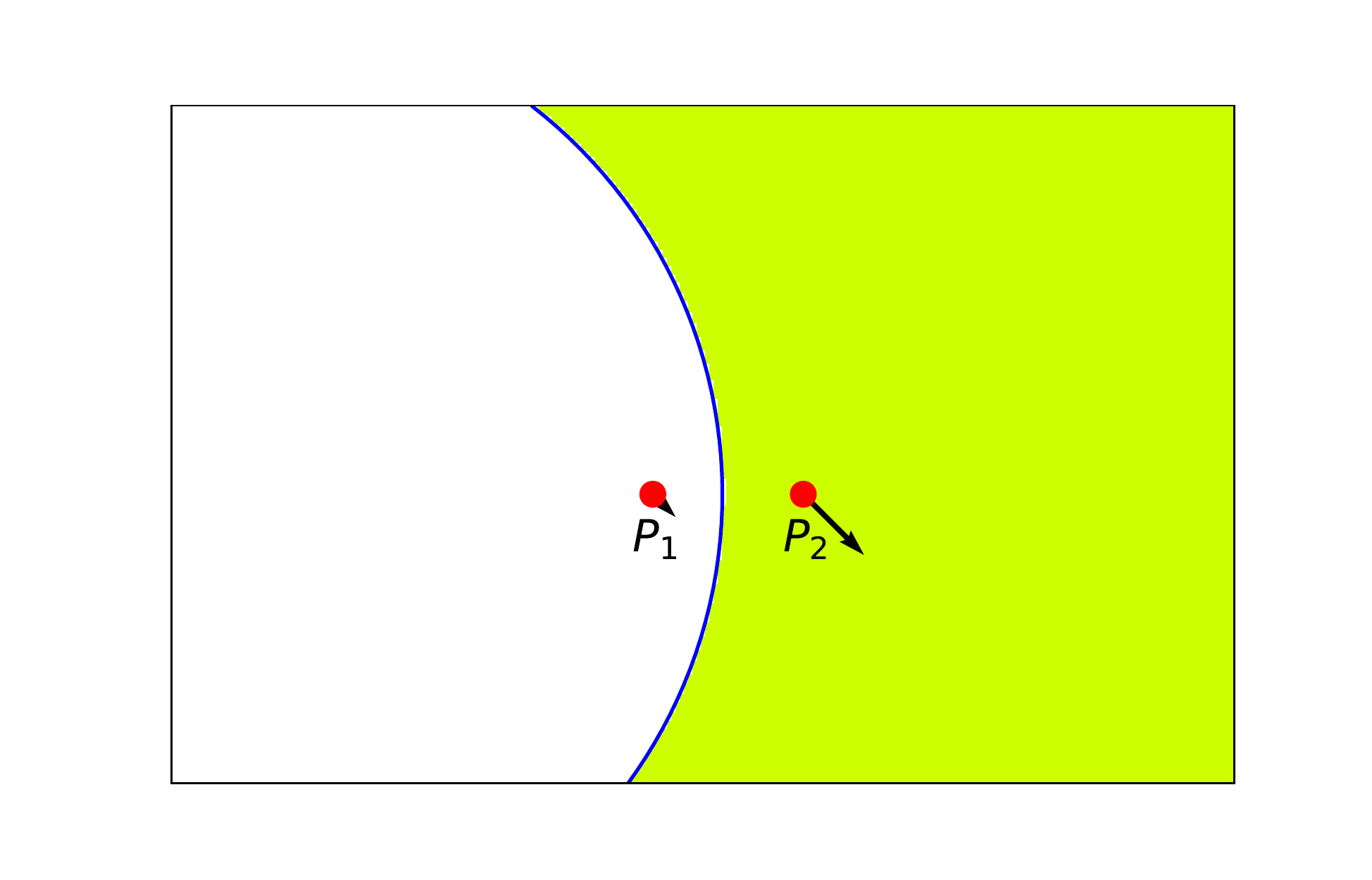}}
\caption{\footnotesize Sketch of the boundary between two players in the presence of air resistance on a 212 meters
               by 136 meters pitch.
               The bottom left corner and the upper right corner have coordinates $(-106,-68)$ and (106,68) respectively.
                The two players P$_1$ and P$_2$ are located at the points $(-10, -10)$ and $(20, -10)$.  The cases are identical with those
                of Figure \ref{fig:BifocalOriginal}. However, the radii of the Apollonius circles are altered as explained in the text
                following equation \eqref{eq:19}.
                The reader can compare the two results to see the changes.}
\label{fig:BifocalAirDrag}
\end{figure}}

Figure \ref{fig:frictionVoronoi} shows the Apollonius diagram with air drag included for the same frames of the  Metrica Sports data.
Comparing the diagrams with those of Figure \ref{fig:apollonius}, one can see the changes in the dominance areas induced by air drag.
{\begin{figure}[h!]
    \centering
    \subfigure[Frame: 98202]{\includegraphics[width=12cm]{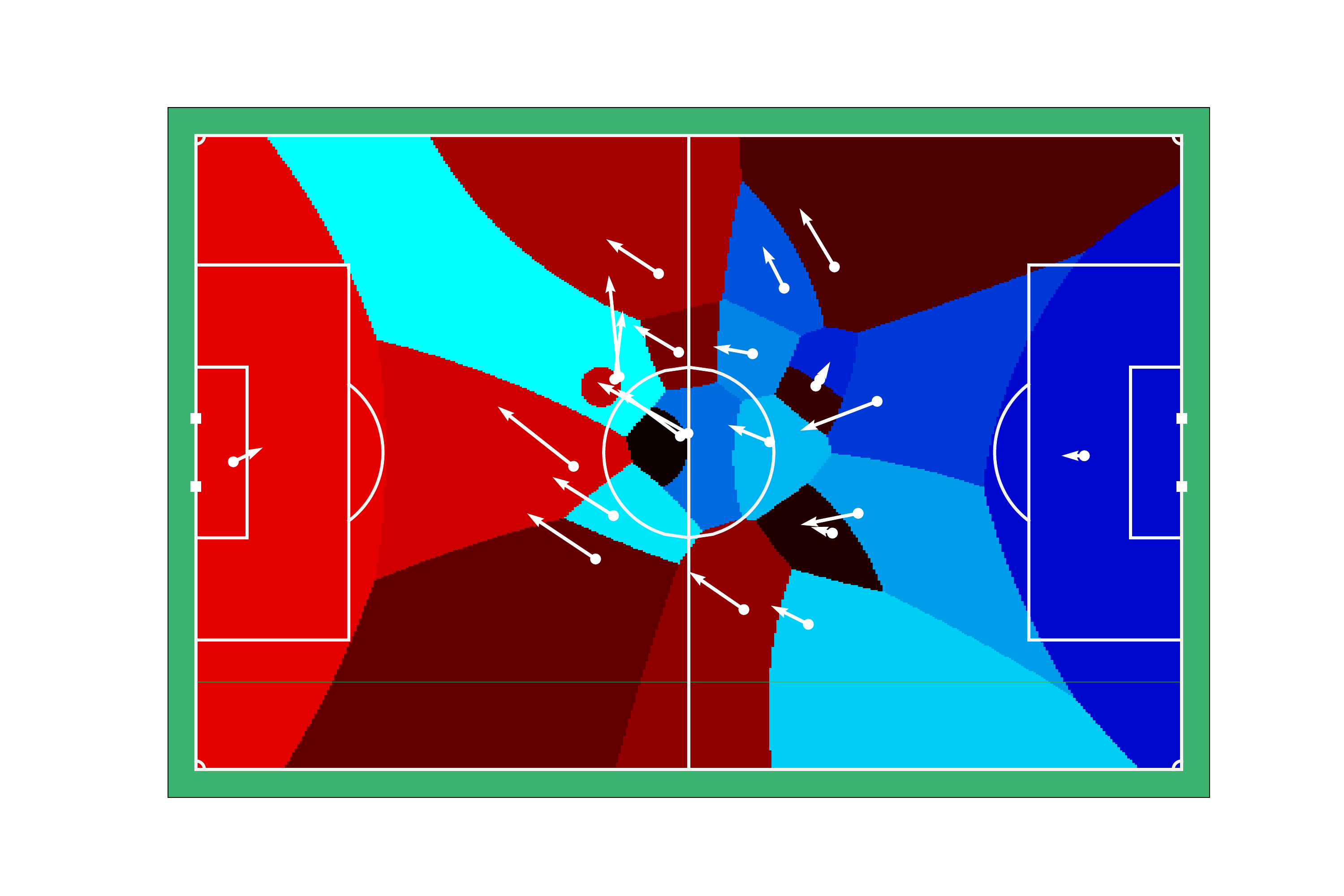}}\\
    \subfigure[Frame: 123000]{\includegraphics[width=12cm]{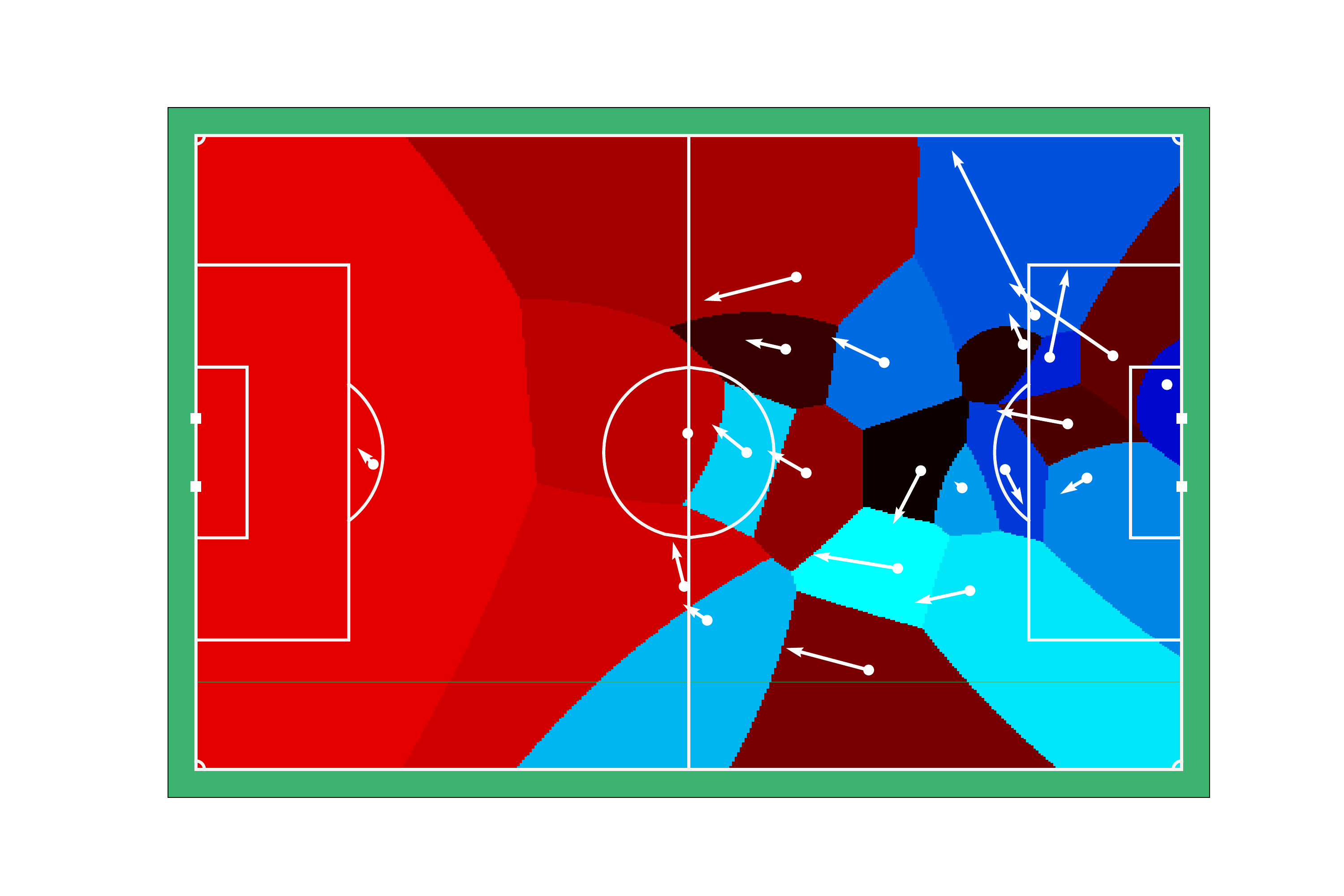}}
\caption{\footnotesize Sketches of Apollonius diagrams in the presence of air drag for the same frames of the Metrica Sports open tracking data.}
\label{fig:frictionVoronoi}
\end{figure}}

\subsection*{\normalsize Inclusion of Air Drag and Internal Dissipation}

In physics courses, it is common to introduce students to a frictional force that is proportional to velocity $-\gamma \, v$, $\gamma>0$. Interestingly,
it has been known for some time that this frictional force can be used to model internal dissipation during sprinting \cite{Furu}. In fact, it is known that more energy is dissipated by internal friction than air drag.  Hence, as an immediate extension to
the previous model, we can imagine that the player is subjected, besides the air drag, to this force too:\footnote{Physically, a player cannot apply a force on theirself. Therefore, from a fundamental point of view this force is not an internal one. The way to state it precisely is as follows: As the player sprints at higher speeds, energy  consumption  affects the ability of the player to push the ground with constant force. Instead,  the player pushes it with a variable effective force of $a-\gamma v$; in turn, the ground reacts and applies the opposite force on the player.}
\begin{align}
   {dv\over dt} =   a - \gamma\, v - k\, v^2  .
\label{eq:20}
\end{align}
The additional term provides us  the quantification of the physiological reason (i.e. the construction of the human body) that restricts  players from keeping a constant rate even in the absence of air drag.\footnote{Incidentally, notice that the inclusion of the internal dissipation implements  suggestion (b) of \cite{TH}.  As explained in the introduction, Taki and Hasegawa thought that as the player's speed increases, the rate of the increase has to decrease. Lacking a model, they did not link this effect to any force; it was just an expectation.}
And, if we wish to activate the term above a critical speed $v_\text{cr}$, then the coefficient $\gamma$ must be discontinuous:
$$
     \gamma = \begin{cases}
                                0,   &\text{if } v<v_\text{cr}, \\
                                \gamma_0,   &\text{if } v\ge v_\text{cr}.
                      \end{cases}
$$
It appears that this force quantifying internal dissipation is of great interest for many different sports in which sprinting occurs. Therefore, it would be a great feat if one can derive it from more fundamental principles of biokinesis and construction of the human body.

Returning to equation \eqref{eq:20}, by completing the square in the right-hand side, the equation can be written in the form
\begin{align}
   {dv\over dt} =   \left( a+{\gamma^2\over4k}\right)   - k\, \left(v+{\gamma\over 2k}\right)^2  .
\label{eq:21}
\end{align}
Hence, it has the exact same form with the equation studied in the previous section
\begin{align*}
   {du\over dt} =   a' - k\, u^2   ,\
\label{eq:22}
\end{align*}
 upon defining
$$
   u = v+ {\gamma\over 2k}, \quad  a'=a+{\gamma^2\over4k}.
$$
Integration of this equation gives
\begin{equation}
    u = u_\text{limit} \, \tanh[ku_\text{limit}  (t-t_0)+\delta] ,
\label{eq:23}
\end{equation}
where
$$
    \delta =   \tanh^{-1} {u_0\over u_\text{limit} } , \quad
    u_\text{limit}^2={a'\over k}.
$$
Then, integration of \eqref{eq:23} gives:
\begin{align*}
    r=& {1\over k} \,   \,  \ln\cosh{[ku_\text{limit}(t-t_0)+\delta]\over\cosh\delta} -  {\gamma\over 2k} \, (t-t_0) ,
\end{align*}
or
\begin{equation}
     r = -{1\over 2k}\, \ln\left[\cosh^2\delta\,  \Big(1-{u^2\over u_\text{limit}^2}\Big)\right] -  {\gamma\over 2k} \, (t-t_0).
\label{eq:24}
\end{equation}
At the point P where the player arrives with maximal speed,
\begin{equation*}
     r_\text{P} = {u_\text{limit}  \, \ln\left[ \cosh^2\delta \,  \Big(1-{U^2\over u_\text{limit}^2}\Big) \right]  \over 2(\delta - \Delta) }\,  (t-t_0)
                -  {\gamma \over  2k}\, (t-t_0) ,
\end{equation*}
where
$$
    \Delta =   \tanh^{-1} {U\over u_\text{limit} } , \quad
    U = V+ {\gamma\over 2k} .
$$
Factoring out the time,
\begin{equation}
     r_\text{P} =  \left[ {u_\text{limit}  \, \ln\left[\cosh^2\delta\, \Big(1-{U^2\over u_\text{limit}^2}\Big)\right]  \over 2(\delta - \Delta) }
                -  \varGamma\right] \,  (t-t_0) .
\label{eq:25}
\end{equation}

Similarly to $a,k$ being a function of the direction the player runs, $\gamma$ should be so too. After all, it relates to the fatigue of the muscles which do work differently along different directions (since $a$ is different for example). We will also assume that the ratio $\varGamma=\gamma/2k$ is the same in any direction.
Once more, with the assumptions presented, the points reached by the player obey an equation of the familiar form
$$
     r = A\, (t-t_0),
$$
where $A$ depends on the player's initial speed $v_0$ and characteristic speed $V$, as well the parameters $\Gamma$ and
$u_\text{limit}$. The last two parameters could be player dependent or could be universal constants. Since we have assumed that $V$ is the maximal
speed that a player can achieve based on their physiology, it is reasonable to assume that $v_\text{limit}$ is a universal constant imposed by nature. As such, since it depends on $\varGamma$, $u_\text{limit}$  should also be a universal constant. Using our previous estimation of $v_\text{limit}$, we can calculate a reasonable value for $\Gamma$ to use in our calculations. If we assume an average elite level player can run from one end of the pitch to the other in 15 seconds, we find that $\Gamma = 2.237$ $\text{m/s}$.

Since $U$ is a particular value of the variable $u$ whose maximum value is $u_\text{limit}$, the ratio $U / u_\text{limit}$ must be
always less than one.
The choice of our parameters  entering  our calculations are consistent with this restriction.

In Figure \ref{fig:BifocalTwinFriction}, we again plot the eight scenarios for two players with these new considerations.
%
{\begin{figure}[h!]
    \centering
    \subfigure[\parbox{5cm}{$v_1 = 8\, \text{m/s}, v_2 = 6\, \text{m/s},\\  \alpha_1 = -135^{\circ}, \alpha_2 = 45^{\circ}$.}]%
                    {\includegraphics[width=6cm]{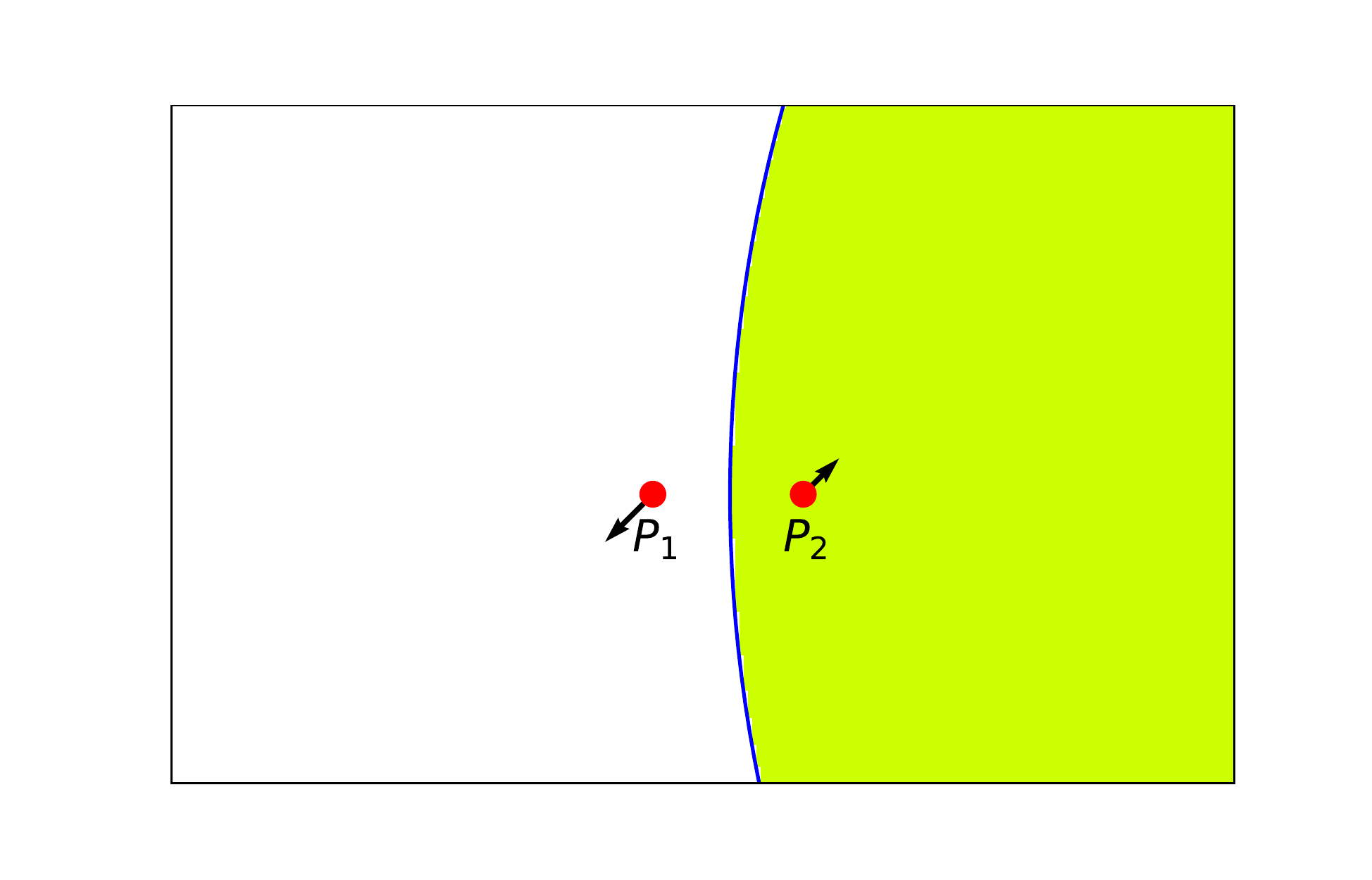}}
    \subfigure[\parbox{5cm}{$v_1 = 8\, \text{m/s}, v_2 = 3\, \text{m/s},\\ \alpha_1 = 0^{\circ}, \alpha_2 = 180^{\circ}$.}]%
                   {\includegraphics[width=6cm]{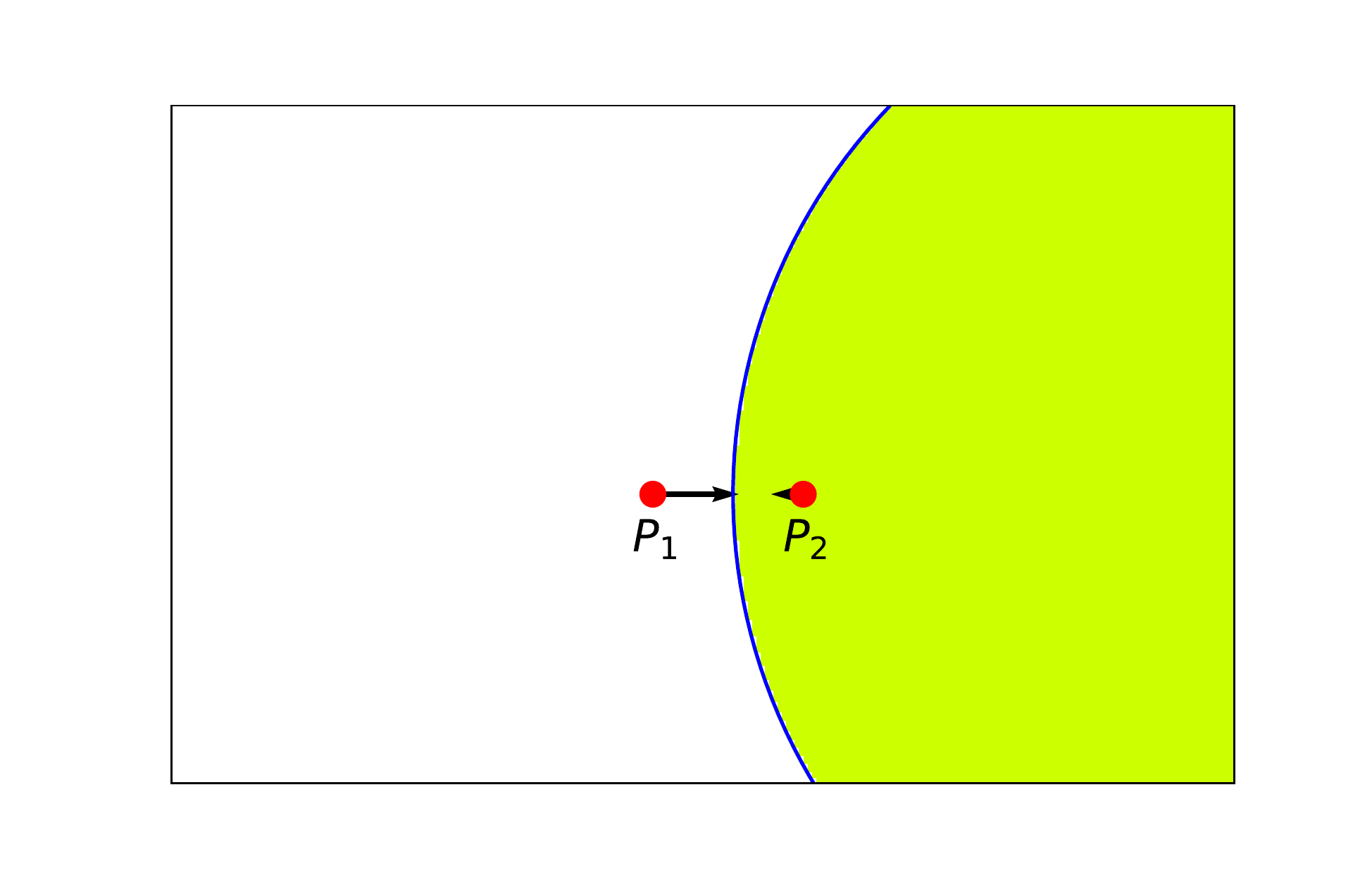}}\\
    \subfigure[\parbox{5cm}{$v_1 \sim 9\, \text{m/s}, v_2 = 0\, \text{m/s},\\ \alpha_1 = 45^{\circ}, \alpha_2 = -50^{\circ}$.}]%
                   {\includegraphics[width=6cm]{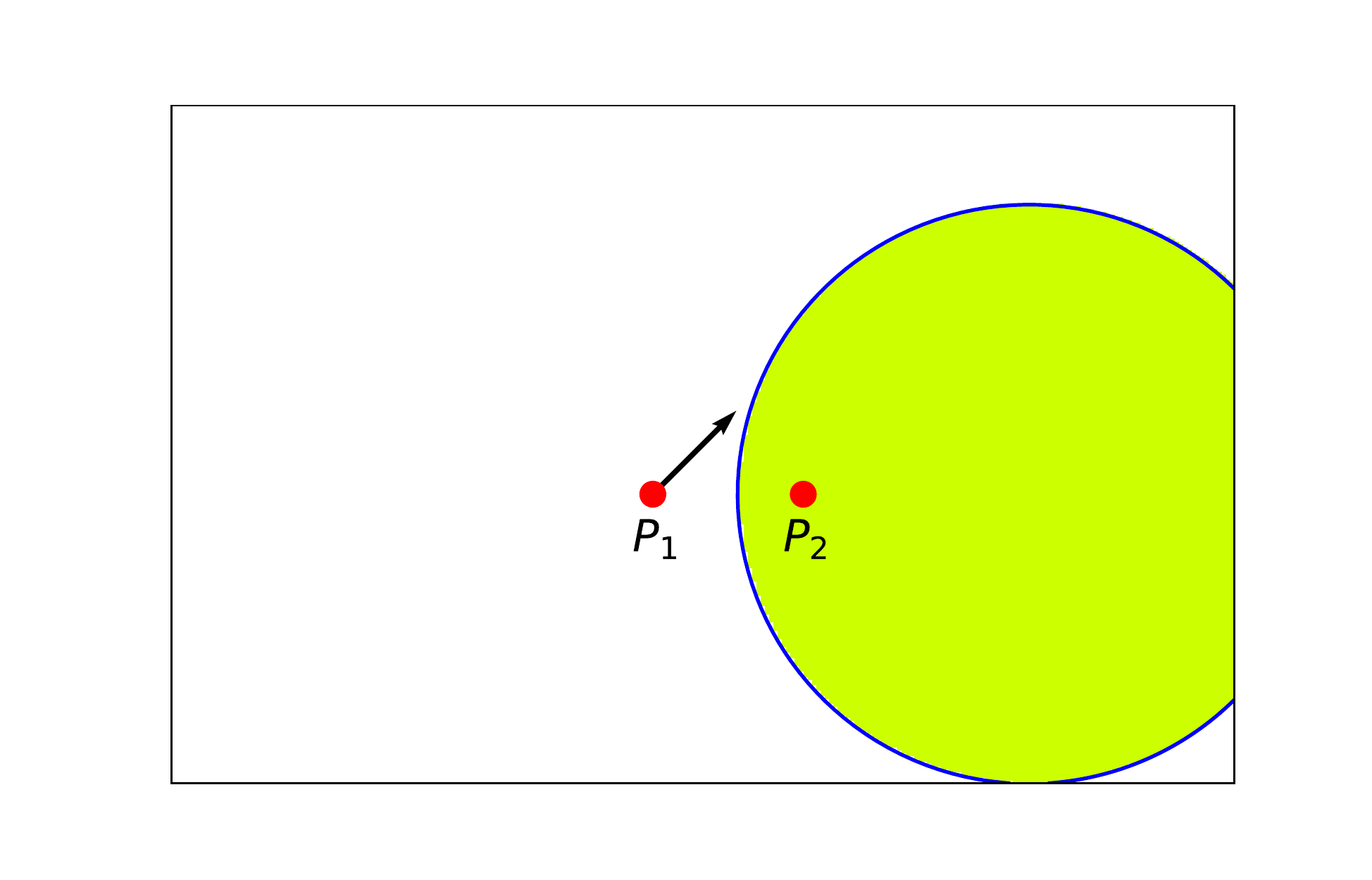}}
    \subfigure[\parbox{5cm}{$v_1 \sim 9\,\text{m/s}, v_2 \sim 9\,\text{m/s},\\ \alpha_1 = 90^{\circ}, \alpha_2 = 90^{\circ}$.}]%
                   {\includegraphics[width=6cm]{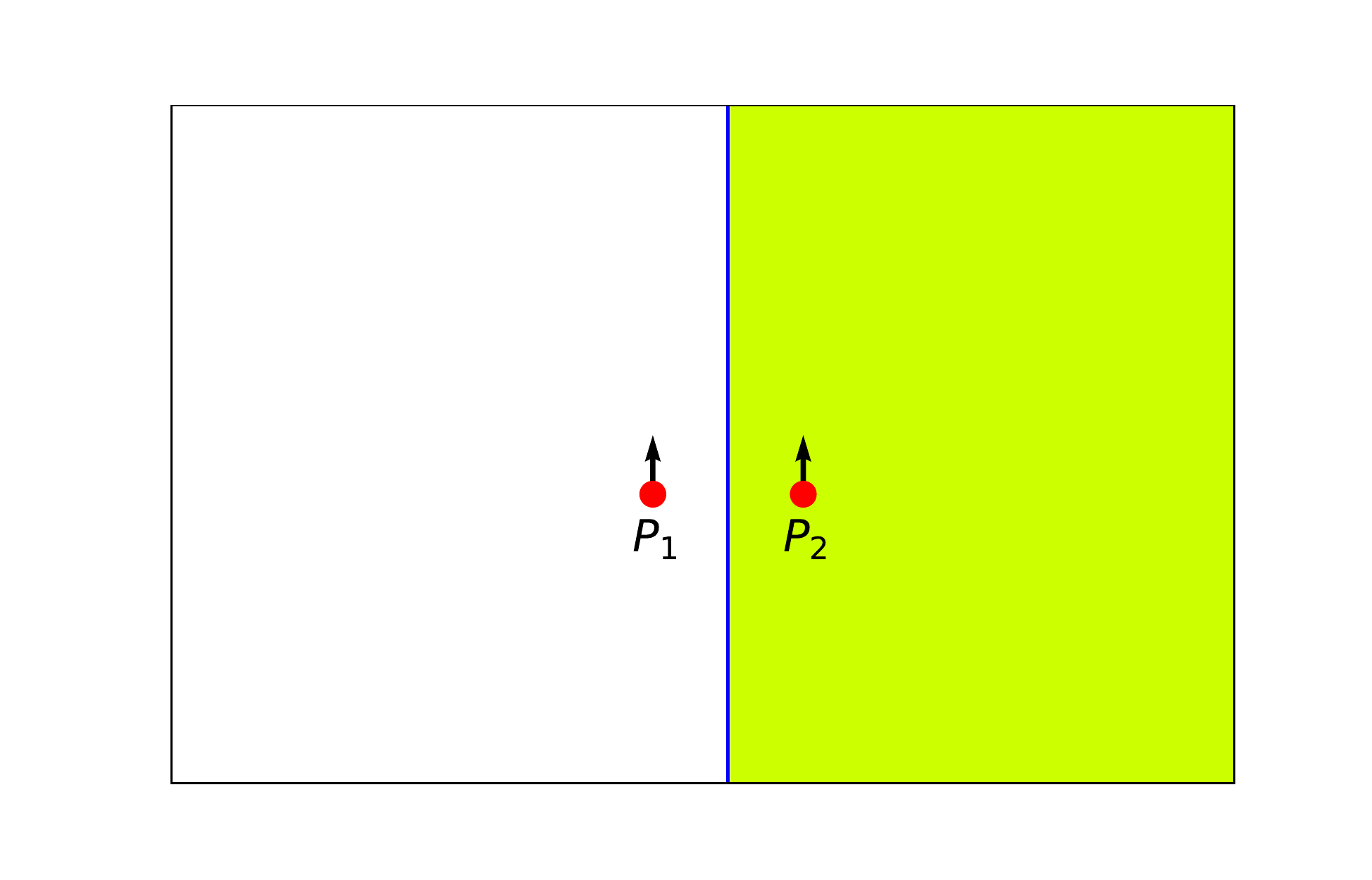}}\\
    \subfigure[\parbox{5cm}{$v_1 \sim 9\, \text{m/s}, v_2 \sim 9\, \text{m/s},\\  \alpha_1 = 45^{\circ}, \alpha_2 = 135^{\circ}$.}]%
                    {\includegraphics[width=6cm]{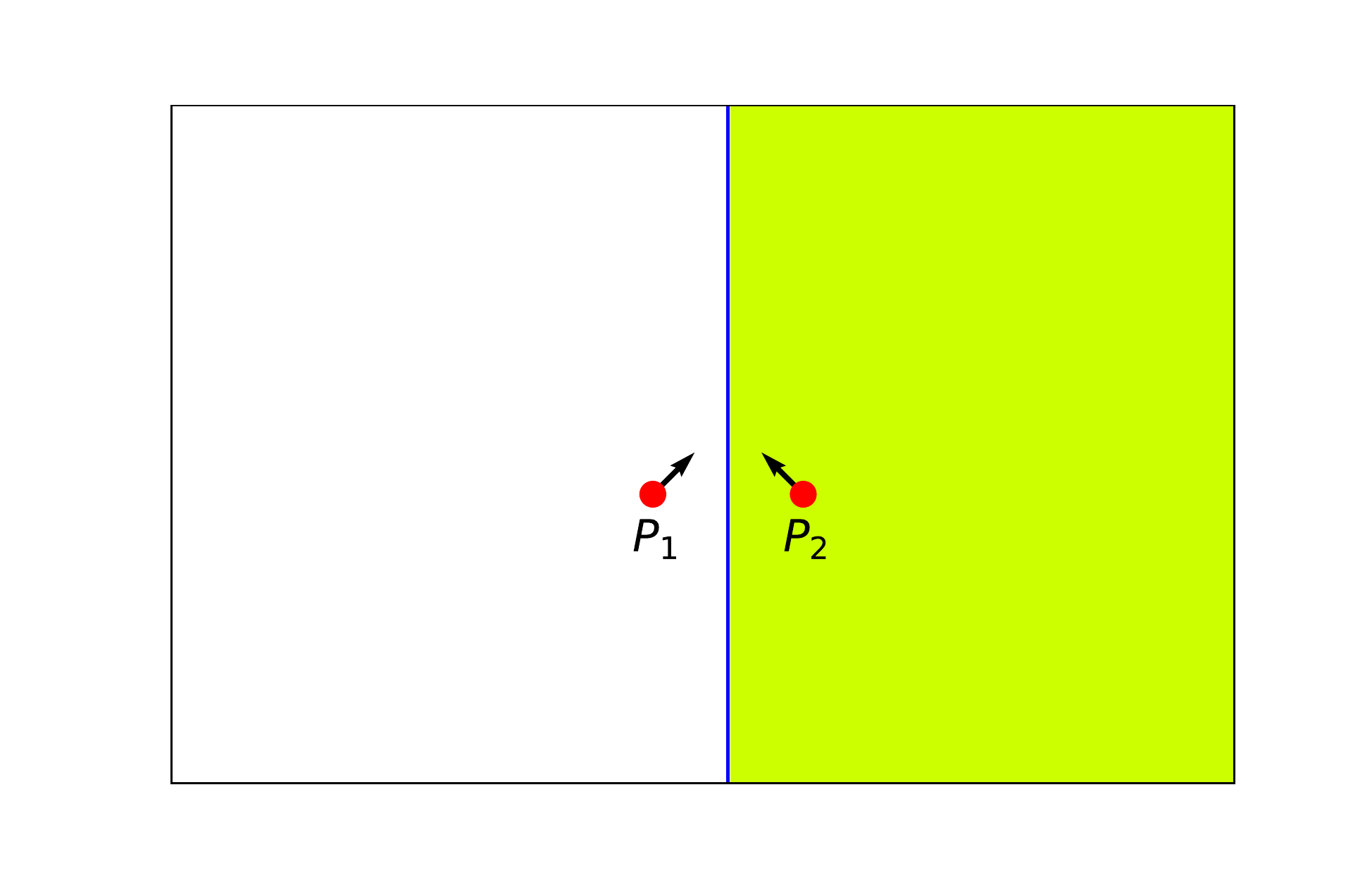}}
    \subfigure[\parbox{5cm}{$v_1 \sim 9\, \text{m/s}, v_2 \sim 9\, \text{m/s},\\ \alpha_1 = 90^{\circ}, \alpha_2 = 180^{\circ}$.}]%
                   {\includegraphics[width=6cm]{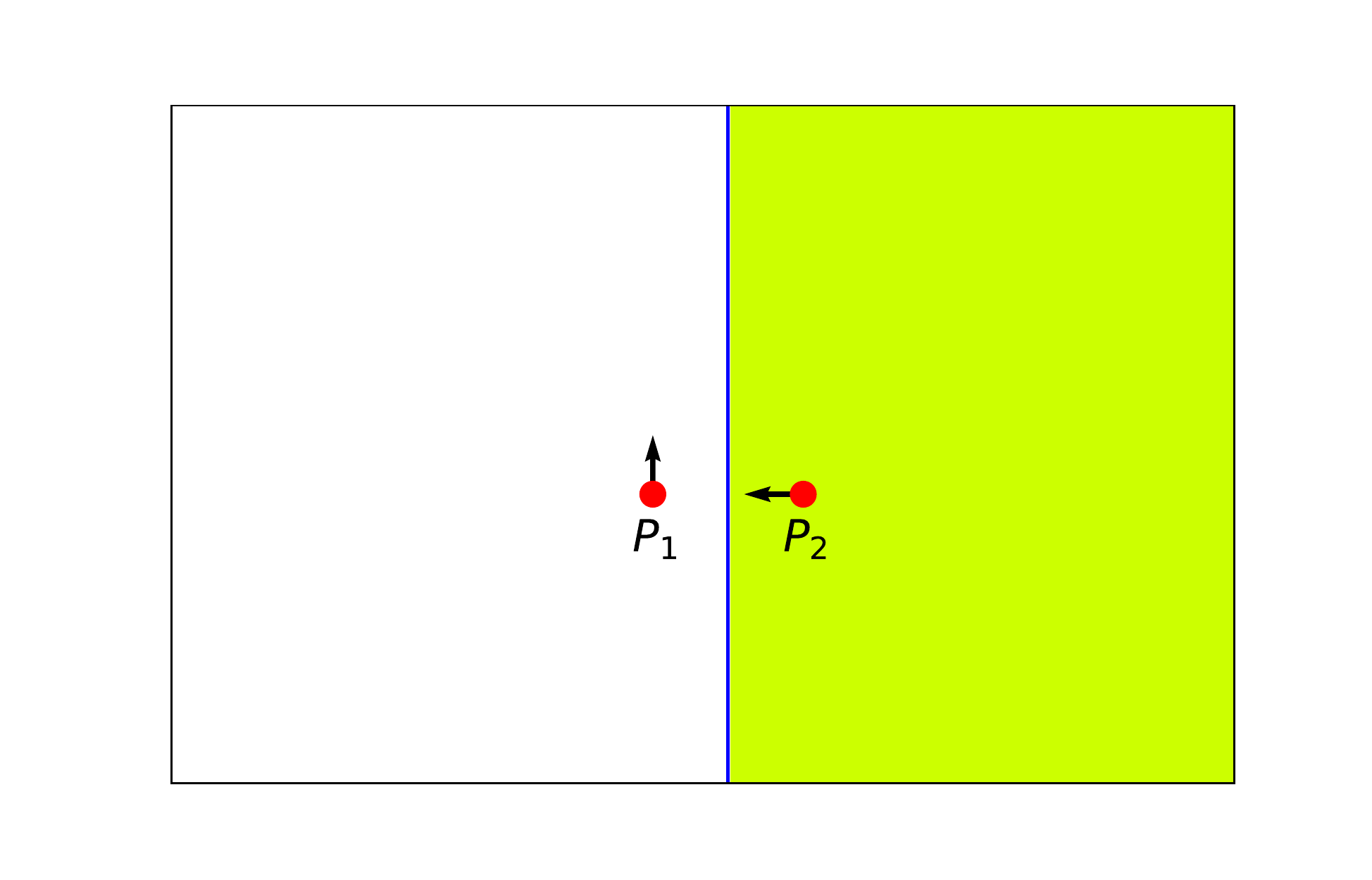}}\\
    \subfigure[\parbox{5cm}{$v_1 = 6\, \text{m/s}, v_2 = 6\, \text{m/s},\\ \alpha_1 = -45^{\circ}, \alpha_2 = 135^{\circ}$.}]%
                   {\includegraphics[width=6cm]{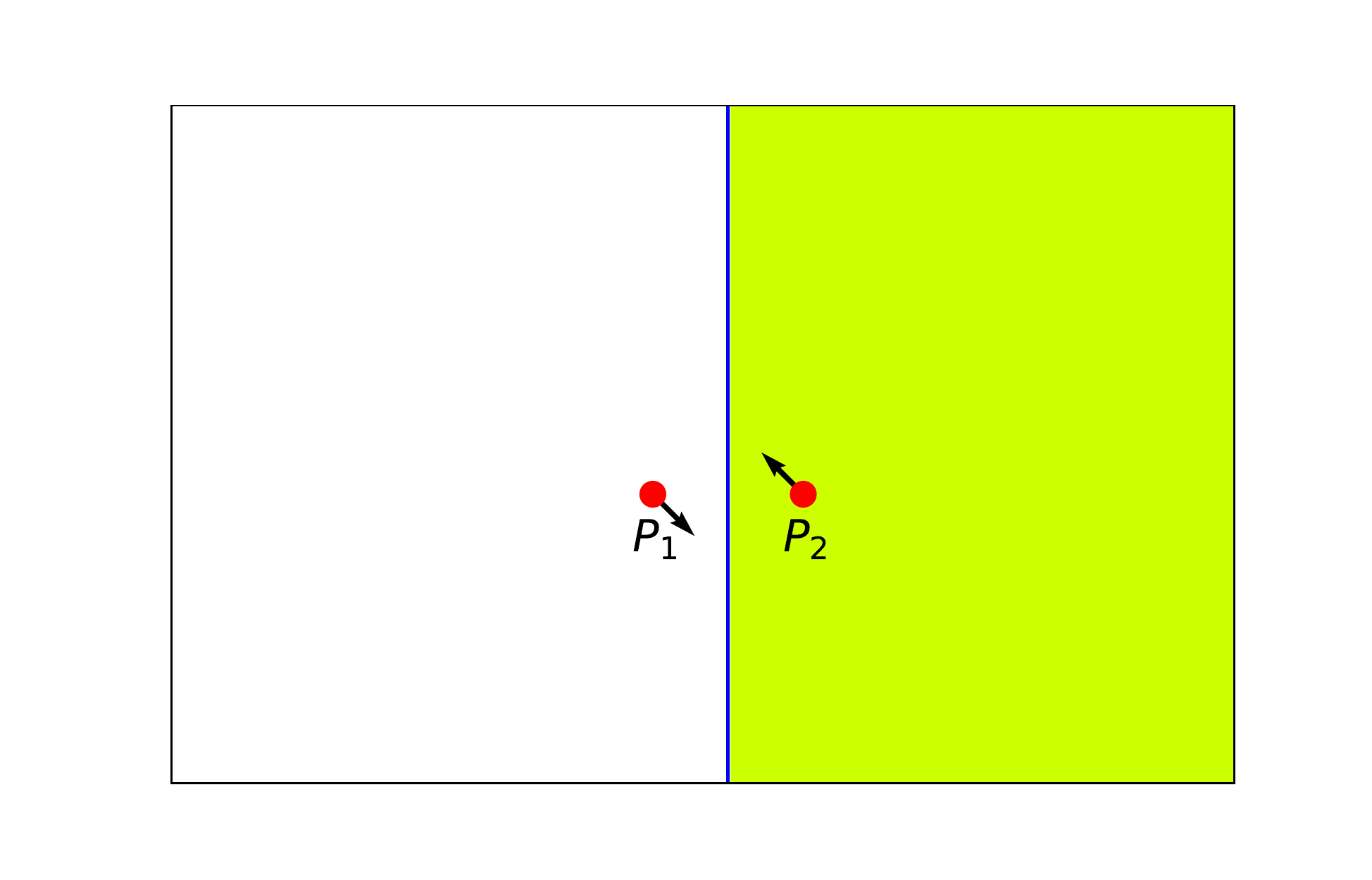}}
    \subfigure[\parbox{5cm}{$v_1=3\,\text{m/s}, v_2=8\,\text{m/s},\\ \alpha_1 = -45^{\circ}, \alpha_2 = -45^{\circ}$.}]%
                   {\includegraphics[width=6cm]{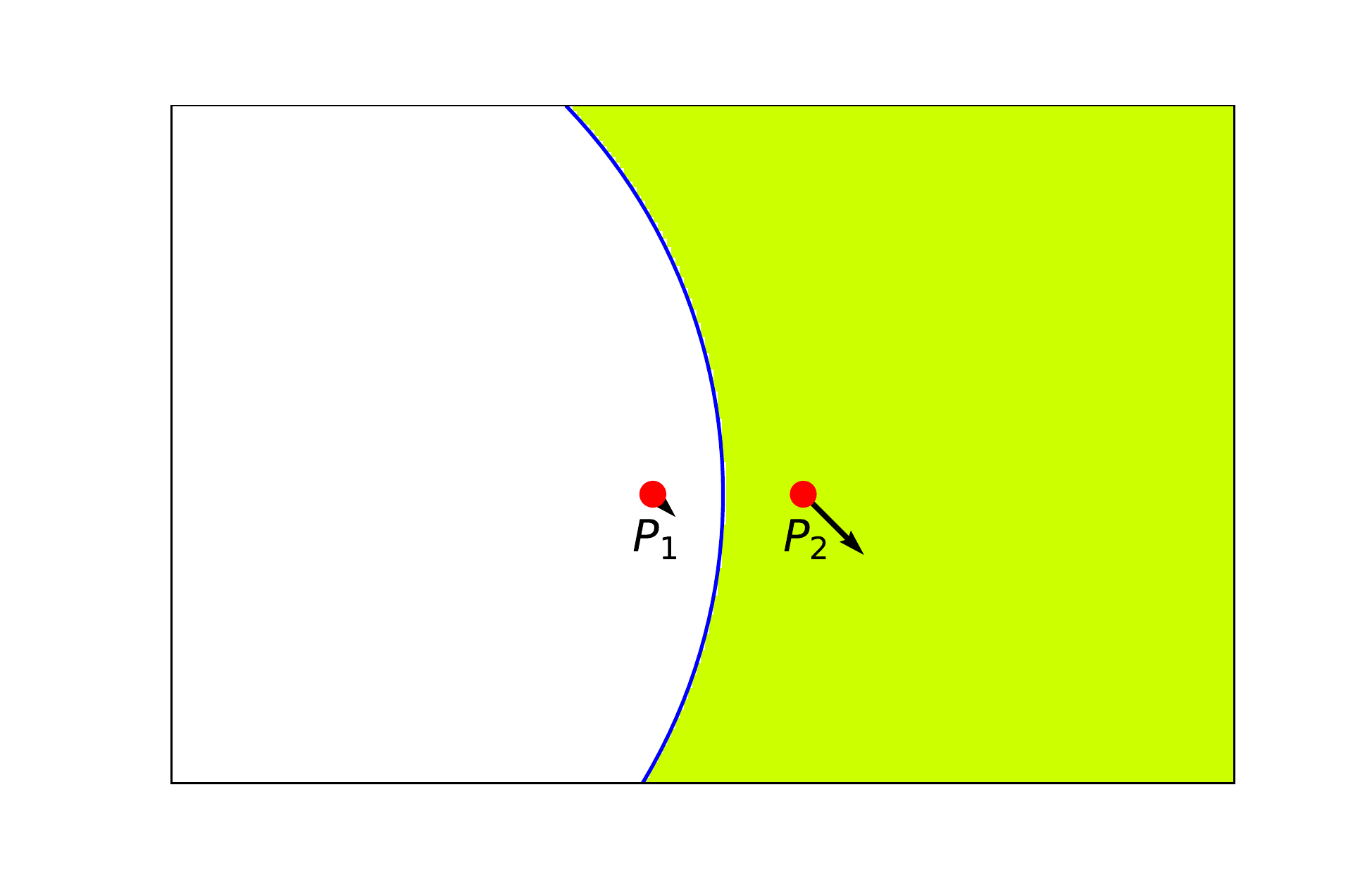}}
\caption{\footnotesize Sketch of the boundary between two players in the presence of air resistance and internal resistance on a
              212 meters  by 136 meters pitch. The bottom left corner and the upper right corner have coordinates $(-106,-68)$ and
              (106,68) respectively.
                The two players P$_1$ and P$_2$ are located at the points $(-10, -10)$ and $(20, -10)$.  The cases are identical with those
                of Figure \ref{fig:BifocalOriginal}.
                The reader can compare the results  with those of Figures \ref{fig:BifocalOriginal} and  \ref{fig:BifocalAirDrag}  to see the
                relative changes.}
\label{fig:BifocalTwinFriction}
\end{figure}}

Next, we plot the boundaries between all twenty-two players for the same two Metrica Sports frames --- see Figure~\ref{fig:twinFrictionVoronoi}.
%
{\begin{figure}[h!]
    \centering
    \subfigure[Frame: 98202]{\includegraphics[width=12cm]{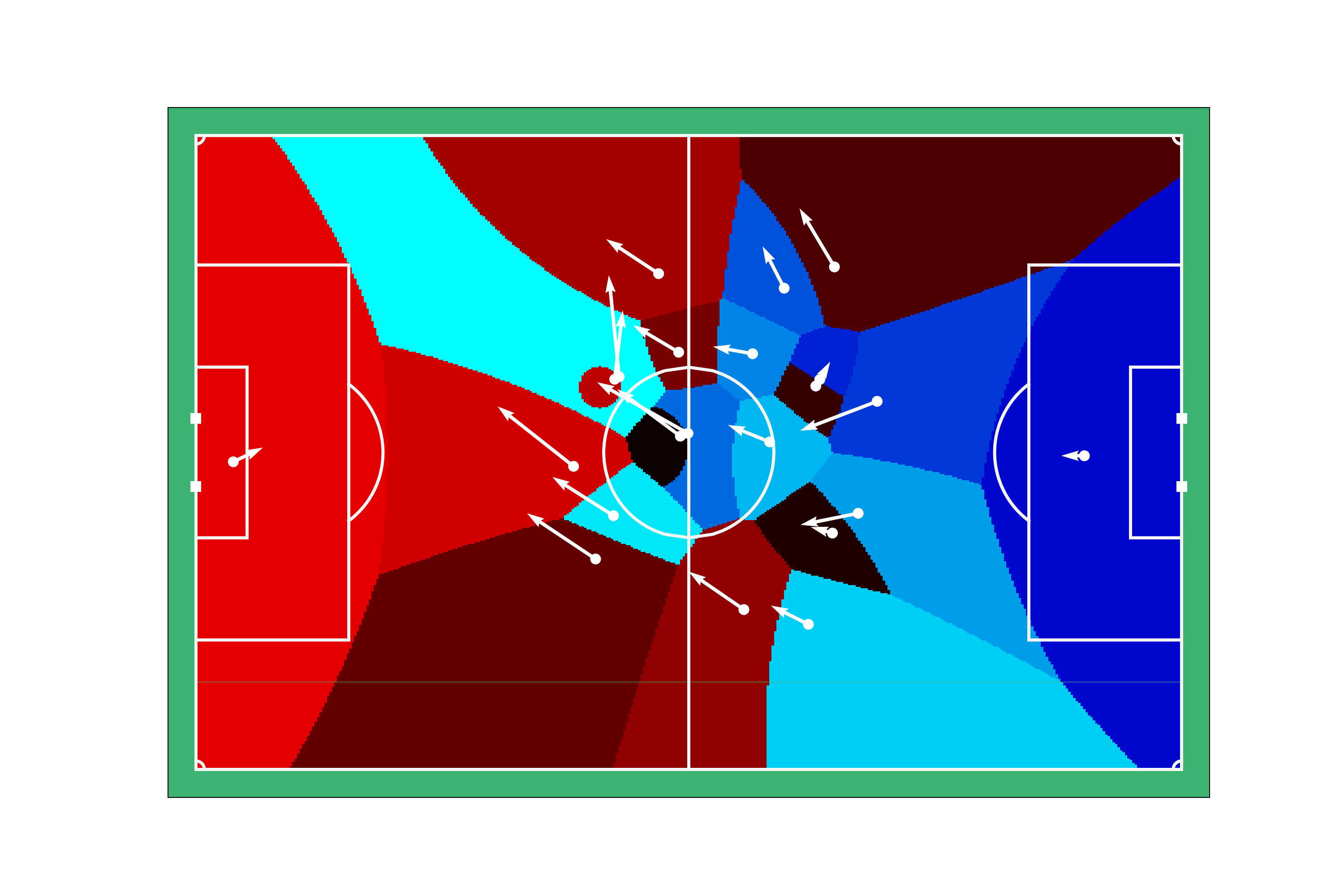}}\\
    \subfigure[Frame: 123000]{\includegraphics[width=12cm]{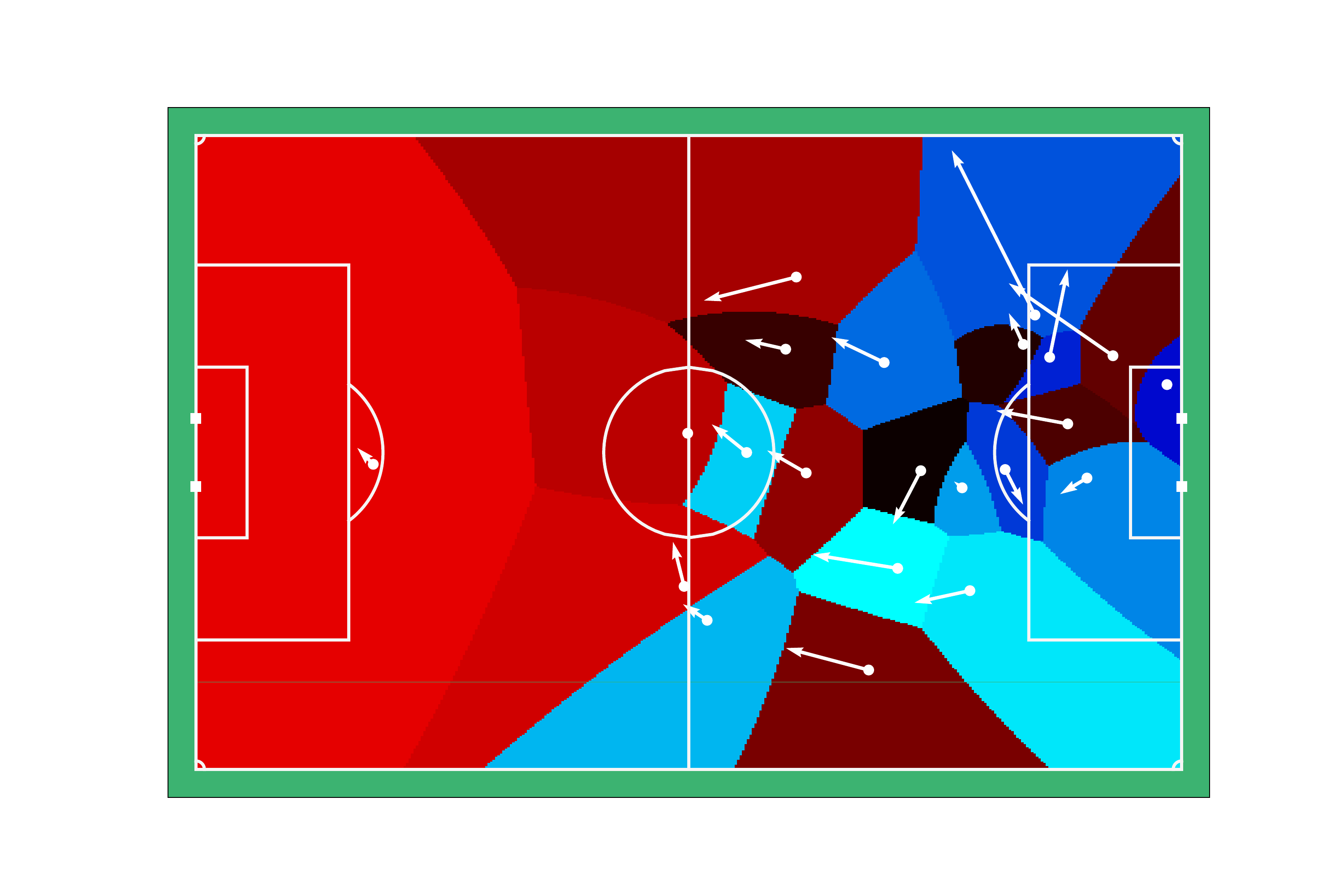}}
\caption{\footnotesize Sketches of Apollonius diagrams in the presence of air drag and internal friction, representing dominance areas for players
               adopted from  Metrica Sports open tracking data.}
\label{fig:twinFrictionVoronoi}
\end{figure}}

\subsection*{\normalsize Introduction of Directional Bias}

For uniformly accelerated motion with acceleration $\vec a$,
 at time $t$, the location of each player will be given by
$$
    \vec r_i = \vec v_i \, (t-t_i) + {1\over 2} \, \vec a_i \, (t-t_i)^2, \quad i=1,2,
$$
Similarly, the velocities of the two players will be
$$
    \vec V_i = \vec v_i + \vec a_i \, (t-t_i) , \quad i=1,2.
$$

We can decompose the above vectorial equations to two algebraic equations: one  radial equation along P$_i$P and one equation along the perpendicular direction. It is tempting to argue that the latter direction determines the reaction time $t_1$; it is connected, at least in part, to the time necessary to make the component $v_i\sin\theta_i$ vanish. Experience indicates that this can happen fast: The player can plant their foot on the ground and receive a push in the proper direction.\footnote{In fact, these abrupt changes often lead to injuries. It would be interesting here to examine the forces on the players' legs and feet. However, this would take us off target.} Hence,  the time delays due to the vanishing of extraneous component of the velocity are expected to be very small. When the angle of the change of direction is quite big (greater than 90$^\circ$), there will be some delay since the player will need to reorient their body and hence will need to spin about their axis.  Finally, recall from \cite{efthimiou} that the time delay parameter is a way to quantify our ignorance about all different aspects influencing the playing not to respond instantaneously. For example, besides the time delay originating from attempting to change the direction of motion (aka make the  component $v_i\sin\theta_i$ vanish), it includes the player's physiological delay and perhaps other environmental factors that influence the motion. Eventually, we will set the time delay parameter $t_i$ to zero but experience indicates that it does not have to be so.

The equations of motion along the radial direction are:
\begin{align*}
     r_i      =&  v_i \, \cos\theta_i \, (t-t_i) + {1\over 2} \, a_i \, (t-t_i)^2 , \\
     V_i     =&  v_i \, \cos\theta_i + a_i \, (t-t_i), \quad i=1,2,
\end{align*}
 With the help of the second relation, we can rewrite the first in the form
\begin{equation}
     r_i =   A_i \, (t-t_i), \quad i=1,2,
 \label{eq:15}
\end{equation}
where
$$
     A_i = {  v_i \, \cos\theta_i+  V_i \over 2 } , \quad i=1,2 .
$$

Result \eqref{eq:15} is similar to those in \cite{efthimiou}. However,  the factor $A_i$ in the original work was \textit{isotropic}; now it has a directional bias. In particular, the factor is skewed along the direction of the player's motion. The points $r_i=\text{const.}$ that the player can reach at time $t$ from their original position lie on a limacon of Pascal.
When $v_i=V_i$, the limacon becomes a cardioid. Figure \ref{fig:limacon} plots the function $A=A(\theta)$ for five different values of the ratio $v/V$.
%
{\begin{figure}[h!]
\centering
\setlength{\unitlength}{1mm}
\begin{picture}(60,55)
\put(0,-3){\includegraphics[width=6cm]{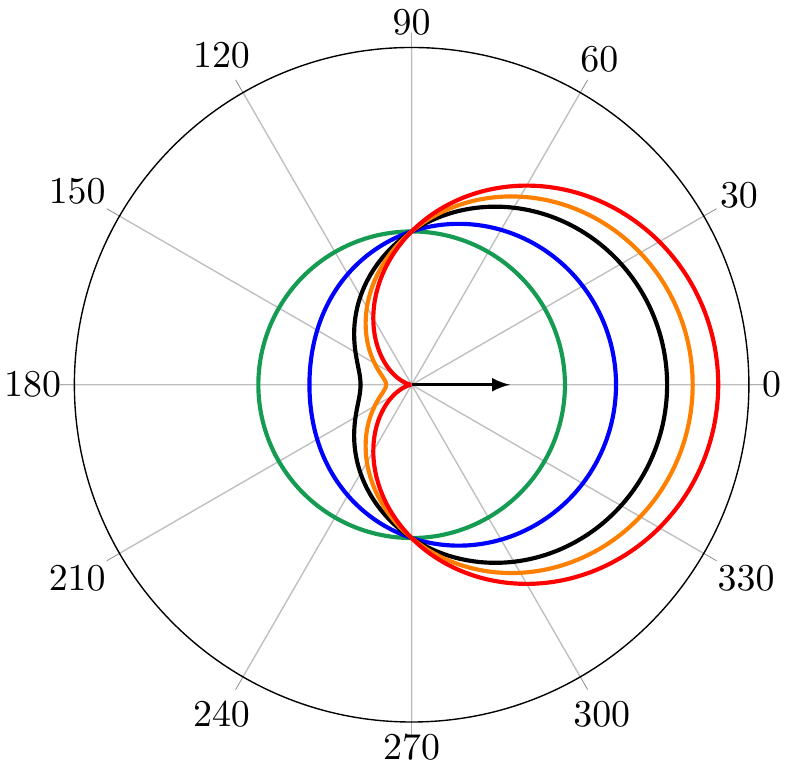}}
\put(38,27.5){\footnotesize $\vec v$}
\end{picture}
\caption{\footnotesize Plot of the function $2A= V+v\, \cos\theta$ in polar coordinates for different magnitudes of $v$. The player is located at the point (0,0) and moving with  initial velocity $\vec v$ as indicated by the corresponding vector. When the player is at rest, $v=0$, the curve is a circle. That is, the player can move along any direction with the same level of ease. However, as their speed $v$ increases, the player starts to have greater difficulty to reach points behind them. At $v=V$, the points on the halfline  behind the player's back are unaccessible.}
\label{fig:limacon}
\end{figure}}

The points that the two players can reach simultaneously at time $t$ from their original positions satisfy
\begin{equation}
     k_1 \, r_1 -k_2 \,  r_2  = t_0 ,
\label{eq:6}
\end{equation}
where  $k_1, k_2$ are the players' slownesses:
\begin{equation}
       k_i = {1\over A_i} = {2 \over \  v_i \, \cos\theta_i + V_i }  \,  \raisebox{2pt}{,} \quad i=1,2 ,
\label{eq:6a}
\end{equation}
and $t_0$ the relative time delay, $t_0=t_2-t_1$.
Assuming that the players' time delays vanish or they are equal, we find
\begin{equation}
     {r_1 \over r_2} =  {v_1 \, \cos\theta_1 + V_1 \over v_2 \, \cos\theta_2 + V_2 }   .
\label{eq:7}
\end{equation}
Because the angle variables $\theta_1, \theta_2$ are measured from different axes, it is better to replace them by
$$
      \phi_i = \theta_i + \alpha_i, \quad i=1,2,
$$
where $\alpha_1,\alpha_2$ are the angles the velocity vectors $\vec v_1, \vec v_2$ make with the line P$_1$P$_2$.  Hence,
\begin{equation}
     {r_1 \over r_2} =  {v_1 \, \cos(\phi_1-\alpha_1) + V_1 \over v_2 \, \cos(\phi_2-\alpha_2) + V_2 }   .
\label{eq:16}
\end{equation}
The pairs $(r_1,\phi_1)$ and $(r_2,\phi_2)$  form a set  of twin polar coordinates with the angles measured from the same line.

In \cite{efthimiou}, we have reviewed the curves resulting from equation \eqref{eq:6} for constant slowness. These are the well known Cartesian
ovals. When $t_0=0$, the Cartesian ovals reduce to Apollonius circles. In the new scenario in which the player's slowness depends on the direction of motion, the Cartesian ovals are in general deformed to new shapes. Unfortunately, to the best of our knowledge, there is no study for such curves. Hence, at the moment, to get a feeling on how the curves look like, we rely on numerical plots.
In Figure \ref{fig:BifocalDirectionalBias}, we plot the boundary between two players (equation \eqref{eq:16})  in the eight scenarios of Figure
\ref{fig:BifocalOriginal}.  In these diagrams, we see that each player has a stronger control in front of theirself than behind theirself, the boundary between only two players can have disconnected regions,\footnote{A set $S=A\cup B$ is made of two disconnected sets $A$ and $B$, if $A\cap B=\varnothing$. In the examples
of Figures \ref{fig:BifocalDirectionalBias} and \ref{fig:BifocalAll}, when the dominance region is made of two sets $A$ and $B$, their intersection is either empty or it contains a point. Hence, for our purposes, we shall call a set $S$ the disconnected union of $A$ and $B$ if the intersection of $A$ and $B$ is a countable set.} and the player with a higher initial speed does not necessarily control a greater area. The reader will certainly appreciate the boundary changes induced by the directional bias in Figure \ref{fig:BifocalDirectionalBias} compared
to Figure \ref{fig:BifocalOriginal}.
%
{\begin{figure}[h!]
    \centering
    \subfigure[\parbox{5cm}{$v_1 = 8\, \text{m/s}, v_2 = 6\, \text{m/s},\\  \alpha_1 = -135^{\circ}, \alpha_2 = 45^{\circ}$.}]%
                    {\includegraphics[width=6cm]{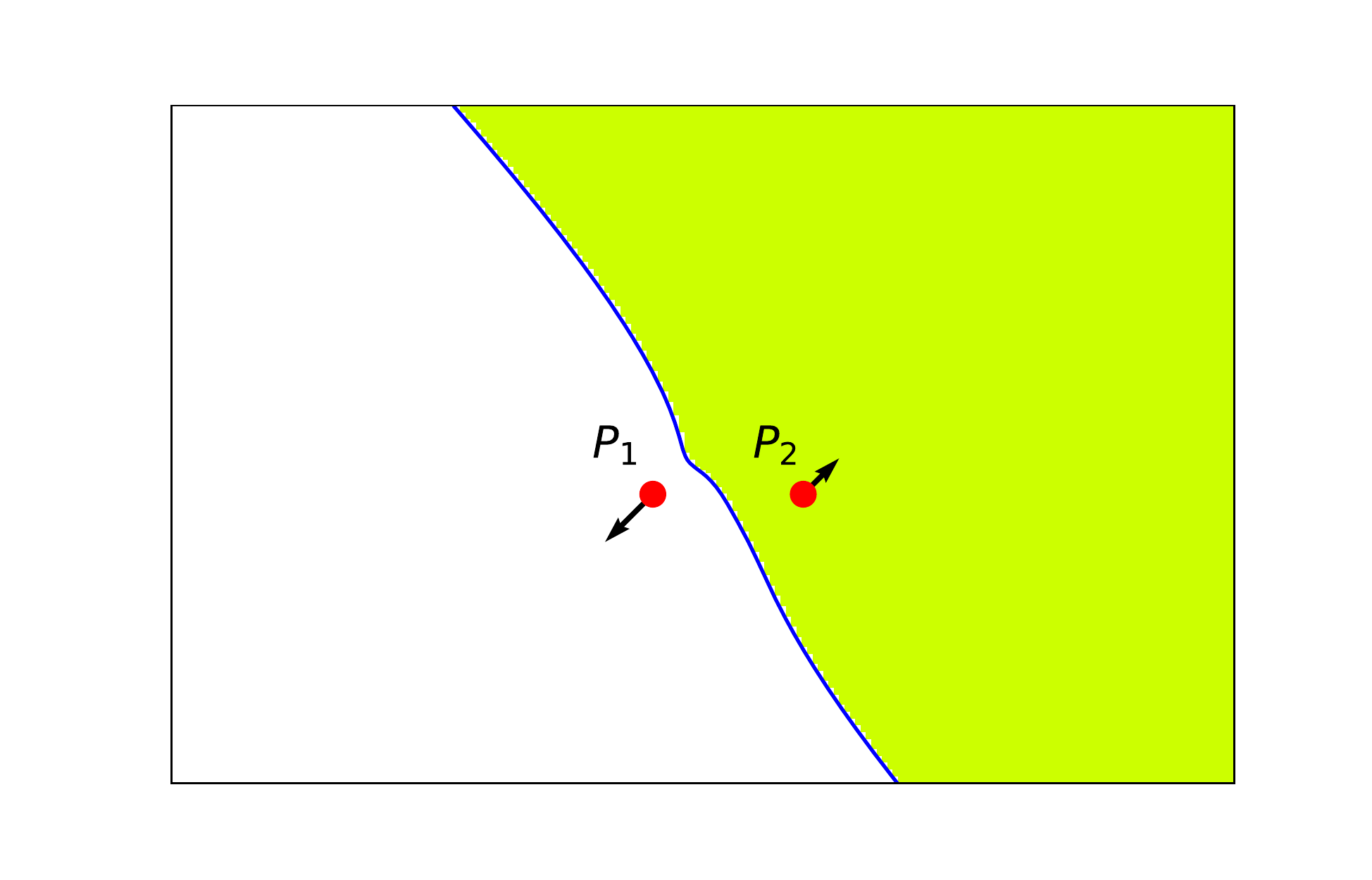}}
    \subfigure[\parbox{5cm}{$v_1 = 8\, \text{m/s}, v_2 = 3\, \text{m/s},\\ \alpha_1 = 0^{\circ}, \alpha_2 = 180^{\circ}$.}]%
                   {\includegraphics[width=6cm]{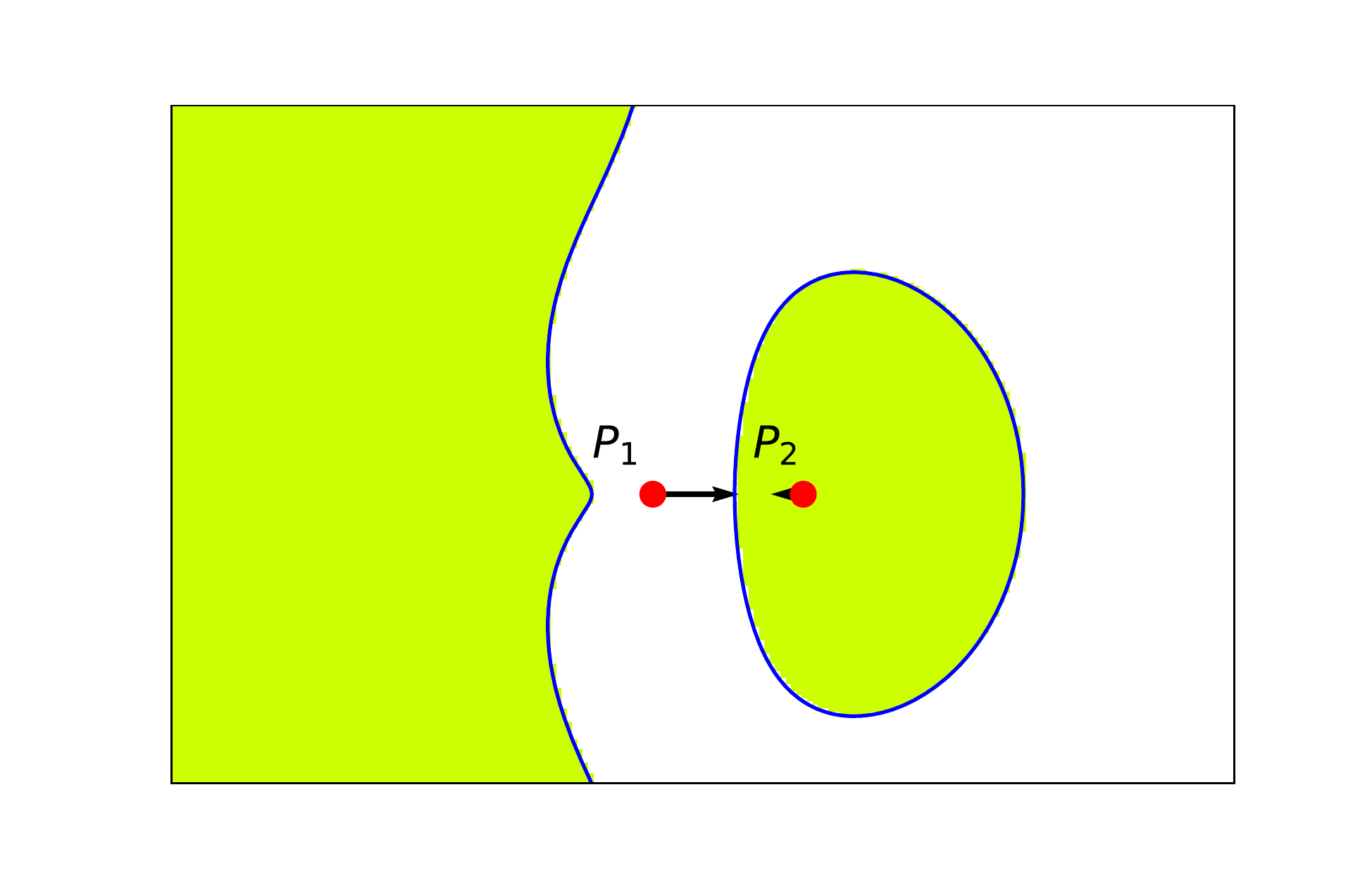}}\\
    \subfigure[\parbox{5cm}{$v_1 \sim 9\, \text{m/s}, v_2 = 0\, \text{m/s},\\ \alpha_1 = 45^{\circ}, \alpha_2 = -50^{\circ}$.}]%
                   {\includegraphics[width=6cm]{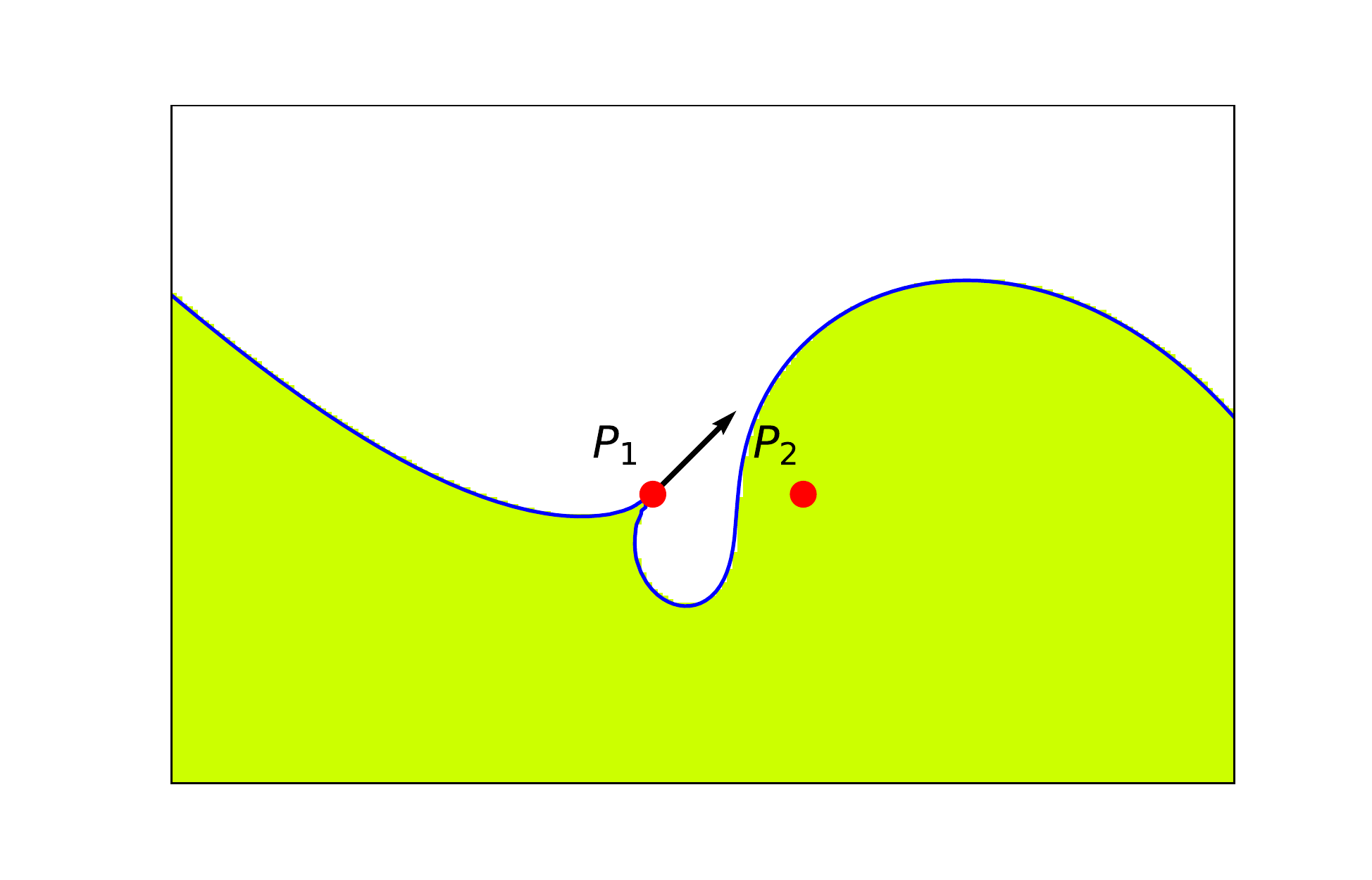}}
    \subfigure[\parbox{5cm}{$v_1 \sim 9\,\text{m/s}, v_2 \sim 9\,\text{m/s},\\ \alpha_1 = 90^{\circ}, \alpha_2 = 90^{\circ}$.}]%
                   {\includegraphics[width=6cm]{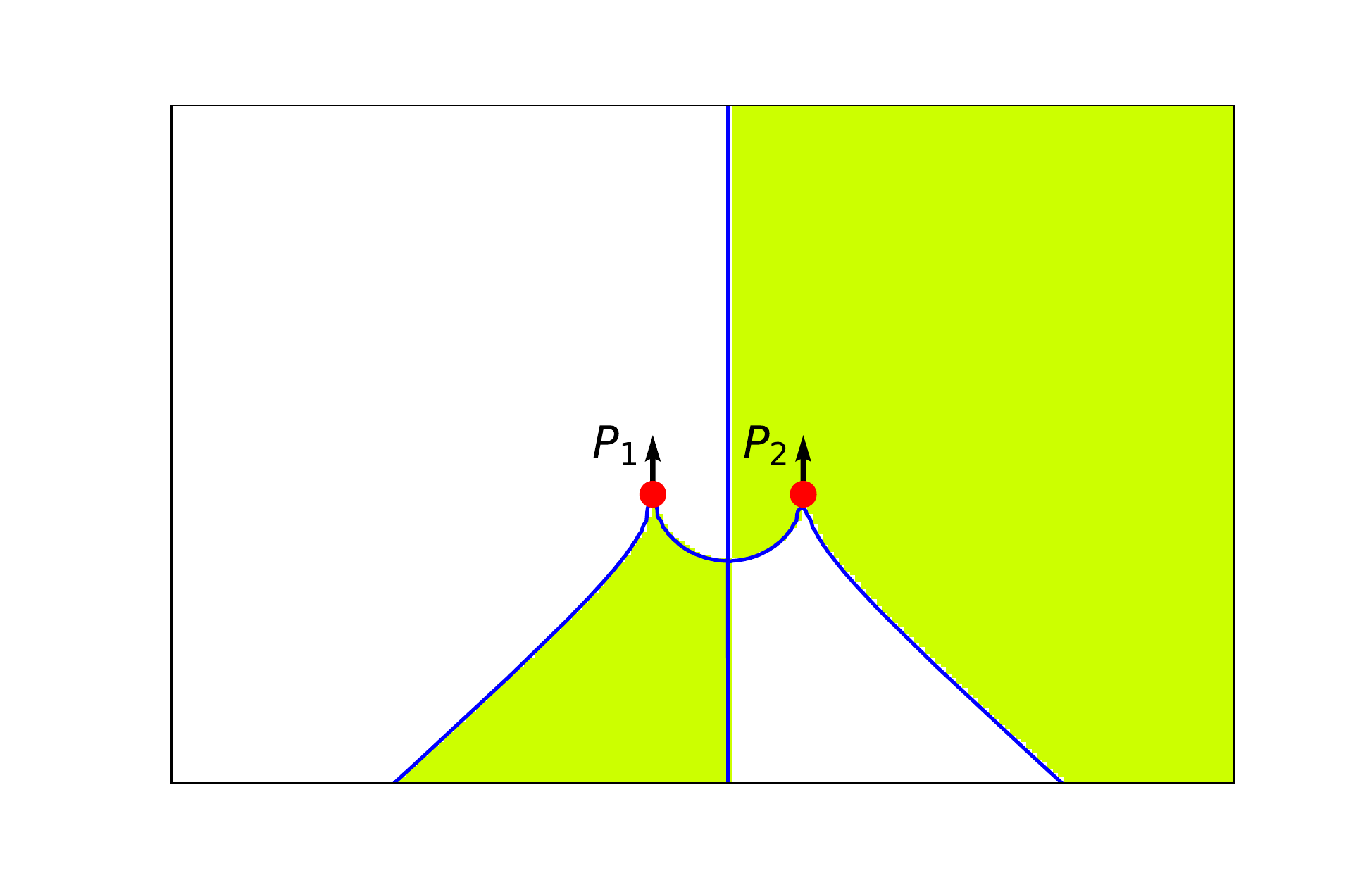}}\\
    \subfigure[\parbox{5cm}{$v_1 \sim 9\, \text{m/s}, v_2 \sim 9\, \text{m/s},\\  \alpha_1 = 45^{\circ}, \alpha_2 = 135^{\circ}$.}]%
                    {\includegraphics[width=6cm]{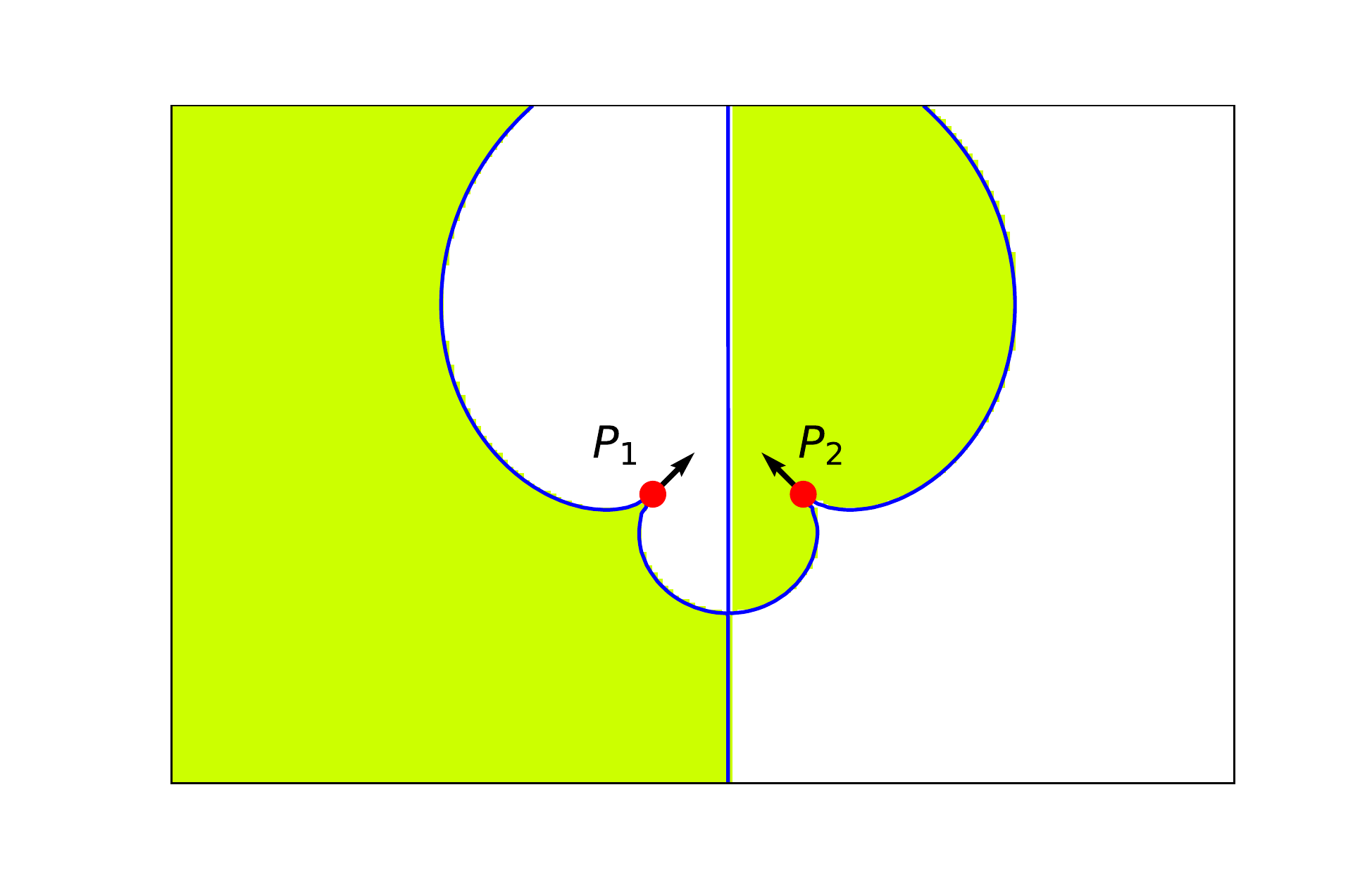}}
    \subfigure[\parbox{5cm}{$v_1 \sim 9\, \text{m/s}, v_2 \sim 9\, \text{m/s},\\ \alpha_1 = 90^{\circ}, \alpha_2 = 180^{\circ}$.}]%
                   {\includegraphics[width=6cm]{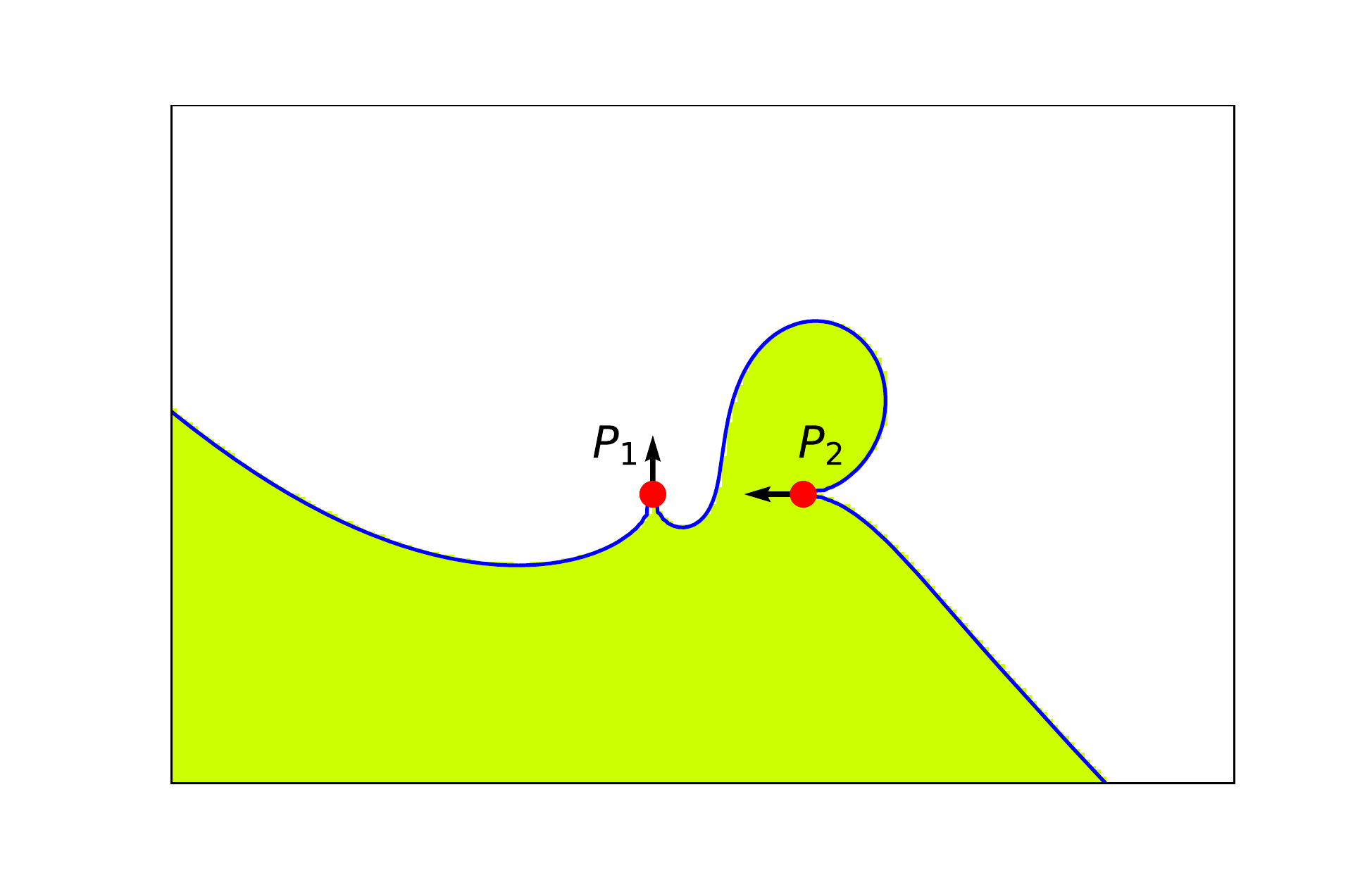}}\\
    \subfigure[\parbox{5cm}{$v_1 = 6\, \text{m/s}, v_2 = 6\, \text{m/s},\\ \alpha_1 = -45^{\circ}, \alpha_2 = 135^{\circ}$.}]%
                   {\includegraphics[width=6cm]{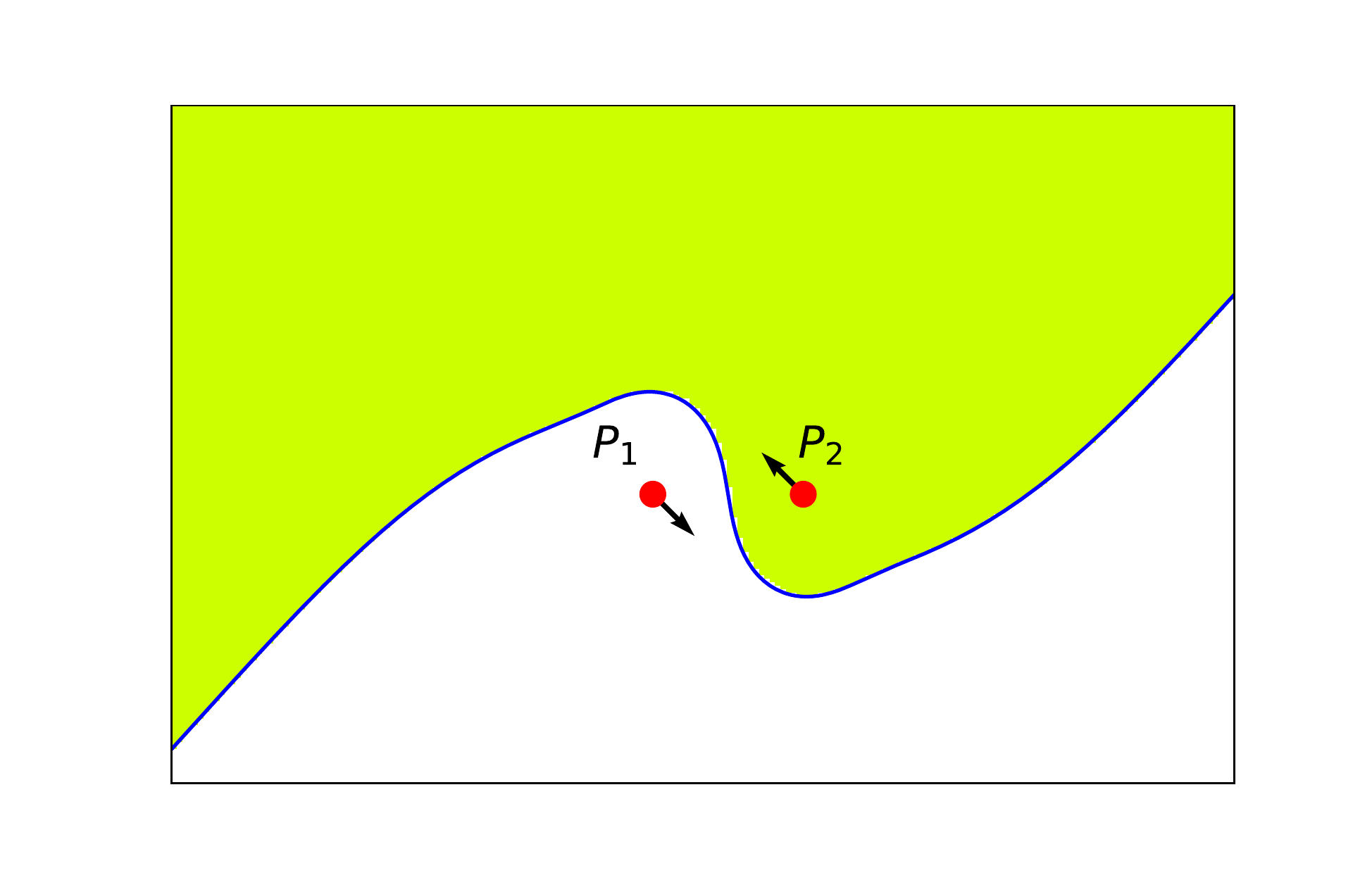}}
    \subfigure[\parbox{5cm}{$v_1=3\,\text{m/s}, v_2=8\,\text{m/s},\\ \alpha_1 = -45^{\circ}, \alpha_2 = -45^{\circ}$.}]%
                   {\includegraphics[width=6cm]{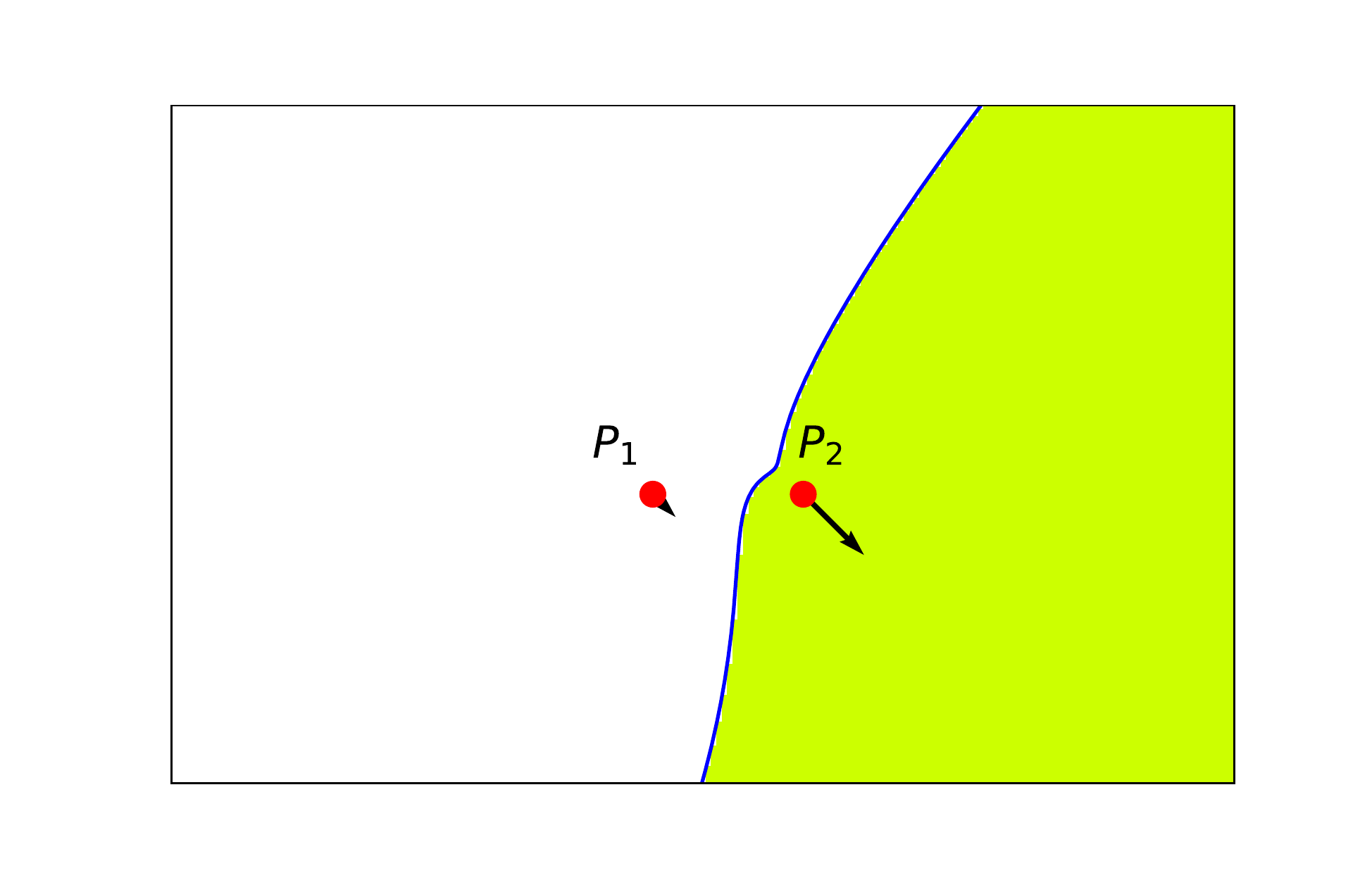}}
\caption{\footnotesize Sketch of the boundary between two players in the presence of directional bias  on a 212 meters by 136 meters pitch.
               The bottom left corner and the upper right corner have coordinates $(-106,-68)$ and $(106,68)$ respectively.
                The two players P$_1$ and P$_2$ are located at the points $(-10, -10)$ and $(20, -10)$.  The cases are identical with those
                of Figure \ref{fig:BifocalOriginal}. However, both the magnitude and direction of the velocity affect the dominance regions.
                The reader can easily compare the two results to see how drastic the changes are.}
\label{fig:BifocalDirectionalBias}
\end{figure}}
%

{\begin{figure}[h!]
    \centering
    \subfigure[\parbox{5cm}{$v_1 = 8\, \text{m/s}, v_2 = 6\, \text{m/s},\\  \alpha_1 = -135^{\circ}, \alpha_2 = 45^{\circ}$.}]%
                    {\includegraphics[width=6cm]{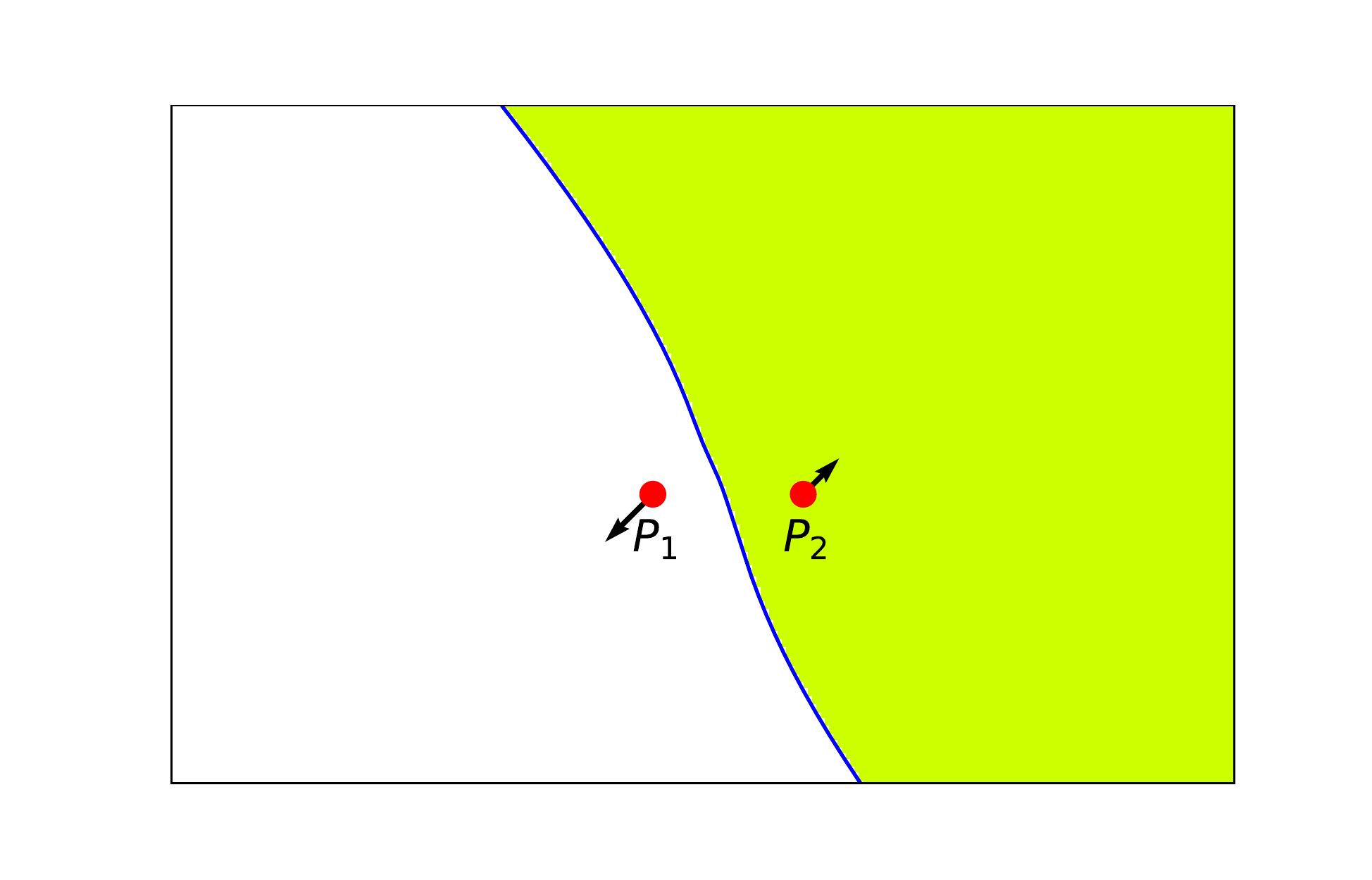}}
    \subfigure[\parbox{5cm}{$v_1 = 8\, \text{m/s}, v_2 = 3\, \text{m/s},\\ \alpha_1 = 0^{\circ}, \alpha_2 = 180^{\circ}$.}]%
                   {\includegraphics[width=6cm]{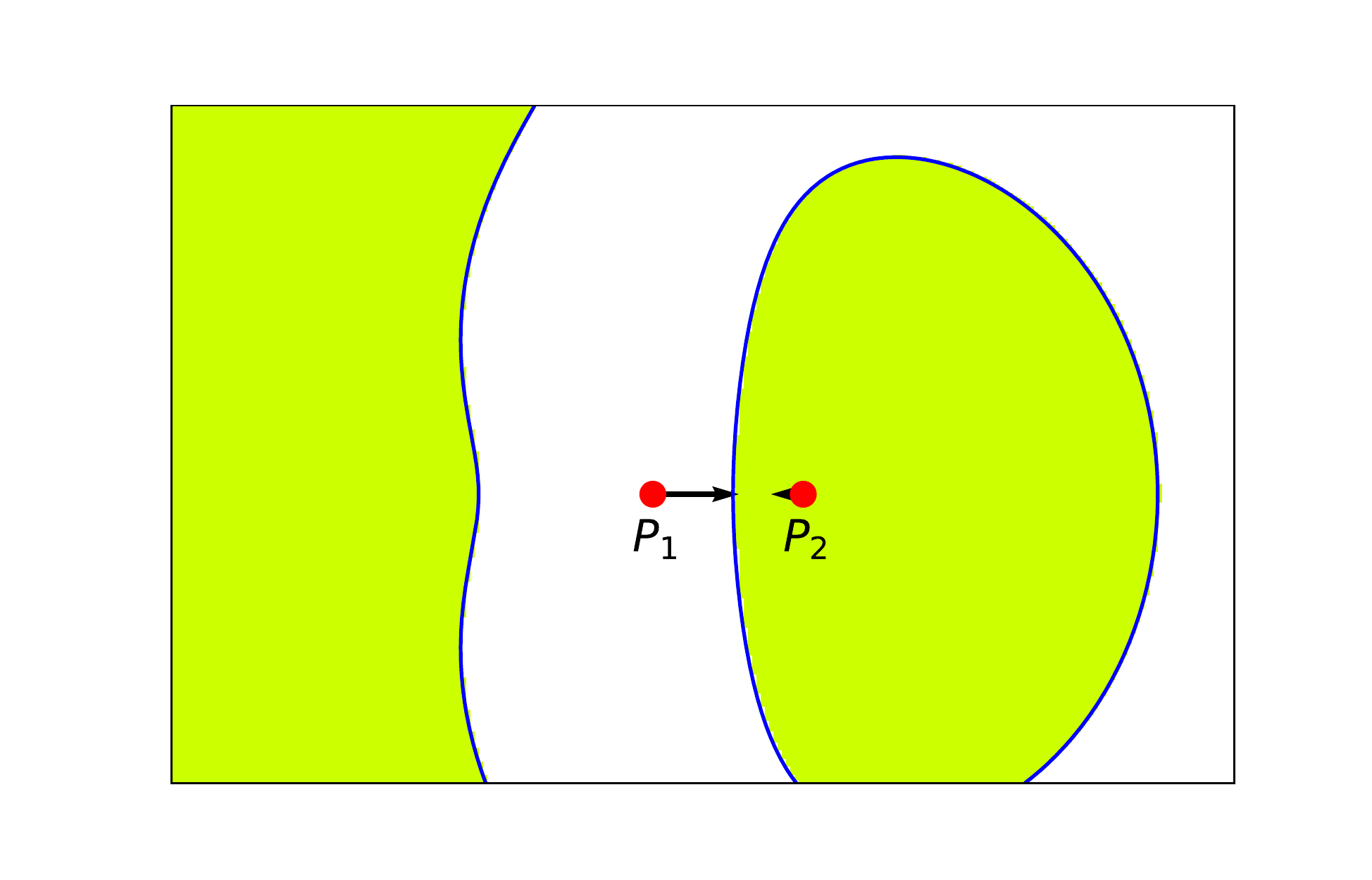}}\\
    \subfigure[\parbox{5cm}{$v_1 \sim 9\, \text{m/s}, v_2 = 0\, \text{m/s},\\ \alpha_1 = 45^{\circ}, \alpha_2 = -50^{\circ}$.}]%
                   {\includegraphics[width=6cm]{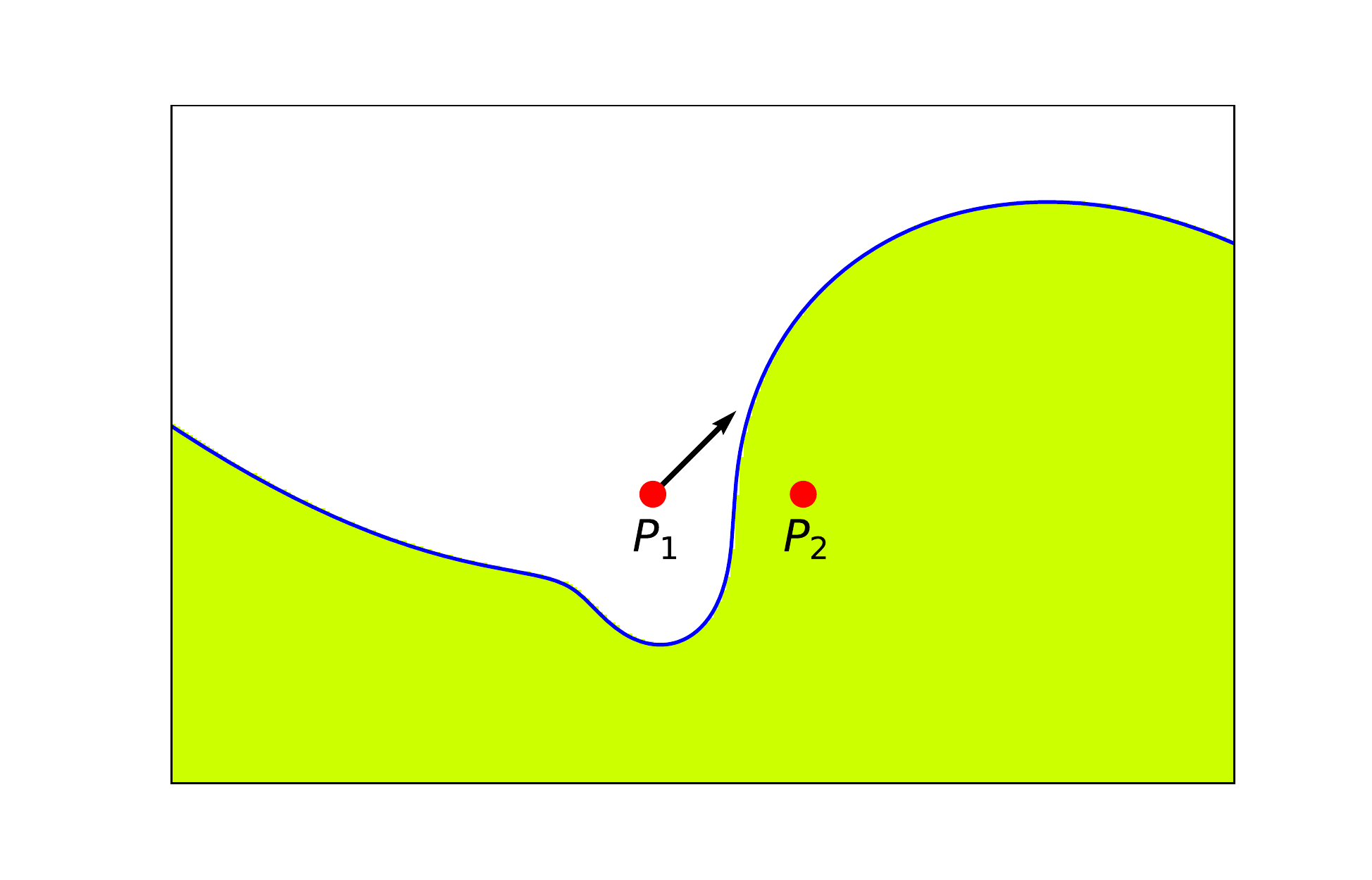}}
    \subfigure[\parbox{5cm}{$v_1 \sim 9\,\text{m/s}, v_2 \sim 9\,\text{m/s},\\ \alpha_1 = 90^{\circ}, \alpha_2 = 90^{\circ}$.}]%
                   {\includegraphics[width=6cm]{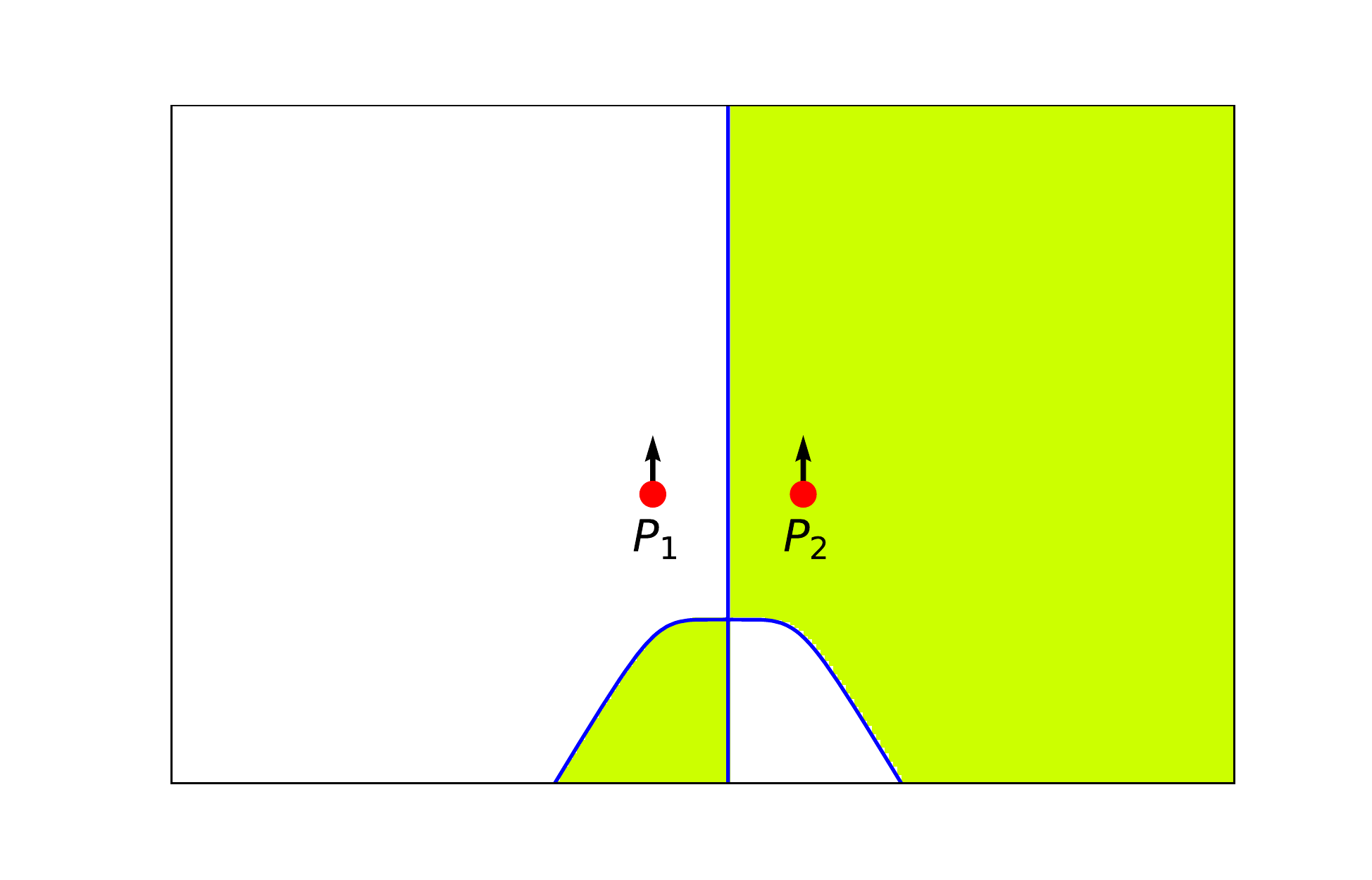}}\\
    \subfigure[\parbox{5cm}{$v_1 \sim 9\, \text{m/s}, v_2 \sim 9\, \text{m/s},\\  \alpha_1 = 45^{\circ}, \alpha_2 = 135^{\circ}$.}]%
                    {\includegraphics[width=6cm]{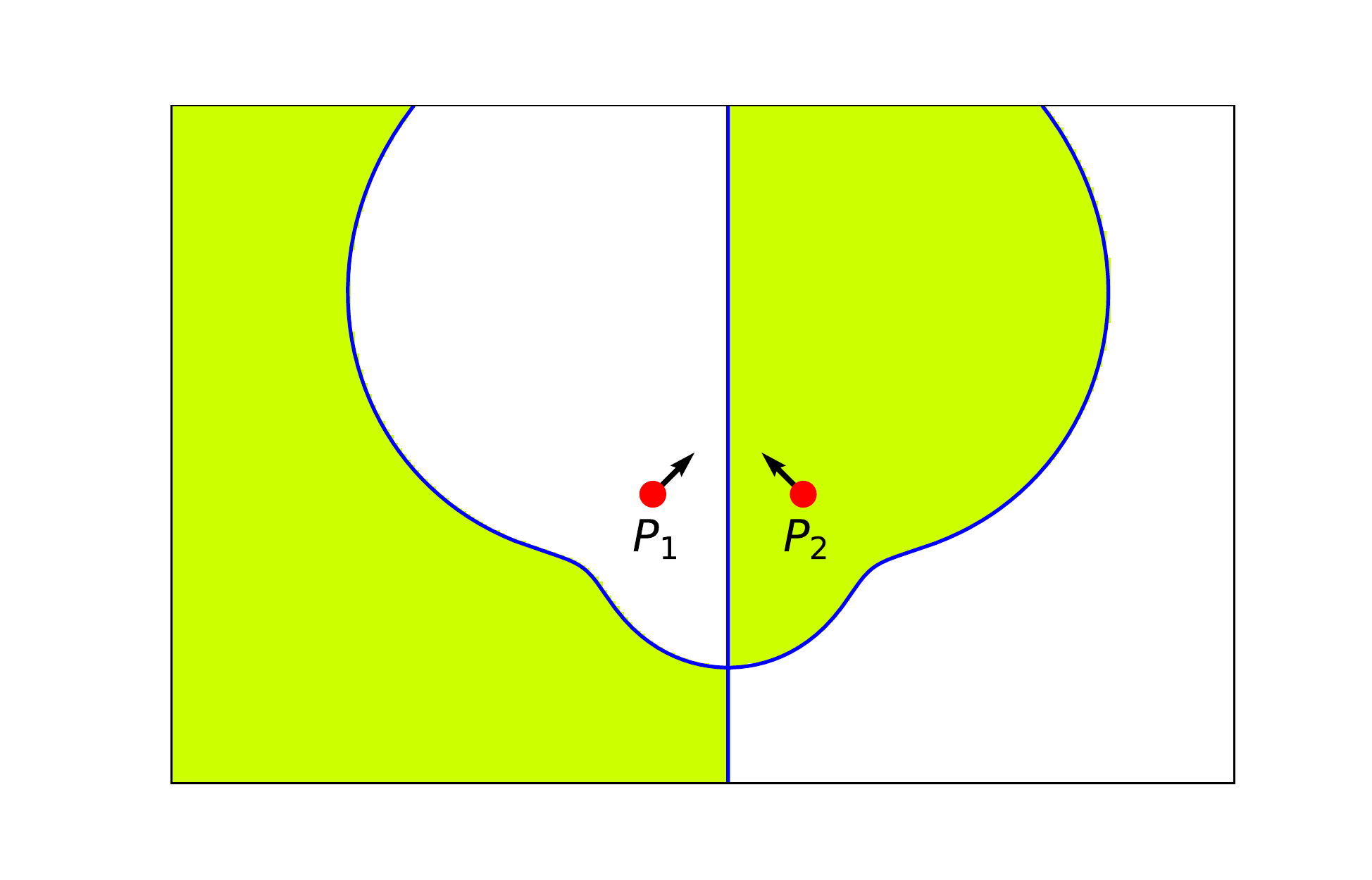}}
    \subfigure[\parbox{5cm}{$v_1 \sim 9\, \text{m/s}, v_2 \sim 9\, \text{m/s},\\ \alpha_1 = 90^{\circ}, \alpha_2 = 180^{\circ}$.}]%
                   {\includegraphics[width=6cm]{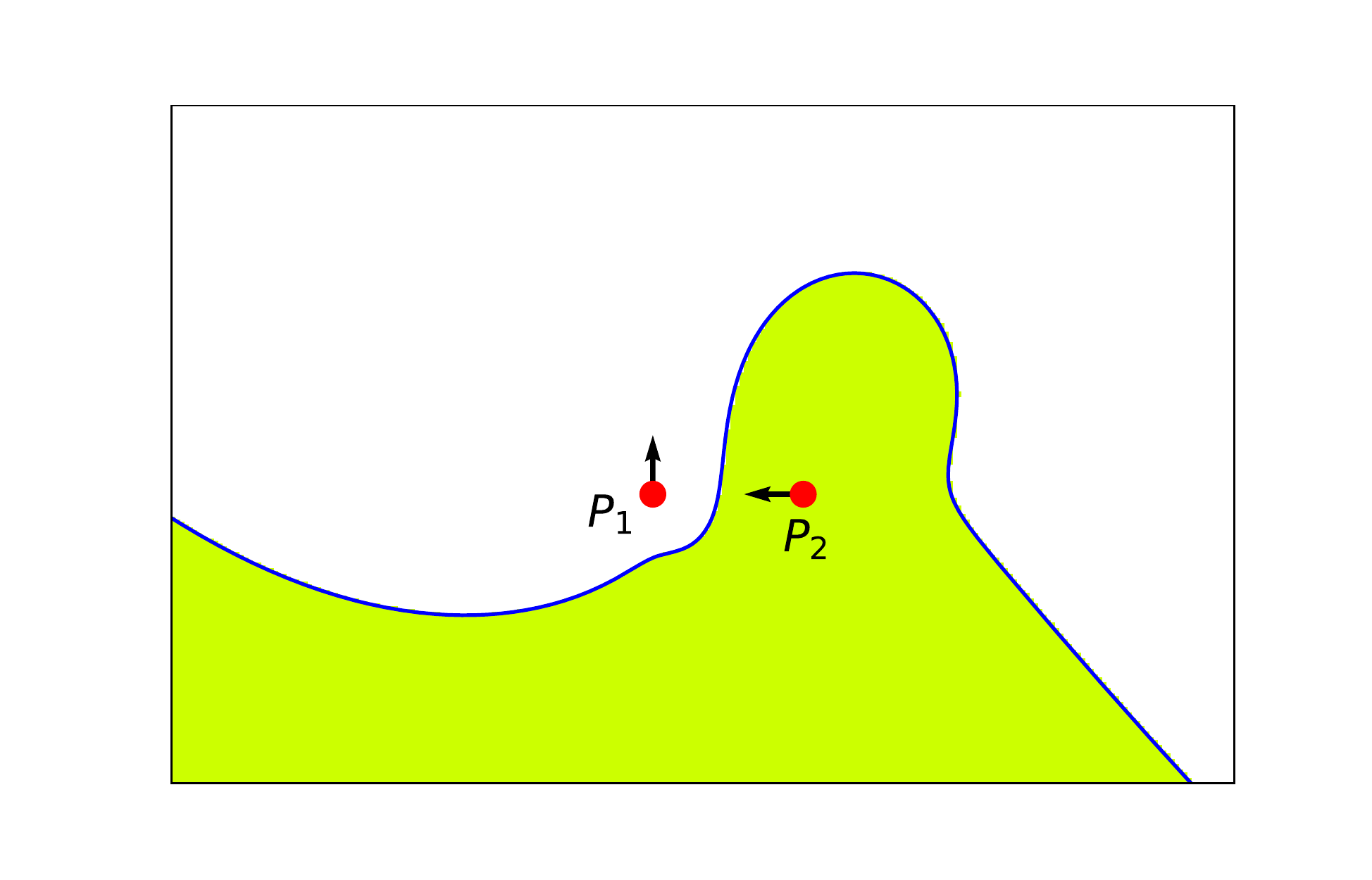}}\\
    \subfigure[\parbox{5cm}{$v_1 = 6\, \text{m/s}, v_2 = 6\, \text{m/s},\\ \alpha_1 = -45^{\circ}, \alpha_2 = 135^{\circ}$.}]%
                   {\includegraphics[width=6cm]{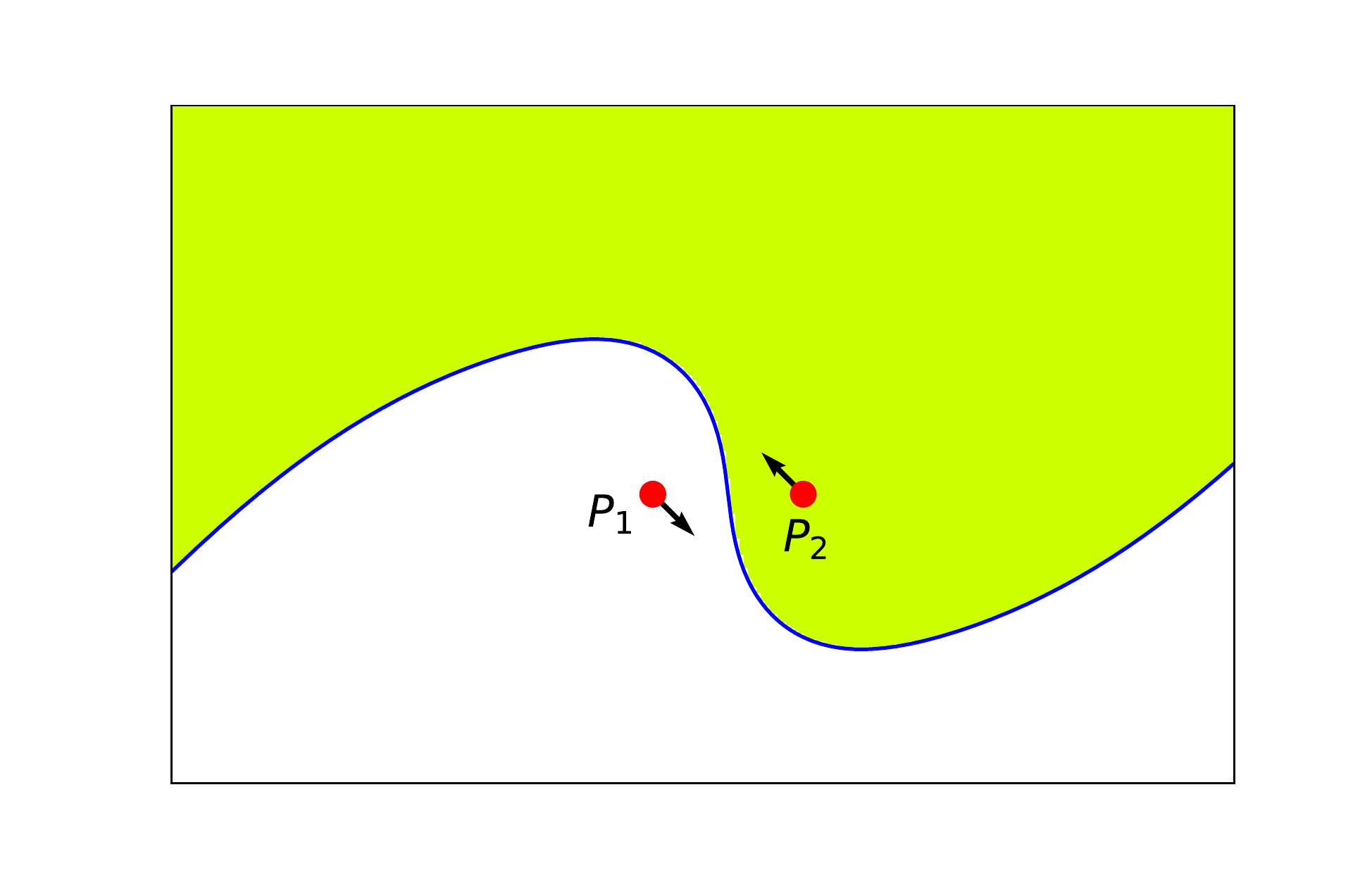}}
    \subfigure[\parbox{5cm}{$v_1=3\,\text{m/s}, v_2=8\,\text{m/s},\\ \alpha_1 = -45^{\circ}, \alpha_2 = -45^{\circ}$.}]%
                   {\includegraphics[width=6cm]{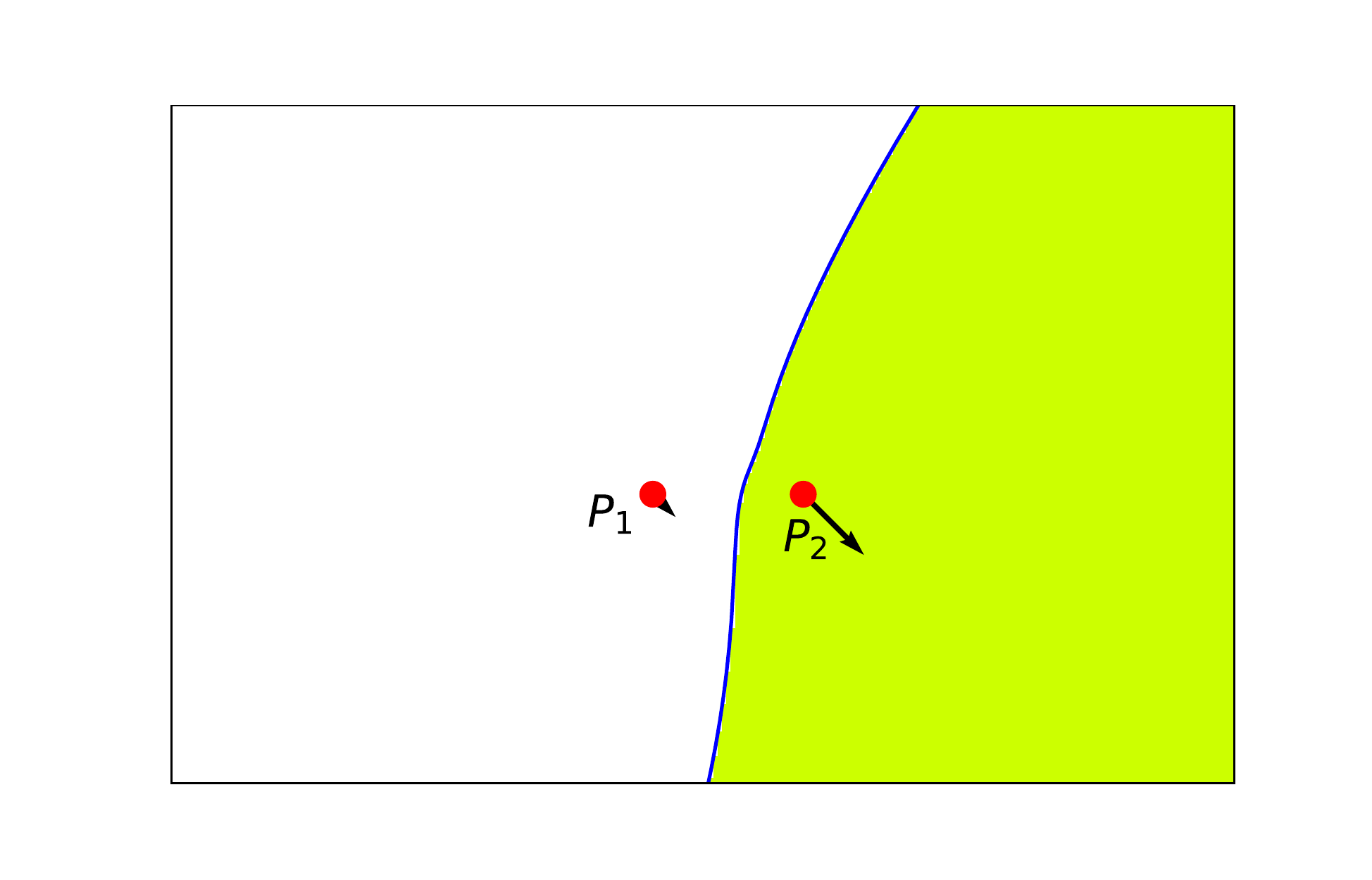}}
\caption{\footnotesize Sketch of the boundary between two players in the presence of directional bias, air drag,  and internal dissipation  on a
                212  meters by 136 meters pitch.  The bottom left corner and the upper right corner have coordinates $(-106,-68)$ and (106,68)
                 respectively.  The two players P$_1$ and P$_2$ are located at the points $(-10, -10)$ and $(20, -10)$. We notice many of the same features from Figure \ref{fig:BifocalDirectionalBias}, but the players still have some control over areas behind themselves and minor alterations to their control in other directions.}
\label{fig:BifocalAll}
\end{figure}}

Next, we plot the boundaries between all twenty-two players from two random frames and record our observations, see Figure~\ref{fig:directionalBiasVoronoi}. It is immediately clear that players control more space in the direction of their velocity. Further, we see that this asymmetry is stronger for players with high initial speeds. We also see that the boundaries between any two players are far more complex and diverse than previous models.
%
{\begin{figure}[h!]
    \centering
    \subfigure[Frame: 98202]{\includegraphics[width=12cm]{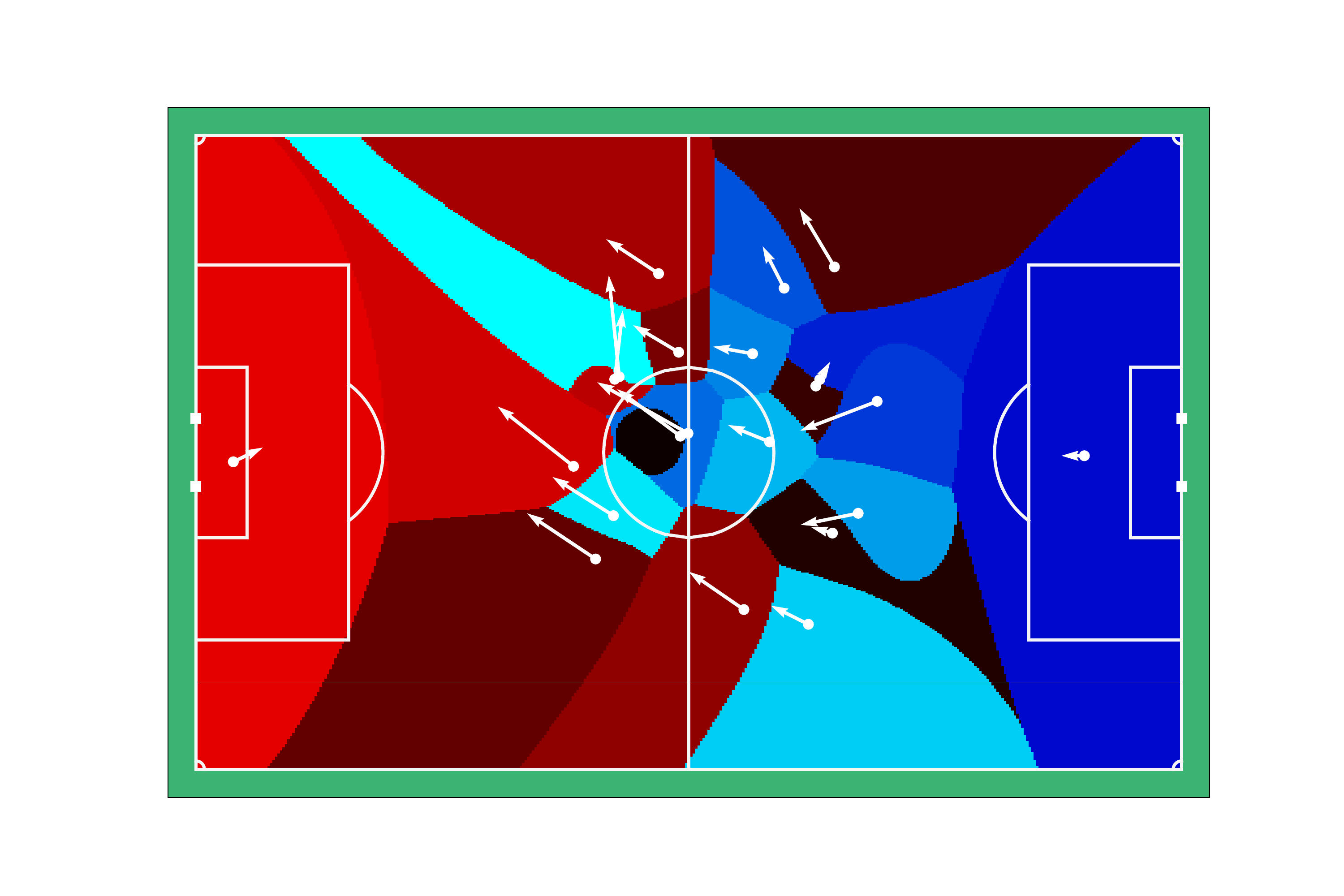}}\\
    \subfigure[Frame: 123000]{\includegraphics[width=12cm]{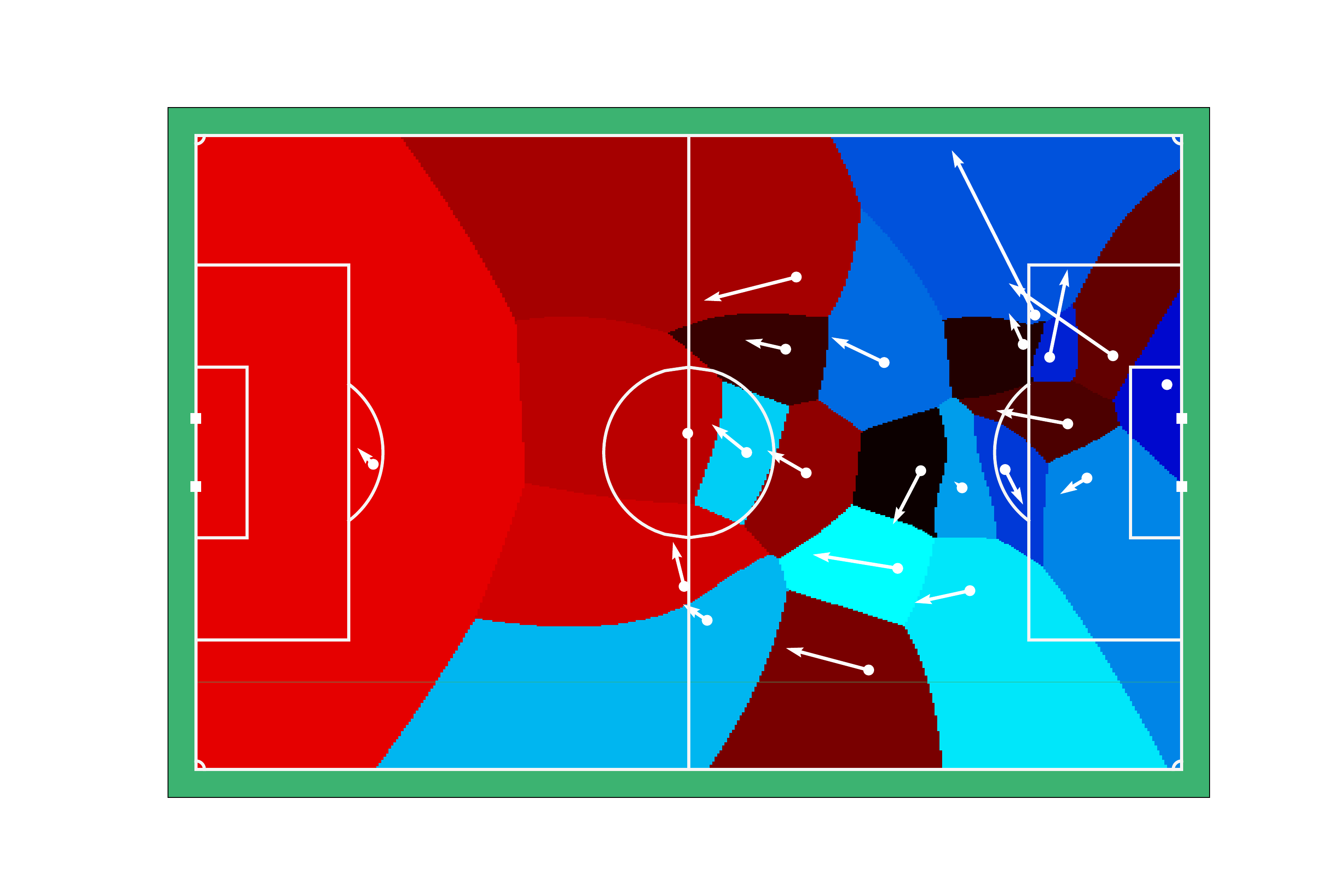}}
\caption{\footnotesize Sketches of Apollonius diagrams in the presence of directional bias for two frames of the
              Metrica Sports open tracking data.}
\label{fig:directionalBiasVoronoi}
\end{figure}}

\subsection*{\normalsize The Boundary in the Presence of Air Drag, Internal Dissipation and Directional Bias}

We can now combine the two models in a unified one. In this case, equation \eqref{eq:15} remains valid
\begin{equation*}
     r =\Bigg[ {u_\text{limit}  \, \ln\left[ \cosh^2\delta \, \Big(1-{U^2\over u_\text{limit}^2}\Big) \right]  \over 2(\delta - \Delta) }
          - \Gamma \Bigg] (t-t_0) ,
\end{equation*}
but
$$
    \delta = \tanh^{-1} {v_0\, \cos\theta + \Gamma \over u_\text{limit} } .
$$
Therefore the borderline is determined by
\begin{equation*}
     k_1 \, r_1 -k_2 \,  r_2  = t_2-t_1 ,
\end{equation*}
with  the players' slownesses being
\begin{equation}
       k_i =  \Bigg[ {u_\text{limit}  \, \ln\left[ \cosh^2\delta \, \Big(1-{U^2\over u_\text{limit}^2}\Big) \right]  \over 2(\delta - \Delta) }
                 - \Gamma \Bigg] ^{-1},        \quad i=1,2 .
\label{eq:40}
\end{equation}
To get a feeling of the result, we again plot the boundaries between two players in the same eight 2-player scenarios --- see  Figure \ref{fig:BifocalAll}.

 Finally, we plot the boundaries between all twenty-two players for two Metrica Sports frames in Figure \ref{fig:twin_direction}.
 %
{\begin{figure}[h!]
    \centering
    \subfigure[Frame: 98202]{\includegraphics[width=12cm]{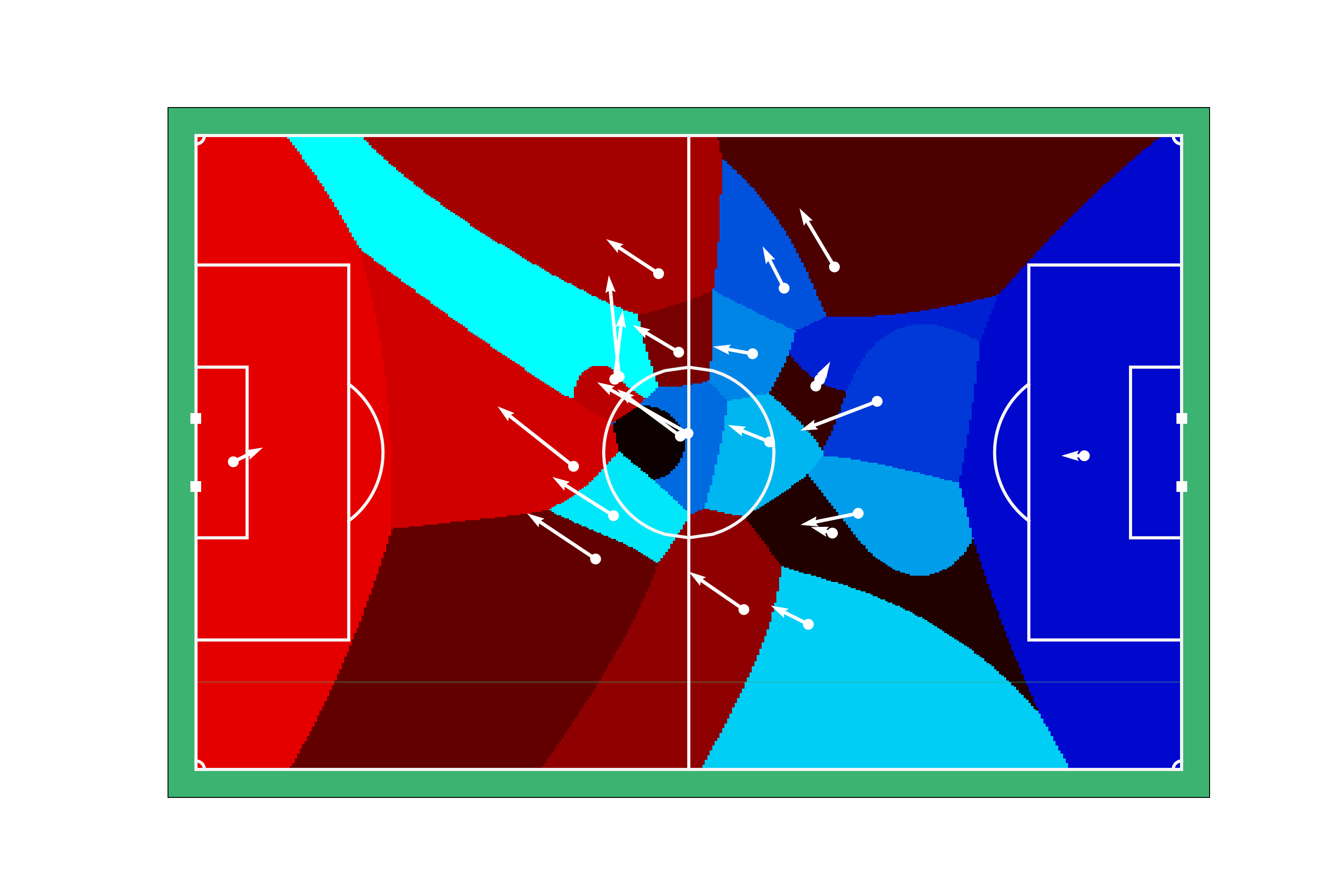}}\\
    \subfigure[Frame: 123000]{\includegraphics[width=12cm]{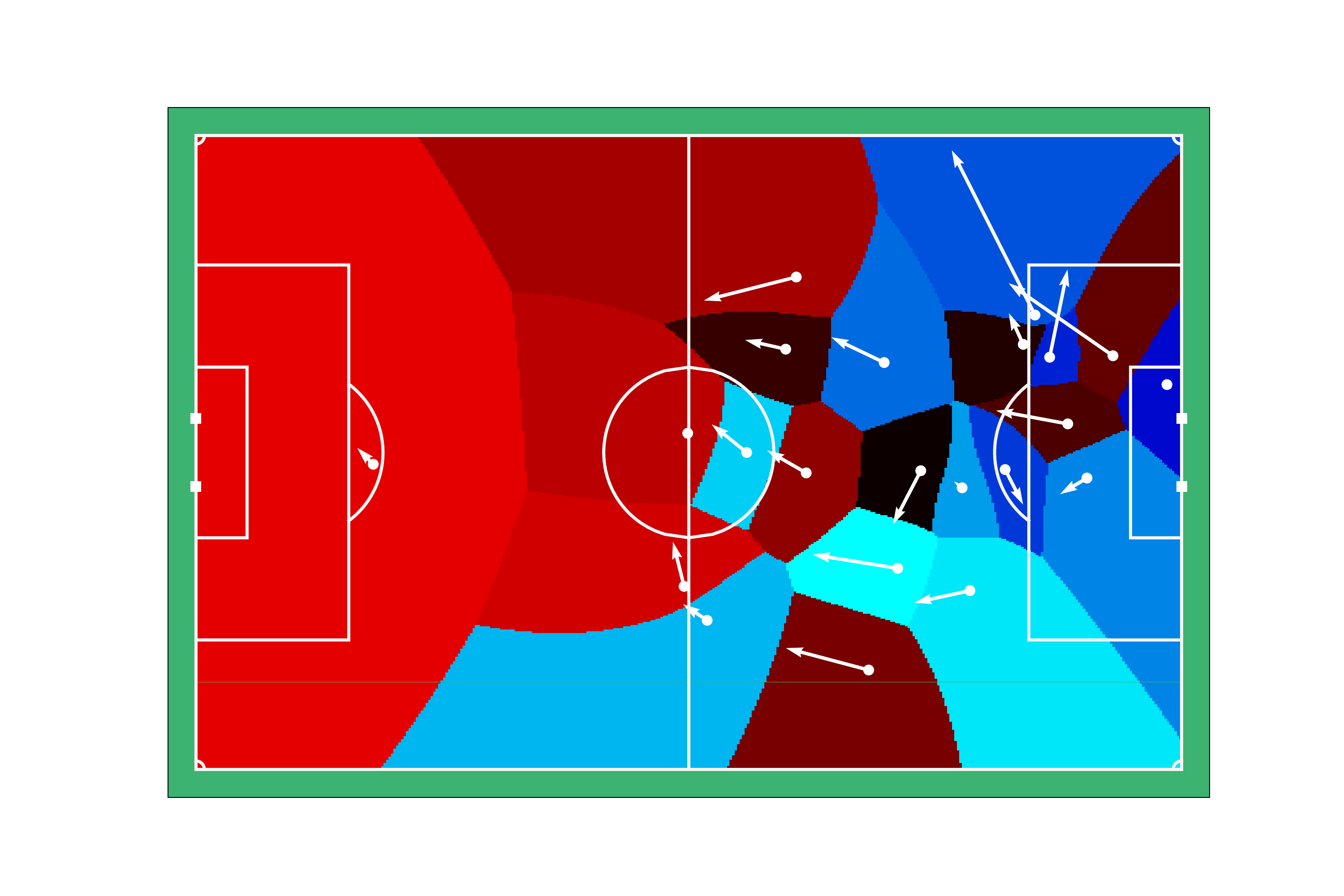}}
    \caption{\footnotesize Sketches of Apollonius diagrams in the presence of directional bias, air drag, and internal friction for two frames of the Metrica Sports open tracking data.}
\label{fig:twin_direction}
\end{figure}}

\section{Conclusions and Discussion}

Our work in this article has focused on building models of dominance regions for soccer players in a match incorporating the effects of directional bias and frictional forces, building on the deterministic models introduced in \cite{efthimiou}.
The concept of asymmetric influence along different directions in the dominance region of the players was based on a reasonable kinematical assumption. Similarly, the introduction of the frictional forces were not arbitrary; they are well known forces that have been verified by biokinetics researchers who have studied sprinting in athletes (\cite{Furu}). Our most polished model combines all effects.

The dominance areas show interesting features not encountered (and, most importantly not expected) in the standard Voronoi diagrams.
In fact, starting with the result of \cite{efthimiou}, the dominance areas between any two players are separated by an
Apollonius circle\footnote{A perpendicular bisector is a special case of an Apollonius circle.} whose
size depends on the ratio of the average speeds  $A_1/A_2$ of the two players, assuming that the players have a negligible difference in reaction times.  When we introduce air drag and internal resistance, the boundaries between players remain Apollonius circles but the radius of each circle resizes since the factors $A$ have a different functional  form compared to the original case.

In general, these models predict that, given that two players have the same characteristic speed,  the player with the greater initial speed is favored. This coincides with the widely accepted idea that it is never a good idea for soccer players to stand still or to be walking. Following the rhythm of the game is a good plan. When players get exhausted, substitutions are warranted.

For the models including frictional forces, the results were obtained under some simplifying assumptions for the acceleration $a$, the drag coefficient $k$ and the internal dissipation constant $\gamma$.  Although there are articles discussing sprinting of soccer players, (see, for example, \cite{Gissis}, \cite{Salvo}, \cite{da Silva}) we have not been able to find precise data in the literature which would allow us to accurately compute the constants $a, k,$ and $\gamma$. So, at the moment the precise behavior of the constants $a, k,$ and $\gamma$ remains an open problem to be settled in the future. Settling it is a relatively easy problem and it can provide hints for modeling several different aspects of the soccer players.

When directional bias is added to the model, the results get modified substantially. Imagine a sprinter who, as they approach the finish line at full speed,  is asked to turn back and reach one of the previous points. Obviously, it is impossible for this to happen fast; the sprinter must decelerate first and then turn back. Keeping this in mind, it is easily understood that when directional bias is added to the model (which already includes the frictional forces), the slower player can reclaim many points in their dominance area. Hence, although it is always a good plan to follow the rhythm of the game, the direction of motion is also of paramount importance and it can be used by either player to their advantage. In other words, there should be a new parameter that should `enter' into the model (and every soccer model): the `Soccer IQ' of the players --- how fast they think and process the visual and auditory data to optimize their motion and be ready to reach vital points on the pitch before opposing players are able to. But that's a different problem and not the goal of this article.

Currently, our model is primarily an observational tool:  It measures/evaluates instantaneous moments of a match as given by the tracking data. However, it is easy to extrapolate that further models can be constructed with it as a foundation which predict or suggest various strategies during a match. We hope to present such works in the near future.

Finally, we close this article by returning to comment furthermore on the simplifying assumptions that we have made.
Although the resulting diagrams are more precise than the widely used standard Voronoi diagrams or other suggestions based on arbitrary assumptions, from a perfectionist's point a view a simplifying assumption is never a satisfying action. Our interest is to continue streamlining these models by using modern human biokinetics research to remove more simplifying assumptions. We are currently working on this and will soon be able to report on the progress.



\begin{thebibliography}{10}

\bibitem{efthimiou}
C. J. Efthimiou, The Voronoi Diagram in Soccer: a theoretical study to measure dominance space, \url{https://arxiv.org/abs/2107.05714}.

\bibitem{TH}
T. Taki and J. Hasegawa,
Visualization of dominance region in team games and its application to teamwork analysis,
Proceedings Computer Graphics International 2000, pp. 227--235,
\url{10.1109/CGI.2000.852338}.

\bibitem{Hill}
 A. V. Hill, The Air-Resistance to a Runner, Proc. R. Soc., \textbf{B102} (1927), pp. 380--385,
\url{https://doi.org/10.1098/rspb.1928.0012}.

\bibitem{Furu}
 K. Furusawa, V. Hill, and J. L. Parkinson, The dynamics of  ``sprint" running,
 Proc. R. Soc., \textbf{B102} (1927), pp. 29--42,
 \url{https://doi.org/10.1098/rspb.1927.0035}.

\bibitem{Kim}
S. Kim, Voronoi Analysis of Soccer Game, Nonlinear Analysis: Modelling and Control, \textbf{9(3)} (2004), 233-240,
\url{10.15388/NA.2004.9.3.15154}.

\bibitem{Fujimura}
A. Fujimura and K. Sugihara, Geometric Analysis and Quantitative Evaluation of Sport Teamwork, Systems and Computers, \textbf{36} (2005), 49--58,
\url{https://doi.org/10.1002/scj.20254}

\bibitem{Sofia}
S. Fonseca, J. Milho, B. Travassos, D. Araujo, Spatial dynamics of team sports exposed by Voronoi diagrams, Human movement science, \textbf{31(6)} (2012) 1652--1659,
\url{https://doi.org/10.1016/j.humov.2012.04.006}.

\bibitem{Ryu}
J. Ryu, S. Yoon, S. Park, T. Kim, S. Yoo, G. Lee, H. Koyama, T. Mochida, Sprinting speed of elite sprinters at the world championships, in ISBS-Conference Proceedings Archive, (2012).

\bibitem{Metrica}
Metrica Sports open tracking data, \url{https://github.com/metrica-sports/sample-data}.

\bibitem{Djaoui}
L. Djaoui, K. Chamari, A. Owen, A. Dellal, Maximal Sprinting Speed of Elite Soccer Players During Training and Matches, Journal of Strength and Conditioning Research \textbf{31} (2017)  1509--1517.
\url{https://doi.org/10.1519/JSC.0000000000001642}.

\bibitem{Gissis}
I. Gissis, C. Papadopoulos, V. I. Kalapotharakos, A. Sotiropoulos, G. Komsis, E. Manolopoulos, Strength and Speed Characteristics of Elite, Subelite, and Recreational Young Soccer Players, Research in Sports Medicine, 14:3, (2006) 205-214,
\url{10.1080/15438620600854769}.

\bibitem{Salvo}
V. Salvo, R. Baron, C. Gonz\'{a}lez-Haro, C. Gormasz, F. Pigozzi, N. Bachl, Sprinting analysis of elite soccer players during European Champions League and UEFA Cup matches, Journal of Sports Sciences, 28:14, (2010) 1489-1494, 
\url{10.1080/02640414.2010.521166}.

\bibitem{da Silva}
J. da Silva, L. Guglielmo, D. Bishop, Relationship Between Different Measures of Aerobic Fitness and Repeated-Sprint Ability in Elite Soccer Players, Journal of Strength and Conditioning Research, 24:8, (2010) 2115-2121,
\url{10.1519/JSC.0b013e3181e34794}



\end{thebibliography}
\end{document}